%% file: HVP_Paper.tex
\pdfoutput=1
\documentclass[a4paper, 11pt, oneside]{article}
\usepackage[a4paper,inner=2cm,outer=2cm,top=3cm,bottom=2.5cm]{geometry}

\usepackage[tbtags]{amsmath}
\usepackage{amssymb}
\usepackage{amsthm}
\usepackage{dsfont}
\usepackage{slashed}
\usepackage{mathrsfs}
\usepackage[tbtags]{mathtools}
\usepackage{color}

\usepackage[export]{adjustbox}

\usepackage[english]{babel}

\usepackage[utf8x]{inputenc}

\usepackage{graphicx,epsfig}

\usepackage{setspace}

\usepackage{booktabs}

\usepackage{float}

\usepackage{dcolumn}

\usepackage{nextpage}

\usepackage{newclude}

\usepackage[small,bf]{caption}
\usepackage{subcaption}

\usepackage{url}

\usepackage[colorlinks=true,
	linkcolor=blue,
	citecolor=blue,
	urlcolor=blue,
	filecolor=black
]{hyperref}

\usepackage[T1]{fontenc}

\usepackage{cancel}

\usepackage{tabularx}

\usepackage{pifont}

\usepackage{etex}

\usepackage{appendix}

\usepackage[numbers,sort&compress]{natbib}

\usepackage{placeins}

\usepackage{ragged2e}

\usepackage{ifmtarg}

\usepackage{tikz}
\usetikzlibrary{shapes,arrows}

\usepackage{bchart}

\newcommand{\EL}{E\L{}}

\renewcommand{\O}{\mathcal{O}}

\renewcommand{\Im}{\text{Im}\,}

\newcommand{\<}{\langle}
\renewcommand{\>}{\rangle}

\newcommand{\mpi}{M_{\pi}}

\newcommand{\mw}{M_\omega}
\newcommand{\beq}{\begin{equation}}
\newcommand{\eeq}{\end{equation}}
\newcommand{\GeV}{\,\text{GeV}}
\newcommand{\MeV}{\,\text{MeV}}
\newcommand{\fm}{\,\text{fm}}

\newcommand{\dilog}{\mathrm{Li}_2}

\newcommand{\remark}[1]{}

\makeatletter
\newcommand{\Cr}[2]{\@ifmtarg{#2}{\mathcal{C}_{#1}}{\mathcal{C}_{#1}\big[#2\big]}}
\makeatother

\newcolumntype{.}{D{.}{.}{2} } 
\newcolumntype{d}{D{.}{.}{2.2} }
\newcolumntype{L}[1]{>{\RaggedRight\hspace{0pt}}p{#1}}
\newcolumntype{R}[1]{>{\RaggedLeft\hspace{0pt}}p{#1}}

\makeatletter
  \def\my@tag@font{\normalsize}
  \def\maketag@@@#1{\hbox{\m@th\normalfont\my@tag@font#1}}
  \let\amsmath@eqref\eqref
  \renewcommand\eqref[1]{{\let\my@tag@font\relax\amsmath@eqref{#1}}}
\makeatother

\makeatletter
\renewcommand\paragraph{\@startsection{paragraph}{4}{\z@}%
  {-3.25ex \@plus -1ex \@minus -0.2ex}%
  {0.01pt}%
  {\bfseries}%
}
\makeatother

\makeatletter
\def\@xfootnote[#1]{%
  \protected@xdef\@thefnmark{#1}%
  \@footnotemark\@footnotetext}
\makeatother

\usepackage{abstract}

\allowdisplaybreaks[1]

\begin{document}

\mbox{}

\vspace{-1.75cm}
\hfill{}\begin{minipage}[t][0cm][t]{5cm}
\raggedleft
\footnotesize
INT-PUB-18-048
\end{minipage}
\vspace{1.25cm}

\bigskip

\begin{center}
{\LARGE{\bf \boldmath Two-pion contribution to hadronic vacuum polarization}}

\vspace{0.5cm}

Gilberto Colangelo${}^a$, Martin Hoferichter${}^{b}$, Peter Stoffer${}^{c}$

\vspace{1em}

\begin{center}
\it
${}^a$Albert Einstein Center for Fundamental Physics, Institute for Theoretical Physics, \\
University of Bern, Sidlerstrasse~5, 3012 Bern, Switzerland \\
\mbox{} \\
${}^b$Institute for Nuclear Theory, University of Washington, Seattle, WA 98195-1550, USA \\
\mbox{}\\
${}^c$Department of Physics, University of California at San Diego, La Jolla, CA 92093, USA
\end{center} 

\end{center}

\vspace{3em}

\hrule

\begin{abstract}
  We present a detailed analysis of $e^+e^-\to\pi^+\pi^-$ data up to
  $\sqrt{s}=1\GeV$ in the framework of dispersion relations. Starting from
  a family of $\pi\pi$ $P$-wave phase shifts, as derived from a previous
  Roy-equation analysis of $\pi\pi$ scattering, we write down an extended
  Omn\`es representation of the pion vector form factor in terms of a few
  free parameters and study to which extent the modern high-statistics data
  sets can be described by the resulting fit function that follows
  from general principles of QCD. We find that statistically acceptable
  fits do become possible as soon as potential uncertainties in the energy
  calibration are taken into account, providing a strong cross check on the
  internal consistency of the data sets, but preferring a mass of the $\omega$ meson
  significantly lower than the current PDG average.
  In addition to a complete
  treatment of statistical and systematic errors propagated from the data,
  we perform a comprehensive analysis of the systematic errors in the
  dispersive representation and derive the consequences for the two-pion
  contribution to hadronic vacuum polarization. In a global fit to both
  time- and space-like data sets we find $a_\mu^{\pi\pi}|_{\leq
    1\GeV}=495.0(1.5)(2.1)\times 10^{-10}$ and $a_\mu^{\pi\pi}|_{\leq
    0.63\GeV}=132.8(0.4)(1.0)\times 10^{-10}$. While the constraints are
  thus most stringent for low energies, we obtain uncertainty estimates
  throughout the whole energy range that should prove valuable in
  corroborating the corresponding contribution to the anomalous magnetic
  moment of the muon. As side products, we obtain improved constraints on
  the $\pi\pi$ $P$-wave, valuable input for future global analyses of
  low-energy $\pi\pi$ scattering, as well as a determination of the pion
  charge radius, $\langle r_\pi^2 \rangle = 0.429(1)(4)\fm^2$.
\end{abstract}

\hrule


\setcounter{tocdepth}{3}
\tableofcontents

\numberwithin{equation}{section}

	\input{sections/Introduction}

	\input{sections/Method}

	\input{sections/Input}

	\input{sections/Fits}

	\input{sections/amu}

	\input{sections/PhaseShifts}

	\input{sections/ChargeRadius}

	\input{sections/Conclusion}

\section*{Acknowledgements}
\addcontentsline{toc}{section}{Acknowledgements}

We thank J.~Bijnens, H.~Czy{\.z}, M.~Davier, A.~Denig, S.~Eidelman, C.~Hanhart, S.~Holz,
A.~Keshavarzi, B.~Kubis, A.~Kup\'s\'c, B.~Malaescu, S.~E.~M\"uller, M.~Procura, C.~Redmer, 
T.~Teubner, G.~Venanzoni, and Z.~Zhang for useful discussions.
Work on this approach had started many years ago in collaboration with
H.~Leutwyler and C.~Smith and was then put aside, but it formed the basis
on which we built the present analysis---we gratefully acknowledge their
contribution.  Financial support by the DOE (Grants No.\ DE-FG02-00ER41132
and DE-SC0009919) and the Swiss National Science Foundation is gratefully
acknowledged.  P.\ S.\ is supported by a grant of the Swiss National
Science Foundation (Project No.\ P300P2\_167751).  G.\ C.\ and P.\ S.\
thank the INT at the University of Washington for its hospitality and the DOE for partial support during
the completion of this work.

\begin{appendices}

\end{appendices}

\renewcommand\bibname{References}
\renewcommand{\bibfont}{\raggedright}
\bibliographystyle{utphysmod}
\phantomsection
\addcontentsline{toc}{section}{References}
\bibliography{Literature}

\end{document}

%% file: sections/Introduction.tex

\section{Introduction}

$\pi\pi$ scattering is one of the simplest hadronic reactions that displays many key features of low-energy QCD~\cite{Colangelo:2001df}, most prominently approximate chiral symmetry, 
its spontaneous breaking, and the explicit breaking due to finite up- and down-quark masses. 
Accordingly, chiral symmetry severely constrains the low-energy scattering amplitude, which can be systematically analyzed in Chiral Perturbation Theory (ChPT)~\cite{Weinberg:1966kf,Weinberg:1978kz,Gasser:1983yg,Gasser:1984gg} and has been worked out up to two-loop order~\cite{Bijnens:1995yn}.
However, $\pi\pi$ scattering is not only unique because of its strong relation to chiral symmetry, but in addition exhibits further remarkable properties that extend beyond the low-energy region where the chiral expansion applies. Most notably, this includes the fact that the process is fully crossing symmetric and that the unitarity relation, up to center-of-mass energies of nearly $\sqrt{s}=1\GeV$,
is totally dominated again by $\pi\pi$ scattering. The resulting constraints from analyticity, unitarity, and crossing symmetry were first formulated systematically in the framework
of Roy equations~\cite{Roy:1971tc} and subsequently analyzed in great detail, ultimately leading to a very precise representation of the $\pi\pi$ phase shifts up to 
roughly $1\GeV$~\cite{Ananthanarayan:2000ht,GarciaMartin:2011cn,Caprini:2011ky}.
Both the methods used in determining these phase shifts and the actual results have had a profound impact on countless more complicated hadronic reactions and decays, such as
$\pi K$ scattering~\cite{Buettiker:2003pp,Pelaez:2018qny}, $\pi N$ scattering~\cite{Ditsche:2012fv,Hoferichter:2015hva}, $\eta\to 3\pi$~\cite{Anisovich:1996tx,Albaladejo:2017hhj,Colangelo:2018jxw}, $\eta'\to\eta\pi\pi$~\cite{Isken:2017dkw}, or $K_{\ell 4}$ decays~\cite{Colangelo:2015kha}. However, arguably the most immediate application concerns pion form factors and here especially the vector
form factor $F_\pi^V$, given that, in marked contrast to the scalar form factor~\cite{Donoghue:1990xh,Ananthanarayan:2004xy,Hoferichter:2012wf,Ropertz:2018stk}, the onset of inelastic corrections is relatively smooth. 

\begin{figure}[t]
	\centering
	\includegraphics[width=3.8cm]{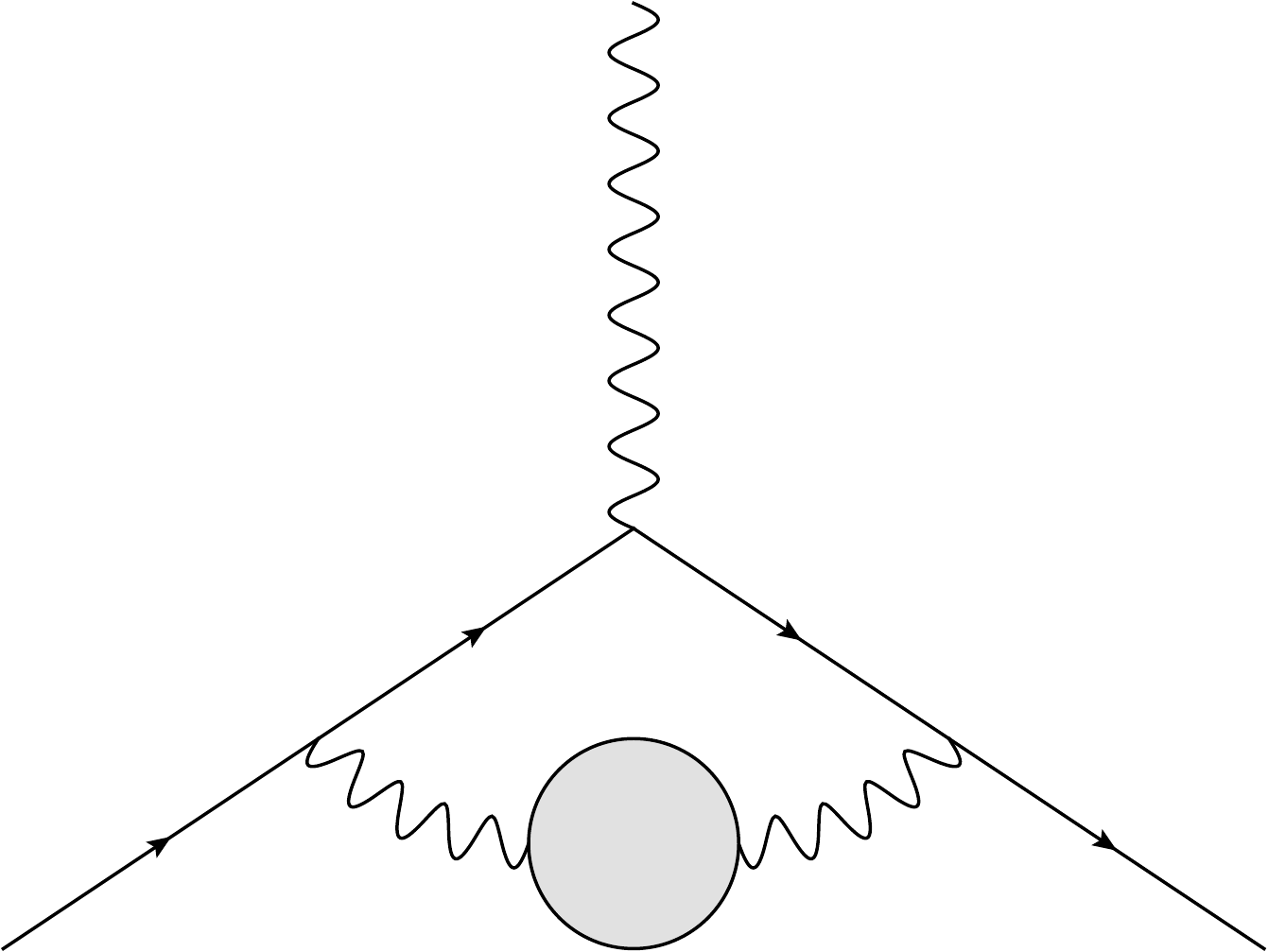}
	\caption{The topology of the leading hadronic contribution to the anomalous magnetic moment of the muon, hadronic vacuum polarization.}
	\label{img:HVP}
\end{figure}

Recent interest in the pion vector form factor (VFF) is mostly driven by the anomalous magnetic moment of the muon $a_\mu=(g-2)_\mu/2$. Its Standard-Model (SM) prediction continues to disagree with
the experimental measurement~\cite{Bennett:2006fi} (corrected for the muon--proton magnetic moment ratio~\cite{Mohr:2015ccw})
\beq
a_\mu^\text{exp}=116\ 592\ 089(63)\times 10^{-11}
\eeq
at the level of $3$--$4\sigma$ and upcoming experiments at Fermilab~\cite{Grange:2015fou} and J-PARC~\cite{Saito:2012zz} will scrutinize and improve upon this result (see also~\cite{Gorringe:2015cma}).
Meanwhile, the uncertainty in the SM value is dominated by hadronic corrections~\cite{Jegerlehner:2009ry,Prades:2009tw,Benayoun:2014tra}, wherein by far the largest individual contribution arises
from $\pi\pi$ intermediate states in hadronic vacuum polarization (HVP), see Fig.~\ref{img:HVP}. It is this contribution that is intimately linked to $F_\pi^V$ and $\pi\pi$ scattering~\cite{DeTroconiz:2001rip,Leutwyler:2002hm,Colangelo:2003yw,deTroconiz:2004yzs}. Similar representations have been used more recently~\cite{Ananthanarayan:2013zua,Ananthanarayan:2016mns,Hoferichter:2016duk,Hanhart:2016pcd}, in particular in the context of our work on a 
dispersive approach to hadronic light-by-light (HLbL) scattering~\cite{Hoferichter:2013ama,Colangelo:2014dfa,Colangelo:2014pva,Colangelo:2015ama,Colangelo:2017qdm,Colangelo:2017fiz},
where the space-like form factor determines the pion-box contribution. Further, $\pi\pi$ scattering plays a crucial role in many hadronic quantities that enter HLbL scattering, e.g.\ in $\gamma^*\gamma^*\to\pi\pi$~\cite{GarciaMartin:2010cw,Hoferichter:2011wk,Moussallam:2013una,Dai:2014zta} or the $\pi^0$~\cite{Niecknig:2012sj,Schneider:2012ez,Hoferichter:2012pm,Hoferichter:2014vra,Hoferichter:2017ftn,Hoferichter:2018dmo,Hoferichter:2018kwz}
and $\eta, \eta'$~\cite{Stollenwerk:2011zz,Hanhart:2013vba,Kubis:2015sga,Xiao:2015uva} transition form factors,
with recent extensions to the $\pi\eta$ system~\cite{Albaladejo:2015aca,Danilkin:2017lyn}. 

Since the early determinations~\cite{DeTroconiz:2001rip,Leutwyler:2002hm,Colangelo:2003yw,deTroconiz:2004yzs} the experimental situation in $e^+e^-\to\pi^+\pi^-$ 
has improved considerably~\cite{Akhmetshin:2001ig,Akhmetshin:2003zn,Achasov:2005rg,Achasov:2006vp,Akhmetshin:2006wh,Akhmetshin:2006bx,Ambrosino:2008aa,Aubert:2009ad,Ambrosino:2010bv,Lees:2012cj,Babusci:2012rp,Ablikim:2015orh,Anastasi:2017eio},
but at the same time the required precision of the HVP contribution to $a_\mu$ has increased further, in particular in view of the anticipated improvement of the experimental measurement of $a_\mu$ by a factor $4$ at the Fermilab experiment.  
In this way, a proper treatment of experimental errors and correlations is becoming absolutely critical. This includes radiative corrections, which need to be taken into account properly
in order to ensure a consistent counting of higher-order HVP iterations~\cite{Calmet:1976kd,Kurz:2014wya} (in principle, the same issue arises in HLbL scattering as well~\cite{Colangelo:2014qya}). 
Most current HVP compilations are based on a direct integration of the experimental data~\cite{Jegerlehner:2017lbd,Davier:2017zfy,Keshavarzi:2018mgv} (see~\cite{Benayoun:2015gxa} for an alternative approach using the hidden-local-symmetry model), wherein conflicting data sets are treated by a local $\chi^2$ inflation.\footnote{We concentrate on time-like approaches here, which are complementary to
efforts based on a space-like representation, as in lattice QCD~\cite{Chakraborty:2016mwy,DellaMorte:2017dyu,Borsanyi:2017zdw,Blum:2018mom,Giusti:2018mdh} or the MUonE proposal~\cite{Abbiendi:2016xup}.} 
The most consequential such tensions currently affect the 
BaBar~\cite{Aubert:2009ad,Lees:2012cj} and KLOE~\cite{Ambrosino:2008aa,Ambrosino:2010bv,Babusci:2012rp,Anastasi:2017eio} data sets for the $\pi\pi$ channel,
and different methods for their combination then give rise to the single largest difference between the HVP compilations of~\cite{Davier:2017zfy} and~\cite{Keshavarzi:2018mgv}. 

In this paper, we return to the description of the $\pi\pi$ contribution to HVP based on a dispersive representation of the VFF. We first clarify the role of radiative corrections, in particular vacuum polarization (VP), and then derive a global fit function that the form factor needs to follow to avoid conflicts with unitarity and analyticity. In addition to two free parameters in the $\pi\pi$ $P$-wave, this Omn\`es-type dispersion relation involves one parameter to account for $\rho$--$\omega$ mixing (plus the $\omega$ mass) and at least one additional parameter to describe inelastic corrections in a conformal expansion. First, we study to which extent the resulting representation can be fit to the modern high-statistics data sets, by using an unbiased fit strategy and including the full experimental covariance matrices where available, to provide a strong check of the internal consistency of each data set. As a second step, we address the systematic uncertainties in the dispersive representation 
and derive the HVP results for various energy intervals. 
Finally, we provide the resulting $\pi\pi$ $P$-wave phase shift and the pion charge radius that follow after determining the free parameters from the fit to the $e^+e^-\to\pi^+\pi^-$ cross section data.

%% file: sections/Method.tex

\section{Dispersive representation of the pion vector form factor}

In this section we review the formalism for a dispersive representation of the pion VFF at the level required for the interpretation of the modern high-statistics $e^+e^-\to\pi^+\pi^-$ data sets. This includes 
the definition of the pion VFF in QCD, the relation to HVP, and conventions regarding radiative corrections, see Sect.~\ref{sec:HVPRadiativeCorrections}; the actual dispersive representation 
including the description of the most important isospin-breaking effect from $\rho$--$\omega$ mixing as well as a term accounting for inelastic contributions, see Sect.~\ref{sec:OmnesRepresentation}; 
and a constraint on the size of these inelastic contributions, the Eidelman--\L{}ukaszuk bound, see Sect.~\ref{sec:ELBound}.

\subsection{Hadronic vacuum polarization and radiative corrections}
\label{sec:HVPRadiativeCorrections}

Hadronic contributions to the anomalous magnetic moment of the muon first arise at $\O(\alpha^2)$ in the expansion in the electromagnetic coupling $\alpha=e^2/(4\pi)$. The leading topology is HVP, shown in Fig.~\ref{img:HVP},
with hadronic input encoded in the QCD two-point function of electromagnetic currents
\beq
	\label{eq:TwoPointFunction}
	\Pi^{\mu\nu} = ie^2\int d^4x e^{iq\cdot x} \< 0 | T \{j^\mu_\mathrm{em}(x) j^\nu_\mathrm{em}(0)\} | 0 \> = (q^2 g^{\mu\nu} - q^\mu q^\nu ) \Pi(q^2),
\eeq
where the Lorentz decomposition follows from gauge invariance, the current is defined by\footnote{As usual in the context of $g-2$, we do not include $e$ in the definition of the current. However, we keep $\Pi^{\mu\nu}$ and $\Pi$ as quantities of $\O(e^2)$ by including the explicit factor of $e^2$ in~\eqref{eq:TwoPointFunction}.}
\beq
	j_\mathrm{em}^\mu := \bar q Q \gamma^\mu q , \quad q = ( u , d, s )^T , \quad Q = \mathrm{diag}\left(\frac{2}{3}, -\frac{1}{3}, -\frac{1}{3}\right),
\eeq
and the sign conventions have been chosen in such a way that the fine-structure constant evolves according to
\beq
	\label{alpha_running}
	\alpha(s)=\frac{\alpha}{1-\Pi^\mathrm{ren}(s)},\qquad \alpha:=\alpha(0), \qquad \Pi^\mathrm{ren}(s) = \Pi(s) - \Pi(0).
\eeq
The renormalized HVP function $\Pi(s)$ is analytic in the complex $s := q^2$ plane and satisfies a dispersion relation\footnote{For simplicity, in the following we drop the superscript ``ren.''}
\beq
	\Pi^\mathrm{ren}(s) = \frac{s}{\pi} \int_{s_\mathrm{thr}}^\infty ds' \frac{\Im \Pi(s')}{s'(s'-s)},
\eeq
where in pure QCD the integral starts at the two-pion threshold, $s_\mathrm{thr} = 4\mpi^2$. Unitarity relates the imaginary part of $\Pi(s)$ to the total hadronic $e^+e^-$ cross section
\beq
	\sigma(e^+e^- \to \mathrm{hadrons}) = \frac{\alpha}{s} \frac{4\pi}{\sigma_e(s)} \Big(  1 + \frac{2 m_e^2}{s} \Big) \Im \Pi(s),
\eeq
where $\sigma_\ell(s) = \sqrt{1-4m_\ell^2/s}$.
The HVP contribution to the anomalous magnetic moment of the muon can then be written as~\cite{Bouchiat:1961,Brodsky:1967sr}
\beq
	a_\mu = \Big( \frac{\alpha m_\mu}{3\pi} \Big)^2
        \int_{s_\mathrm{thr}}^\infty ds \frac{\hat K(s)}{s^2}
        R_\mathrm{had}(s),
\label{eq:amuHVP}
\eeq
where the kernel function is
\begin{align}
	\begin{split}
		\hat K(s) &= \frac{3s}{m_\mu^2} \bigg[ 
		 \frac{x^2}{2} (2-x^2) + \frac{(1+x^2)(1+x)^2}{x^2} \Big( \log(1+x) - x + \frac{x^2}{2} \Big) + \frac{1+x}{1-x} x^2 \log x \bigg] , \\
		x &= \frac{1-\sigma_\mu(s)}{1+\sigma_\mu(s)},
	\end{split}
\label{eq:Ks}
\end{align}
and $R_\mathrm{had}$ is related to the hadronic cross section by
\beq
	\label{eq:Rratio}
	R_\mathrm{had}(s) = \frac{3s}{4\pi\alpha^2} \frac{s\sigma_e(s)}{s+2m_e^2} \sigma(e^+e^-\to\mathrm{hadrons}).
\eeq
We stress that the usual $R$ ratio, defined as the ratio of hadronic to
muonic $e^+e^-$ cross sections, is not what enters the dispersive
representation of the HVP contribution: our representation
(\ref{eq:amuHVP}), with (\ref{eq:Ks}) and (\ref{eq:Rratio}) as input, is exact. $R_\mathrm{had}(s)$ and $R(s)$ coincide for a tree-level muonic cross section and in the limit $s\gg
m_\mu^2$, where of course also the electron mass does not play any role,
but for clarity, we have provided above the expression of the HVP
contribution without any approximations.

The contribution of the two-pion intermediate state can be expressed in terms of the pion VFF
\beq
	\< \pi^\pm(p') | j_\mathrm{em}^\mu(0) | \pi^\pm(p) \> =\pm (p'+p)^\mu F_\pi^V((p'-p)^2)
\eeq
according to
\beq
	\sigma(e^+e^-\to\pi^+\pi^-) = \frac{\pi \alpha^2}{3s} \sigma_\pi^3(s) \big| F_\pi^V(s) \big|^2 \frac{s+2m_e^2}{s \sigma_e(s)} \, , \quad \sigma_\pi(s) = \sqrt{1-\frac{4\mpi^2}{s}}.
\eeq

As becomes apparent already from~\eqref{alpha_running}, for a consistent counting of higher orders in $\alpha$ it is critical that radiative corrections be properly taken into account, otherwise this would induce corrections at the same order as HVP iterations or HLbL scattering. The prevalent convention is that the leading-order HVP include, 
in the sum over intermediate states in the unitarity relation, not only the hadronic channels but the (one-)photon-inclusive ones. 
In particular, the lowest-lying intermediate state is no longer the two-pion state, but the $\pi^0\gamma$ state, i.e.\ $s_\mathrm{thr} = M_{\pi^0}^2$.
In this way, the HVP input corresponds to infrared-finite photon-inclusive cross sections including both real and virtual corrections, but to avoid double counting
in next-to-leading-order iterations and beyond
each contribution is required to be one-particle-irreducible. This convention has important consequences for the definition of the pion VFF and the corresponding $e^+e^-\to\pi^+\pi^-$ cross section.
That is, the cross section to be inserted in~\eqref{eq:Rratio} has to be inclusive of final-state radiation (FSR), but both VP and initial-state-radiation (ISR) effects need to be subtracted. This defines the bare cross section
\begin{align}
\label{bare_cross_section}
	\begin{split}
		\sigma^{(0)}(e^+e^-\to\gamma^*\to\mathrm{hadrons}(\gamma)) &= \left|\frac{\alpha}{\alpha(s)}\right|^2 \sigma(e^+e^-\to\gamma^*\to\mathrm{hadrons}(\gamma)) \\
			&= \big| 1 - \Pi^\mathrm{SM}(s) \big|^2 \sigma(e^+e^-\to\gamma^*\to\mathrm{hadrons}(\gamma)),
	\end{split}
\end{align}
where the running of $\alpha$, see~\eqref{alpha_running}, is determined by the full renormalized VP function in the SM, e.g.\ including the lepton-loop contribution
\beq
	\Pi_\ell(s) = \frac{2\alpha}{\pi} \int_0^1 dx \, x(1-x) \log\Big[ 1 - x (1-x) \frac{s}{m_\ell^2} \Big].
\eeq
While by means of the above equations the subtraction of VP effects may be taken into account afterwards, the correction of ISR and ISR/FSR interference effects is 
performed with Monte Carlo generators in the context of each experiment~\cite{Rodrigo:2001kf,Czyz:2002np,Czyz:2003ue,Czyz:2004rj}.

Accordingly, the two-pion contribution should be understood as the
photon-inclusive two-pion channel. This, however, is not directly compatible
with our goal to treat the pion VFF dispersively, because
this is usually done in the isospin limit, i.e.\ with photon emission
switched off. 
In order to be able to apply our dispersive treatment of the
VFF, we therefore need to extract from the data on the
photon-inclusive process the cross section $\sigma(e^+ e^- \to \pi^+
\pi^-)$ in the isospin limit, i.e.\ with $m_u=m_d$ and
$\alpha=0$. While taking this limit is unproblematic at first sight, 
subtleties arise once one realizes that experiments exist only in our
isospin-broken world and that any input quantity has to be taken and defined
away from the isospin limit. Phrased differently, the actual
question one faces is whether it is possible to establish a procedure to
extract from the measured photon-inclusive cross section $\sigma(e^+ e^-
\to \pi^+ \pi^- (\gamma))$ the one in the isospin limit, where the VFF in
pure QCD appears. 

A similar question shows up also in other contexts. One case
that has been discussed in detail in the literature is the problem of the
extraction of the purely strong pion decay constant from the measurement of
the decay rate $\Gamma(\pi \to \mu \nu_\mu (\gamma))$. Strictly speaking, an
unambiguous and uniquely defined extraction is not possible: any practical
and operative definition of a purely strong decay constant extracted from
experiment is necessarily convention dependent. More precisely, it depends
on how one defines the strong isospin limit and on a matching scale, as
explained in detail in~\cite{Gasser:2003hk,Gasser:2010wz} from the perspective
of an effective-field-theory, perturbative approach, and
in~\cite{Carrasco:2015xwa} from a non-perturbative point of view. 
For the pion decay constant it has been shown~\cite{Gasser:2010wz} that the
scale dependence is very weak, mainly thanks to the smallness of $\alpha$
and the logarithmic dependence on the scale, and that the extraction of the
pion decay constant from experiment is indeed accurate at the claimed
accuracy, of course barring the choice of absurd values of the matching
scale. 

An example where the scale dependence in defining a purely strong quantity
cannot be neglected concerns proton--proton scattering. Here, the scale-independent photon-inclusive 
scattering length $a_{pp}^C=-7.8063(26)\fm$~\cite{Bergervoet:1988zz} differs significantly 
from the scale-dependent photon-subtracted one~\cite{Jackson:1950zz,Kong:1999sf}, 
depending on the choice of scale e.g.\ $a_{pp}=-17.3(4)\fm$~\cite{Miller:1990iz}.
In this case, the size of the effect is enhanced by the interference of the Coulomb interaction with the
short-distance part of the nuclear force, and virtual photons could only be
subtracted consistently everywhere, including the running of operators, if the
underlying theory were known~\cite{Gegelia:2003ta,Gasser:2010wz}.
This situation should be contrasted with perturbative systems, e.g.\
the extraction of the pion--nucleon scattering lengths from pionic atoms~\cite{Baru:2010xn,Baru:2011bw,Hoferichter:2012bz},
where in principle the same ambiguities related to the removal of QED effects appear,
but, without such an enhancement mechanism,   
the resulting scale dependence can be neglected at the level of the experimental accuracy. 

For the case of the $e^+e^- \to \pi^+ \pi^- (\gamma)$ cross section the
situation is completely analogous to that of $F_\pi$: in principle, the
purely strong VFF cannot be extracted in an unambiguous way from data, but one
may hope that a convention-dependent extraction (and corresponding
definition) of such a strong VFF only shows a very weak dependence on this 
arbitrariness and can be taken as a good approximation to a purely strong
VFF. The problem has been analyzed in the
literature mainly with the help of scalar QED and extensions thereof that include resonance
exchanges~\cite{Gluza:2002ui,Czyz:2003ue,Czyz:2004rj,Czyz:2004nq,Bystritskiy:2005ib,Ambrosino:2005wk,Actis:2010gg,Muller:2009pj}. 
In these models there is no ambiguity in the extraction of the
cross section $\sigma(e^+ e^- \to \pi^+ \pi^-)$, but this happens at the
price of losing model independence. In either case, these studies indicate that at the
present level of accuracy scalar QED describes reasonably well the behavior
of the observed FSR: the relation established within this model between
the cross section without photon emission and the fully inclusive one
is likely to be sufficiently accurate for our
purposes, although a more detailed confrontation with actual data would be
desirable.\footnote{A similar factorization assumption has recently been 
  used in order to extract the $\eta \to 3 \pi$ differential decay rate in
  the isospin limit from the measured one~\cite{Colangelo:2018jxw}.}   
To first order in $\alpha$ the relation reads as follows~\cite{Gluza:2002ui,Czyz:2004rj,Bystritskiy:2005ib} 
\begin{align}
\label{FSR}
	\begin{split}
		\sigma(e^+e^-&\to\pi^+\pi^-(\gamma)) = \Big[ 1 + \frac{\alpha}{\pi} \eta(s) \Big] \sigma(e^+e^-\to\pi^+\pi^-), \\
			\eta(s) &= \frac{3(1+\sigma_\pi^2(s))}{2\sigma_\pi^2(s)} - 4 \log\sigma_\pi(s) + 6 \log \frac{1+\sigma_\pi(s)}{2} + \frac{1+\sigma_\pi^2(s)}{\sigma_\pi(s)} F(\sigma_\pi(s)) \\
				&\quad - \frac{(1-\sigma_\pi(s))\big(3+3\sigma_\pi(s)-7\sigma_\pi^2(s)+5\sigma_\pi^3(s)\big)}{4\sigma_\pi^3(s)} \log\frac{1+\sigma_\pi(s)}{1-\sigma_\pi(s)}, \\
			F(x) &= -4\dilog(x)+4\dilog(-x)+2\log x \log\frac{1+x}{1-x} + 3 \dilog\Big(\frac{1+x}{2} \Big) - 3\dilog\Big(\frac{1-x}{2} \Big) + \frac{\pi^2}{2}, \\
			\dilog(x) &= - \int_0^x dt \frac{\log(1-t)}{t}.
	\end{split}
\end{align}
As for the pion VFF, this step to extract it from experiment as an object
defined in QCD, i.e.\
\begin{align}
	\label{eq:BareCrossSectionVSFormFactor}
	\sigma^{(0)}(e^+e^-\to\gamma^*\to\pi^+\pi^-) = \frac{\pi \alpha^2}{3s} \sigma_\pi^3(s) \big| F_\pi^V(s) \big|^2 \frac{s+2m_e^2}{s \sigma_e(s)},
\end{align}
is absolutely essential for our purposes: otherwise the dispersive
constraints to be discussed in the next section would not apply. 

There are other issues related to radiative corrections which have also
been discussed in the literature, in particular $\rho$--$\omega$ and
$\rho$--$\gamma$ mixing~\cite{Jegerlehner:2011ti}, in the context of a
Bethe--Salpeter approach for the coupled-channel system of $e^+e^-$,
$\pi^+\pi^-$, and $3\pi$, see~\cite{Hanhart:2012wi,Hanhart:inprep}.
The main result is that additional corrections from $\rho$--$\gamma$
mixing~\cite{Jegerlehner:2011ti} only become relevant if an attempt is made
to define external $\rho$ states, as required for estimates of
isospin-breaking corrections in the interpretation of
$\tau\to\pi\pi\nu_\tau$ data to be used as input for
HVP~\cite{Cirigliano:2001er,Cirigliano:2002pv}, but in the
$e^+e^-\to\pi^+\pi^-$ channel full consistency is ensured as long as the
same pure-QCD form factor $F_\pi^V$ that determines the bare cross
section~\eqref{eq:BareCrossSectionVSFormFactor} defines, self-consistently,
the $\pi^+\pi^-$ contribution to the VP function $\Pi^\mathrm{SM}$ in its
extraction from experiment~\eqref{bare_cross_section}. In practice, we find
that the VP routines applied in the modern experiments are sufficiently
close to such a fully self-consistent solution that we can use the bare
cross sections as provided by experiment.\footnote{In fact, if the normalization is determined from $e^+e^-\to\mu^+\mu^-$, the resulting cross section is automatically bare because VP drops out
in the ratio. This applies to the BaBar~\cite{Aubert:2009ad,Lees:2012cj} and KLOE12~\cite{Babusci:2012rp} data sets.}

Accordingly, the physical FSR-inclusive cross section takes the form
\beq
\sigma(e^+e^-\to\gamma^*\to\pi^+\pi^-(\gamma)) = \Big[ 1 + \frac{\alpha}{\pi} \eta(s) \Big]  \frac{\pi \big|\alpha(s)\big|^2}{3s} \sigma_\pi^3(s) \big| F_\pi^V(s) \big|^2 \frac{s+2m_e^2}{s \sigma_e(s)},
\eeq
where the VP function has been expressed in terms of the running coupling $\alpha(s)$. 
Unfortunately, the common procedure in the literature amounts to absorbing a factor $\alpha(s)/\alpha$ into the definition of the form factor, see~\cite{Akhmetshin:2001ig,Akhmetshin:2003zn,Achasov:2005rg,Achasov:2006vp,Akhmetshin:2006wh,Akhmetshin:2006bx,Ambrosino:2008aa,Aubert:2009ad,Ambrosino:2010bv,Lees:2012cj,Babusci:2012rp,Ablikim:2015orh,Anastasi:2017eio},
but in these conventions we could not formulate the dispersive constraints. For this reason, we do not use the results for $F_\pi^V(s)$ provided by each $e^+e^-$ experiment, but rather the bare cross
section in order to reconstruct the actual QCD form factor.

\subsection{Omn\`es representation of the form factor}
\label{sec:OmnesRepresentation}

In the following, we present the dispersive representation of the pion VFF $F_\pi^V(s)$ as put forward in~\cite{Leutwyler:2002hm,Colangelo:2003yw}. In particular, we treat the form factor in pure QCD 
and include the most important strong isospin-breaking effect from the mixing into the $3\pi$ channel.
In the isospin limit, $F_\pi^V(s)$ is an analytic function of $s$, apart from a branch cut in the complex $s$-plane that lies on the real axis, $s \in [4\mpi^2, \infty)$, and is dictated by unitarity. 
The form factor is real on the real axis below the branch point $4\mpi^2$, hence it fulfills the Schwarz reflection principle. We parametrize the pion VFF as a product of three functions,
\beq
	\label{eq:PionVFF}
	F_\pi^V(s) = \Omega_1^1(s) G_\omega(s) G_\mathrm{in}^N(s),
\eeq
where
\begin{align}
	\label{eq:OmnesFunction}
	\Omega_1^1(s) = \exp\left\{ \frac{s}{\pi} \int_{4\mpi^2}^\infty ds^\prime \frac{\delta_1^1(s^\prime)}{s^\prime(s^\prime-s)} \right\}
\end{align}
is the usual Omn\`es function~\cite{Omnes:1958hv} with $\delta_1^1(s)$ the isospin $I=1$ elastic $\pi\pi$ phase shift in the isospin-symmetric limit. The Omn\`es function alone is the solution for the VFF in the isospin limit and disregarding inelastic contributions to the unitarity relation. Therefore, the quotient function $F_\pi^V(s) / \Omega_1^1(s)$ is analytic in the complex $s$-plane apart from a cut on the real axis starting at $s=9\mpi^2$.

The factor $G_\omega$ accounts for $\rho$--$\omega$ mixing, the most important isospin-breaking effect, which becomes enhanced by the small mass difference between the $\rho$ and $\omega$ resonances.
The full parametrization 
\beq
\label{Gomega}
	G_\omega(s) = 1 + \frac{s}{\pi} \int_{9\mpi^2}^\infty ds^\prime \frac{\Im g_\omega(s^\prime)}{s^\prime(s^\prime-s)} \left( \frac{1 - \frac{9\mpi^2}{s^\prime}}{1 - \frac{9\mpi^2}{\mw^2}} \right)^4
\eeq
with
\beq
\label{gomega}
	g_\omega(s) = 1 + \epsilon_\omega \frac{s}{(\mw - \frac{i}{2} \Gamma_\omega)^2 - s}
\eeq
implements the correct threshold behavior of the discontinuity, i.e.\ the right-hand cut starting at $9\mpi^2$ opens with the fourth power of the center-of-mass momentum~\cite{Leutwyler:2002hm}. 
In practice, it would even be possible to replace $G_\omega(s)$ by $g_\omega(s)$ with almost no observable effect in the energy range of interest, in particular, due to the strong localization around the $\omega$ resonance, the imaginary part of $g_\omega(s)$ below threshold is tiny.
We still use the dispersively-improved variant~\eqref{Gomega} to remove this unphysical imaginary part altogether and have the threshold behavior correct, but stress that
if the difference became relevant, this form would not suffice to go beyond the narrow-width approximation. For that also the 
spectral shape would need to be improved~\cite{Hoferichter:2014vra,Hoferichter:2018kwz}.
For completeness, we remark that while in general $P$-wave phase space predicts a behavior proportional to $(s-n^2\mpi^2)^{3(n-1)/2}$, the leading term vanishes for $n=3$, 
giving rise to the extra power in~\eqref{Gomega}.

The remaining function $G_\mathrm{in}^N(s)$ is analytic in the complex $s$-plane with a cut on the real axis starting at $s=16\mpi^2$. It takes into account all further inelastic contributions to the unitarity relation. We describe it by a conformal polynomial
\beq
\label{Gin_def}
	G_\mathrm{in}^N(s) = 1 + \sum_{k=1}^N c_k ( z^k(s) - z^k(0) ),
\eeq
where the conformal variable is
\beq
	z(s) = \frac{\sqrt{s_\mathrm{in} - s_c} - \sqrt{s_\mathrm{in} - s}}{\sqrt{s_\mathrm{in} - s_c} + \sqrt{s_\mathrm{in} - s}}
\eeq
and we consider inelasticities only above $s_\mathrm{in} = (M_{\pi^0} + \mw)^2$, since $4\pi$ inelasticities are extremely weak below, see Sect.~\ref{sec:ELBound}. 
The conformal polynomial generates a branch-cut singularity at $s = s_\mathrm{in}$ and in the variant~\eqref{Gin_def} does not modify the charge, $G_\mathrm{in}^N(0) = 1$. 
We also require the cut to reproduce $P$-wave behavior at the inelastic threshold, i.e.\ close to $s_\mathrm{in}$ the function $G_\mathrm{in}^N(s)$ has to behave like $(s_\mathrm{in}-s)^{3/2}$
\begin{align}
	\begin{split}
		G_\mathrm{in}^N(s) &= \mathrm{const.} + \sum_{k=1}^N c_k z^k(s) = \mathrm{const.} + \sum_{k=1}^N c_k \left( \frac{(\sqrt{s_\mathrm{in} - s_c} - \sqrt{s_\mathrm{in} - s})^2}{s-s_c} \right)^k \\
			&= \mathrm{poly.} - 2  \sum_{k=1}^N k c_k \frac{\sqrt{s_\mathrm{in} - s}}{\sqrt{s_\mathrm{in}-s_c}} + \O\left((s_\mathrm{in}-s)^{3/2}\right) \, .
	\end{split}
\end{align}
Hence, in order to have a vanishing coefficient of the $\sqrt{s_\mathrm{in}-s}$ term, we impose
\begin{align}
	\label{eq:ConformalParameterPWaveConstraint}
	c_1 = - \sum_{k=2}^N k \, c_k \, .
\end{align}
In summary, our parametrization of the form factor fulfills all requirements of analyticity and unitarity,
including explicitly the $2\pi$ and $3\pi$ channels and inelastic corrections in a conformal polynomial with threshold dictated by phenomenology.
We expect this representation to be accurate as long as the conformal polynomial provides an efficient description of inelastic effects, conservatively estimated below $\sqrt{s}=1\GeV$. 
As main input, we require the elastic $\pi\pi$ $P$-wave phase shift $\delta_1^1(s)$, see Sect.~\ref{sec:PhaseShiftInput}, 
while the isospin-breaking and inelastic corrections are parametrized in terms of the $\omega$ parameters ($\epsilon_\omega$, $\mw$, and $\Gamma_\omega$)
and $c_k$ and $s_c$ in the conformal polynomial, respectively.

\subsection{Inelastic contributions and Eidelman--\L{}ukaszuk bound}
\label{sec:ELBound}

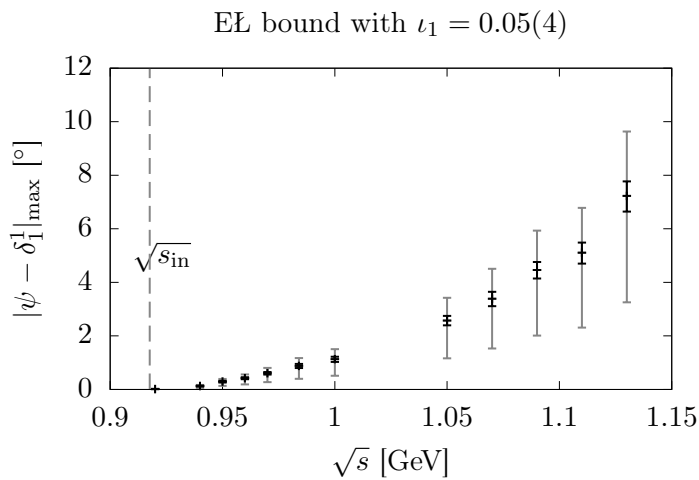
\begin{figure}[t]
	\centering
	\input{plots/EL}
	\caption{\EL{} bound on the difference between the phases of the pion VFF and the elastic $\pi\pi$ $P$-wave. The bound uses the data compilation of~\cite{Eidelman:2003uh} for the cross section ratio $r$, and an elasticity parameter calculated with $\iota_1=0.05(4)$. The smaller black error bars indicate the uncertainty due to $r$, the larger gray error bars the uncertainty due to $\iota_1$.}
	\label{img:ELBound}
\end{figure}

In~\cite{Lukaszuk:1973jd,Eidelman:2003uh}, a generalization of Watson's theorem~\cite{Watson:1954uc} was derived that amounts to a constraint on the difference between the phase of the VFF and the elastic $\pi\pi$ scattering phase shift, the Eidelman--\L{}ukaszuk (\EL) bound:
\beq
	\label{eq:ELBound}
	\left( \frac{1-\eta_1}{2} \right)^2 + \eta_1 \sin^2(\psi-\delta_1^1) \le \frac{1-\eta_1^2}{4} \, r, \qquad 0 \le \eta_1 \le 1,
\eeq
where $\psi$ denotes the phase of the form factor, $F_\pi^V(s) = |F_\pi^V(s)| e^{i\psi(s)}$, $\eta_1$ is the $\pi\pi$ elasticity parameter, defined by the expression for the $\pi\pi$ $P$-wave amplitude
\beq
	t_1^1(s) = \frac{\eta_1(s) e^{2i\delta_1^1(s)} - 1}{2i\sigma_\pi(s)},
\eeq
and $r$ is the ratio of non-$2\pi$ to $2\pi$ hadronic cross sections
\beq
	r = \frac{\sigma^{I=1}(e^+e^-\to\mathrm{hadrons})}{\sigma(e^+e^-\to\pi^+\pi^-)} - 1
\eeq
in the isospin $I=1$ channel. For $r<1$, the bound~\eqref{eq:ELBound} implies $\eta_1\geq (1-r)/(1+r)$, resulting in~\cite{Lukaszuk:1973jd}
\beq
	\label{eq:ELBoundAlt}
	\sin^2(\psi-\delta_1^1) \le \frac{1}{2} \Big( 1-\sqrt{1-r^2} \Big).
\eeq
With a given input for the elasticity parameter $\eta_1$, the bound~\eqref{eq:ELBound} usually provides a much stronger constraint than~\eqref{eq:ELBoundAlt},
but the latter shows that a non-trivial bound arises as soon as $r>0$ irrespective of $\eta_1$.

We use a representation of the elasticity parameter from the $\pi\pi$ Roy-equation analysis~\cite{Ananthanarayan:2000ht,Caprini:2011ky}
\beq
	\label{eq:ElasticityParameter}
	\eta_1(s) = \frac{s_a^3 - \iota_1 (s-4\mpi^2)^{3/2} (s-s_\mathrm{in})^{3/2}}{s_a^3 + \iota_1 (s-4\mpi^2)^{3/2} (s-s_\mathrm{in})^{3/2}}
\eeq
with $s_a=(1\GeV)^2$ and $\iota_1 = 0.05(4)$. With the experimental input on $r$ from~\cite{Eidelman:2003uh}, we obtain the bound on the phase difference shown in Fig.~\ref{img:ELBound}.
Using the parametrization~\eqref{eq:ElasticityParameter} for small values of $\iota_1$, the bound can be conveniently written as
\beq
	\Delta := |\psi - \delta_1^1|^2 \le \iota_1 r \frac{(s-4\mpi^2)^{3/2} (s-s_\mathrm{in})^{3/2}}{s_a^3} + \O(\iota_1^2) ,
\eeq
where the negligible $\O(\iota_1^2)$ term is negative as long as $r^2<3$.
The \EL{} bound provides an important constraint on the parameters $c_k$ of the conformal polynomial that we use to describe the inelastic contributions.

We note that in contrast to the value of $\iota_1 = 0.05(5)$ from~\cite{Ananthanarayan:2000ht,Caprini:2011ky}, we vary the parameter in a slightly smaller range in order to exclude a vanishing value of $\iota_1$, which would correspond to $\eta_1=1$, while a non-zero value of $r$ always implies $\eta_1 < 1$.\footnote{Even if $r>0$ resulted only from hadronic states that do not couple directly to two pions, an inelasticity would be produced by the coupling through a virtual photon.} In principle, very small values of $\iota_1$ could be excluded by considering particular channels, such as 
the $\pi^0\omega$ intermediate state~\cite{Niecknig:2012sj} that motivates the functional form~\eqref{eq:ElasticityParameter}. At slightly higher energy, the $\bar K K$ channel opens, which gives a 
rather small contribution to the inelasticity~\cite{Buettiker:2003pp,Niecknig:2012sj}, but also shows that at some point $\eta_1=1$ is excluded by data. Here, we motivate the lower bound on $\iota_1$ directly through the fits to the $e^+e^-$ data: if $\iota_1$ is chosen too small, the conformal polynomial becomes constrained too much, resulting in an unacceptable fit quality. In this way, 
the $e^+e^-$ data themselves imply that the inelasticity cannot be arbitrarily small. Our range covers those values for which the fits are still acceptable.

%% file: plots/EL.tex
\begingroup
  \makeatletter
  \providecommand\color[2][]{%
    \GenericError{(gnuplot) \space\space\space\@spaces}{%
      Package color not loaded in conjunction with
      terminal option `colourtext'%
    }{See the gnuplot documentation for explanation.%
    }{Either use 'blacktext' in gnuplot or load the package
      color.sty in LaTeX.}%
    \renewcommand\color[2][]{}%
  }%
  \providecommand\includegraphics[2][]{%
    \GenericError{(gnuplot) \space\space\space\@spaces}{%
      Package graphicx or graphics not loaded%
    }{See the gnuplot documentation for explanation.%
    }{The gnuplot epslatex terminal needs graphicx.sty or graphics.sty.}%
    \renewcommand\includegraphics[2][]{}%
  }%
  \providecommand\rotatebox[2]{#2}%
  \@ifundefined{ifGPcolor}{%
    \newif\ifGPcolor
    \GPcolorfalse
  }{}%
  \@ifundefined{ifGPblacktext}{%
    \newif\ifGPblacktext
    \GPblacktexttrue
  }{}%
  \let\gplgaddtomacro\g@addto@macro
  \gdef\gplbacktext{}%
  \gdef\gplfronttext{}%
  \makeatother
  \ifGPblacktext
    \def\colorrgb#1{}%
    \def\colorgray#1{}%
  \else
    \ifGPcolor
      \def\colorrgb#1{\color[rgb]{#1}}%
      \def\colorgray#1{\color[gray]{#1}}%
      \expandafter\def\csname LTw\endcsname{\color{white}}%
      \expandafter\def\csname LTb\endcsname{\color{black}}%
      \expandafter\def\csname LTa\endcsname{\color{black}}%
      \expandafter\def\csname LT0\endcsname{\color[rgb]{1,0,0}}%
      \expandafter\def\csname LT1\endcsname{\color[rgb]{0,1,0}}%
      \expandafter\def\csname LT2\endcsname{\color[rgb]{0,0,1}}%
      \expandafter\def\csname LT3\endcsname{\color[rgb]{1,0,1}}%
      \expandafter\def\csname LT4\endcsname{\color[rgb]{0,1,1}}%
      \expandafter\def\csname LT5\endcsname{\color[rgb]{1,1,0}}%
      \expandafter\def\csname LT6\endcsname{\color[rgb]{0,0,0}}%
      \expandafter\def\csname LT7\endcsname{\color[rgb]{1,0.3,0}}%
      \expandafter\def\csname LT8\endcsname{\color[rgb]{0.5,0.5,0.5}}%
    \else
      \def\colorrgb#1{\color{black}}%
      \def\colorgray#1{\color[gray]{#1}}%
      \expandafter\def\csname LTw\endcsname{\color{white}}%
      \expandafter\def\csname LTb\endcsname{\color{black}}%
      \expandafter\def\csname LTa\endcsname{\color{black}}%
      \expandafter\def\csname LT0\endcsname{\color{black}}%
      \expandafter\def\csname LT1\endcsname{\color{black}}%
      \expandafter\def\csname LT2\endcsname{\color{black}}%
      \expandafter\def\csname LT3\endcsname{\color{black}}%
      \expandafter\def\csname LT4\endcsname{\color{black}}%
      \expandafter\def\csname LT5\endcsname{\color{black}}%
      \expandafter\def\csname LT6\endcsname{\color{black}}%
      \expandafter\def\csname LT7\endcsname{\color{black}}%
      \expandafter\def\csname LT8\endcsname{\color{black}}%
    \fi
  \fi
    \setlength{\unitlength}{0.0500bp}%
    \ifx\gptboxheight\undefined%
      \newlength{\gptboxheight}%
      \newlength{\gptboxwidth}%
      \newsavebox{\gptboxtext}%
    \fi%
    \setlength{\fboxrule}{0.5pt}%
    \setlength{\fboxsep}{1pt}%
\begin{picture}(5400.00,3780.00)%
    \gplgaddtomacro\gplbacktext{%
      \csname LTb\endcsname
      \put(682,704){\makebox(0,0)[r]{\strut{}$0$}}%
      \put(682,1107){\makebox(0,0)[r]{\strut{}$2$}}%
      \put(682,1509){\makebox(0,0)[r]{\strut{}$4$}}%
      \put(682,1912){\makebox(0,0)[r]{\strut{}$6$}}%
      \put(682,2314){\makebox(0,0)[r]{\strut{}$8$}}%
      \put(682,2717){\makebox(0,0)[r]{\strut{}$10$}}%
      \put(682,3119){\makebox(0,0)[r]{\strut{}$12$}}%
      \put(814,484){\makebox(0,0){\strut{}$0.9$}}%
      \put(1652,484){\makebox(0,0){\strut{}$0.95$}}%
      \put(2490,484){\makebox(0,0){\strut{}$1$}}%
      \put(3327,484){\makebox(0,0){\strut{}$1.05$}}%
      \put(4165,484){\makebox(0,0){\strut{}$1.1$}}%
      \put(5003,484){\makebox(0,0){\strut{}$1.15$}}%
      \put(982,1710){\makebox(0,0)[l]{\strut{}$\sqrt{s_\mathrm{in}}$}}%
    }%
    \gplgaddtomacro\gplfronttext{%
      \csname LTb\endcsname
      \put(198,1911){\rotatebox{-270}{\makebox(0,0){\strut{}$|\psi-\delta_1^1|_\mathrm{max}$ [${}^\circ$]}}}%
      \put(2908,154){\makebox(0,0){\strut{}$\sqrt{s}$ [GeV]}}%
      \put(2908,3449){\makebox(0,0){\strut{}E\L{} bound with $\iota_1 = 0.05(4)$}}%
    }%
    \gplbacktext
    \put(0,0){\includegraphics{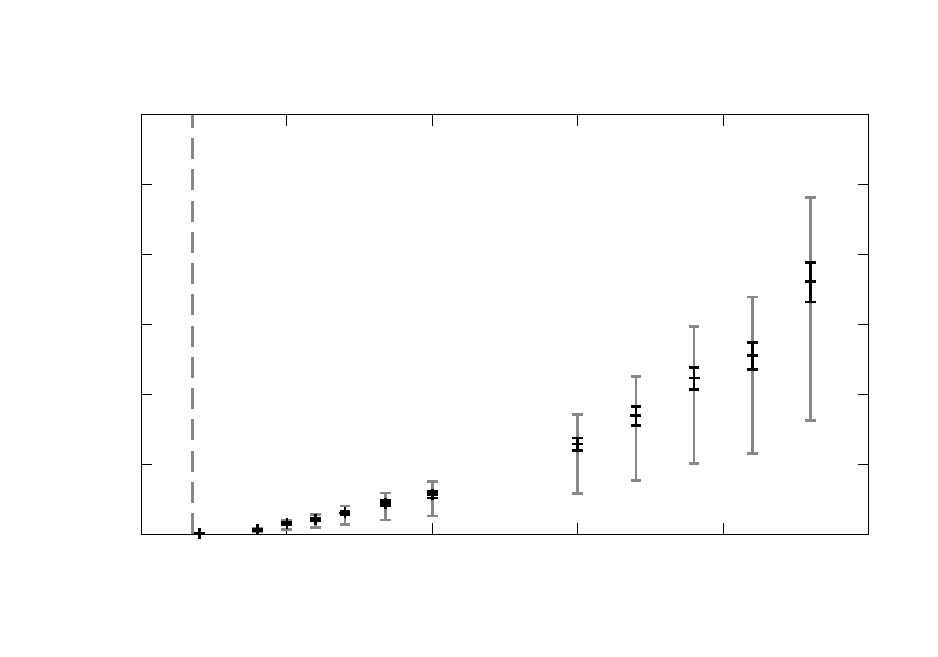}}%
    \gplfronttext
  \end{picture}%
\endgroup

%% file: sections/Input.tex

\section{Input for the phase shift}
\label{sec:PhaseShiftInput}

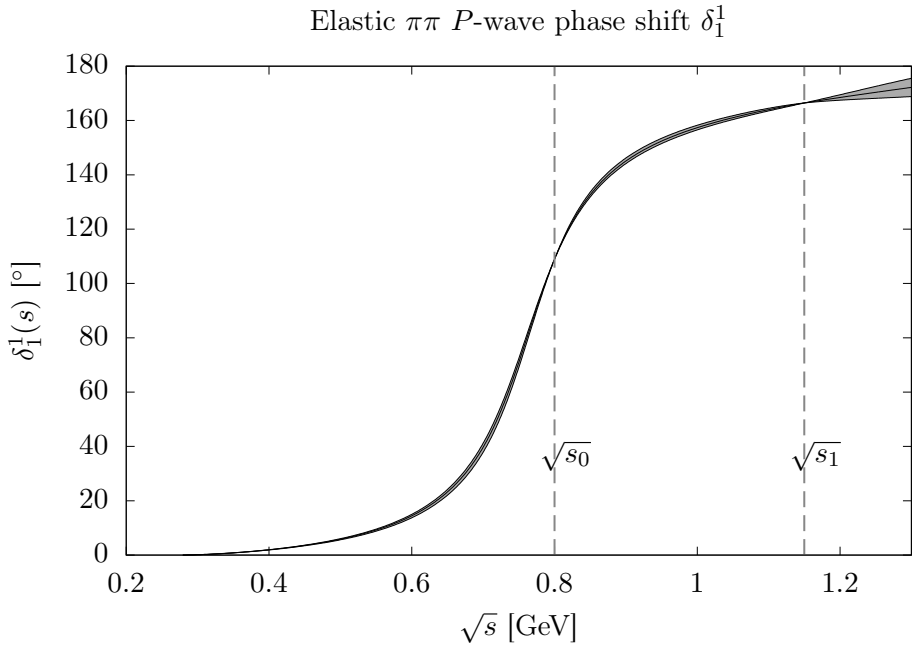
\begin{figure}[t]
	\centering
	\input{plots/d11}
	\caption{Elastic $P$-wave $\pi\pi$ scattering phase shift $\delta_1^1$ from the solution of the Roy equations~\cite{Caprini:2011ky}. We show the central phase solution with an uncertainty band generated by varying all parameters apart from the phase values at $s_0$ and $s_1$.}
	\label{img:Delta11}
\end{figure}

\begin{figure}[t]
	\centering
	\input{plots/d11-he}
	\caption{Continuation of the elastic $P$-wave $\pi\pi$ scattering phase shift $\delta_1^1$ to energies above the validity of the Roy equations~\cite{Caprini:2011ky}, used to estimate the impact of uncertainties due to the phase above $1.3\GeV$.}
	\label{img:Delta11HE}
\end{figure}
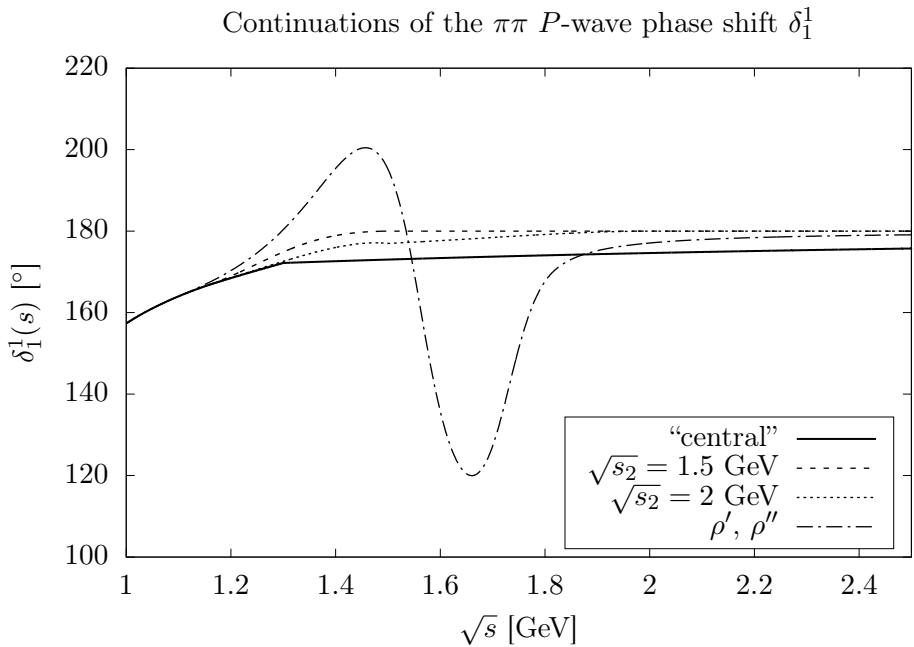

The central input in our representation of the pion VFF is the elastic $P$-wave $\pi\pi$ scattering phase shift. We use the solution of the Roy-equation analysis~\cite{Ananthanarayan:2000ht,Caprini:2011ky}.
The parametrization of the phase shift of~\cite{Caprini:2011ky} depends on 27 parameters, but most of them concern elasticity parameters or input from Regge
theory for the asymptotic region, both in the $P$-wave and the other amplitudes related by crossing symmetry. 
The (elastic) $P$-wave phase shift itself, below $1.15\GeV$, only involves two free parameters, which can be identified with its values
at $s_0 = (0.8\GeV)^2$ and at $s_1=(1.15\GeV)^2$, whose current estimates are~\cite{Caprini:2011ky}
\begin{align}
	\label{eq:Delta11ParametersRoy}
	\delta_1^1(s_0) = 108.9(2.0)^\circ, \quad \delta_1^1(s_1) = 166.5(2.0)^\circ.
\end{align}
This counting of the degrees of freedom in the solution of the Roy equations depends on the so-called matching point $s_{\mathrm m}$, for
$s_{\mathrm m}=(1.15\GeV)^2$ as adopted in~\cite{Caprini:2011ky} the mathematical properties of the equations dictate that there be exactly
two free parameters~\cite{Gasser:1999hz,Wanders:2000mn}.
In our description of the VFF, the values of the phase at $s_0$ and $s_1$ enter as fit parameters, while the values~\eqref{eq:Delta11ParametersRoy}, derived from previous analyses of the VFF, only serve for comparison and as starting values in the fit. 
All the remaining 25 parameters of the Roy solution will be varied within the ranges given in~\cite{Caprini:2011ky} and treated as a source of systematic uncertainties in our description. The central solution for the phase shift is shown in Fig.~\ref{img:Delta11} together with an uncertainty band generated by varying these 25 parameters.

At energies above $1.3\GeV$, the $\pi\pi$ phase shift is not as well known as in the low-energy region shown in Fig.~\ref{img:Delta11}. However, this uncertainty will not have a strong impact on the low-energy description of the form factor. We estimate this uncertainty by studying different prescriptions for the high-energy continuation of the phase shift. Asymptotically, we assume that the phase shift reaches~\cite{Chernyak:1977as,Chernyak:1980dj,Efremov:1978rn,Efremov:1979qk,Farrar:1979aw,Lepage:1979zb,Lepage:1980fj,Leutwyler:2002hm}
\begin{align}
	\lim_{s\to\infty}\delta_1^1(s) = \pi,
\end{align}
so that the Omn\`es function behaves asymptotically as $\Omega_1^1(s) \asymp s^{-1}$.
For our central phase solution, we use the simple prescription~\cite{Moussallam:1999aq}
\begin{align}
	\delta_1^1(s)^\mathrm{asymp} = \left\{ \begin{matrix}
									\delta_1^1(s) & \text{ if } s<s_a, \\
									\pi + ( \delta_1^1(s_a) - \pi ) \frac{2}{1 + (s/s_a)^{3/4}} & \text{ if } s \ge s_a \,\;\;
								 \end{matrix} \right.
\end{align}
with $s_a = (1.3\GeV)^2$, and we compare to the prescription
\begin{align}
	\delta_1^1(s)^\mathrm{asymp} = \left\{ \begin{matrix}
									\delta_1^1(s) & \text{ if } s < s_1, \\
									\delta_1^1(s) + ( \pi - \delta_1^1(s) ) f(s) & \text{ if } s_1 \le s < s_b, \\
									\delta_1^1(s) + ( \pi - \delta_1^1(s_b) ) f(s) & \text{ if } s_b \le s < s_2, \\
									\pi & \text{ if } s \ge s_2,
								 \end{matrix} \right.
								 \quad f(s) = \frac{(s-s_1)^2(3s_2-2s-s_1)}{(s_2-s_1)^3}
\end{align}
with $s_b = (1.5\GeV)^2$ and $s_1 = (1.15\GeV)^2$, and the point $s_2$, where the phase reaches $\pi$, is varied in a range $\sqrt{s_2} = 1.5\ldots 2\GeV$. Alternatively, we use the phase of~\cite{Schneider:2012ez} that estimates the effects of the excited resonances $\rho'(1450)$ and $\rho''(1700)$ from their impact on $\tau^-\to\pi^-\pi^0\nu_\tau$~\cite{Fujikawa:2008ma}. The different continuations of the phase $\delta_1^1$ that we use to estimate the uncertainties from the energy region above $1.3\GeV$ are shown in Fig.~\ref{img:Delta11HE}.

\begin{figure}[t]
	\centering
	\input{plots/omega11}
	\caption{Squared absolute value of the Omn\`es function $|\Omega_1^1(s)|^2$, calculated with the elastic $P$-wave $\pi\pi$ scattering phase shift $\delta_1^1$. The uncertainty band is generated by varying all parameters apart from the phase values at $s_0$ and $s_1$, which are kept fixed at their central values~\eqref{eq:Delta11ParametersRoy} from the Roy solution~\cite{Caprini:2011ky}.}
	\label{img:Omega11}
\end{figure}
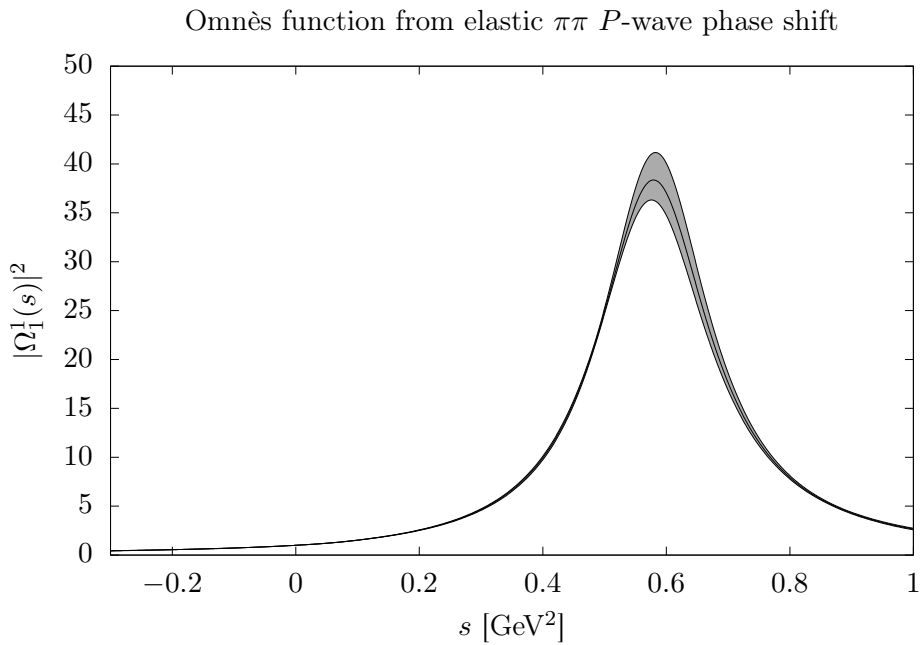

From the $\pi\pi$ scattering phase shift $\delta_1^1$ we calculate the Omn\`es function~\eqref{eq:OmnesFunction}, with squared absolute value shown in Fig.~\ref{img:Omega11}. The uncertainty band is again generated by varying all the parameters of the Roy solution apart from the phase values at $s_0$ and $s_1$, which for this plot we have fixed at the central values~\eqref{eq:Delta11ParametersRoy}. Note that although the Omn\`es function already closely resembles the pion VFF, the uncertainty of $|\Omega_1^1|^2$ shown in the plot will not translate directly to $|F_\pi^V|^2$, because for the description of the VFF 
$\delta_1^1(s_0)$ and $\delta_1^1(s_1)$ will not be fixed but enter as fit parameters.

The fact that only two free parameters are allowed in the description of the phase shift emphasizes the stringent constraints that follow from $\pi\pi$ Roy equations, ensuring in each step that the solution
for the phase shift is consistent with analyticity, unitarity, and crossing symmetry. This is the crucial advantage over using a phenomenological parameterization of the phase shift instead, based on which
a confrontation of the VFF data with these general QCD properties would not be possible. 

%% file: plots/d11.tex
\begingroup
  \makeatletter
  \providecommand\color[2][]{%
    \GenericError{(gnuplot) \space\space\space\@spaces}{%
      Package color not loaded in conjunction with
      terminal option `colourtext'%
    }{See the gnuplot documentation for explanation.%
    }{Either use 'blacktext' in gnuplot or load the package
      color.sty in LaTeX.}%
    \renewcommand\color[2][]{}%
  }%
  \providecommand\includegraphics[2][]{%
    \GenericError{(gnuplot) \space\space\space\@spaces}{%
      Package graphicx or graphics not loaded%
    }{See the gnuplot documentation for explanation.%
    }{The gnuplot epslatex terminal needs graphicx.sty or graphics.sty.}%
    \renewcommand\includegraphics[2][]{}%
  }%
  \providecommand\rotatebox[2]{#2}%
  \@ifundefined{ifGPcolor}{%
    \newif\ifGPcolor
    \GPcolorfalse
  }{}%
  \@ifundefined{ifGPblacktext}{%
    \newif\ifGPblacktext
    \GPblacktexttrue
  }{}%
  \let\gplgaddtomacro\g@addto@macro
  \gdef\gplbacktext{}%
  \gdef\gplfronttext{}%
  \makeatother
  \ifGPblacktext
    \def\colorrgb#1{}%
    \def\colorgray#1{}%
  \else
    \ifGPcolor
      \def\colorrgb#1{\color[rgb]{#1}}%
      \def\colorgray#1{\color[gray]{#1}}%
      \expandafter\def\csname LTw\endcsname{\color{white}}%
      \expandafter\def\csname LTb\endcsname{\color{black}}%
      \expandafter\def\csname LTa\endcsname{\color{black}}%
      \expandafter\def\csname LT0\endcsname{\color[rgb]{1,0,0}}%
      \expandafter\def\csname LT1\endcsname{\color[rgb]{0,1,0}}%
      \expandafter\def\csname LT2\endcsname{\color[rgb]{0,0,1}}%
      \expandafter\def\csname LT3\endcsname{\color[rgb]{1,0,1}}%
      \expandafter\def\csname LT4\endcsname{\color[rgb]{0,1,1}}%
      \expandafter\def\csname LT5\endcsname{\color[rgb]{1,1,0}}%
      \expandafter\def\csname LT6\endcsname{\color[rgb]{0,0,0}}%
      \expandafter\def\csname LT7\endcsname{\color[rgb]{1,0.3,0}}%
      \expandafter\def\csname LT8\endcsname{\color[rgb]{0.5,0.5,0.5}}%
    \else
      \def\colorrgb#1{\color{black}}%
      \def\colorgray#1{\color[gray]{#1}}%
      \expandafter\def\csname LTw\endcsname{\color{white}}%
      \expandafter\def\csname LTb\endcsname{\color{black}}%
      \expandafter\def\csname LTa\endcsname{\color{black}}%
      \expandafter\def\csname LT0\endcsname{\color{black}}%
      \expandafter\def\csname LT1\endcsname{\color{black}}%
      \expandafter\def\csname LT2\endcsname{\color{black}}%
      \expandafter\def\csname LT3\endcsname{\color{black}}%
      \expandafter\def\csname LT4\endcsname{\color{black}}%
      \expandafter\def\csname LT5\endcsname{\color{black}}%
      \expandafter\def\csname LT6\endcsname{\color{black}}%
      \expandafter\def\csname LT7\endcsname{\color{black}}%
      \expandafter\def\csname LT8\endcsname{\color{black}}%
    \fi
  \fi
    \setlength{\unitlength}{0.0500bp}%
    \ifx\gptboxheight\undefined%
      \newlength{\gptboxheight}%
      \newlength{\gptboxwidth}%
      \newsavebox{\gptboxtext}%
    \fi%
    \setlength{\fboxrule}{0.5pt}%
    \setlength{\fboxsep}{1pt}%
\begin{picture}(7200.00,5040.00)%
    \gplgaddtomacro\gplbacktext{%
      \csname LTb\endcsname
      \put(814,704){\makebox(0,0)[r]{\strut{}$0$}}%
      \put(814,1112){\makebox(0,0)[r]{\strut{}$20$}}%
      \put(814,1521){\makebox(0,0)[r]{\strut{}$40$}}%
      \put(814,1929){\makebox(0,0)[r]{\strut{}$60$}}%
      \put(814,2337){\makebox(0,0)[r]{\strut{}$80$}}%
      \put(814,2746){\makebox(0,0)[r]{\strut{}$100$}}%
      \put(814,3154){\makebox(0,0)[r]{\strut{}$120$}}%
      \put(814,3562){\makebox(0,0)[r]{\strut{}$140$}}%
      \put(814,3971){\makebox(0,0)[r]{\strut{}$160$}}%
      \put(814,4379){\makebox(0,0)[r]{\strut{}$180$}}%
      \put(946,484){\makebox(0,0){\strut{}$0.2$}}%
      \put(2011,484){\makebox(0,0){\strut{}$0.4$}}%
      \put(3076,484){\makebox(0,0){\strut{}$0.6$}}%
      \put(4141,484){\makebox(0,0){\strut{}$0.8$}}%
      \put(5206,484){\makebox(0,0){\strut{}$1$}}%
      \put(6271,484){\makebox(0,0){\strut{}$1.2$}}%
      \put(4034,1459){\makebox(0,0)[l]{\strut{}$\sqrt{s_0}$}}%
      \put(5898,1459){\makebox(0,0)[l]{\strut{}$\sqrt{s_1}$}}%
    }%
    \gplgaddtomacro\gplfronttext{%
      \csname LTb\endcsname
      \put(198,2541){\rotatebox{-270}{\makebox(0,0){\strut{}$\delta_1^1(s)$ [${}^\circ$]}}}%
      \put(3874,154){\makebox(0,0){\strut{}$\sqrt{s}$ [GeV]}}%
      \put(3874,4709){\makebox(0,0){\strut{}Elastic $\pi\pi$ $P$-wave phase shift $\delta_1^1$}}%
    }%
    \gplbacktext
    \put(0,0){\includegraphics{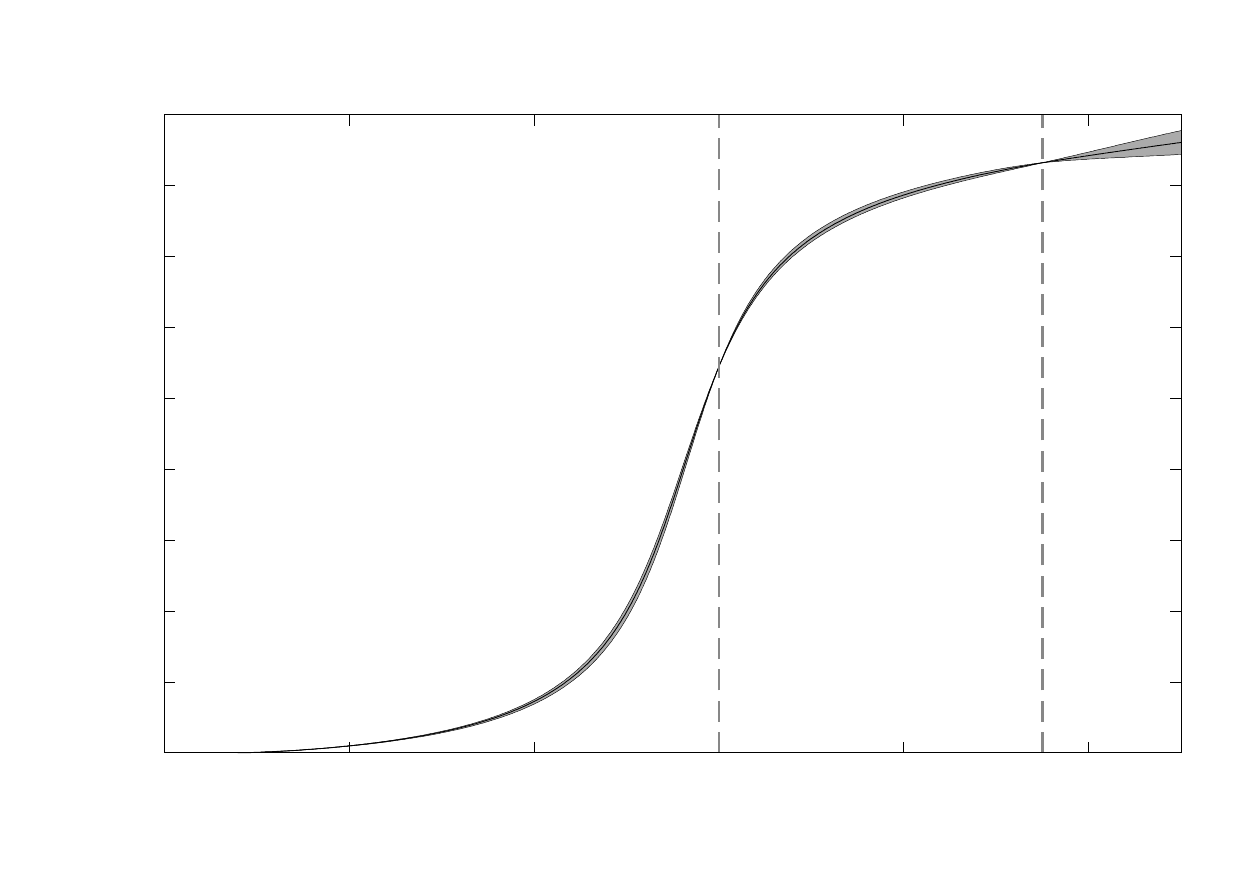}}%
    \gplfronttext
  \end{picture}%
\endgroup

%% file: plots/d11-he.tex
\begingroup
  \makeatletter
  \providecommand\color[2][]{%
    \GenericError{(gnuplot) \space\space\space\@spaces}{%
      Package color not loaded in conjunction with
      terminal option `colourtext'%
    }{See the gnuplot documentation for explanation.%
    }{Either use 'blacktext' in gnuplot or load the package
      color.sty in LaTeX.}%
    \renewcommand\color[2][]{}%
  }%
  \providecommand\includegraphics[2][]{%
    \GenericError{(gnuplot) \space\space\space\@spaces}{%
      Package graphicx or graphics not loaded%
    }{See the gnuplot documentation for explanation.%
    }{The gnuplot epslatex terminal needs graphicx.sty or graphics.sty.}%
    \renewcommand\includegraphics[2][]{}%
  }%
  \providecommand\rotatebox[2]{#2}%
  \@ifundefined{ifGPcolor}{%
    \newif\ifGPcolor
    \GPcolorfalse
  }{}%
  \@ifundefined{ifGPblacktext}{%
    \newif\ifGPblacktext
    \GPblacktexttrue
  }{}%
  \let\gplgaddtomacro\g@addto@macro
  \gdef\gplbacktext{}%
  \gdef\gplfronttext{}%
  \makeatother
  \ifGPblacktext
    \def\colorrgb#1{}%
    \def\colorgray#1{}%
  \else
    \ifGPcolor
      \def\colorrgb#1{\color[rgb]{#1}}%
      \def\colorgray#1{\color[gray]{#1}}%
      \expandafter\def\csname LTw\endcsname{\color{white}}%
      \expandafter\def\csname LTb\endcsname{\color{black}}%
      \expandafter\def\csname LTa\endcsname{\color{black}}%
      \expandafter\def\csname LT0\endcsname{\color[rgb]{1,0,0}}%
      \expandafter\def\csname LT1\endcsname{\color[rgb]{0,1,0}}%
      \expandafter\def\csname LT2\endcsname{\color[rgb]{0,0,1}}%
      \expandafter\def\csname LT3\endcsname{\color[rgb]{1,0,1}}%
      \expandafter\def\csname LT4\endcsname{\color[rgb]{0,1,1}}%
      \expandafter\def\csname LT5\endcsname{\color[rgb]{1,1,0}}%
      \expandafter\def\csname LT6\endcsname{\color[rgb]{0,0,0}}%
      \expandafter\def\csname LT7\endcsname{\color[rgb]{1,0.3,0}}%
      \expandafter\def\csname LT8\endcsname{\color[rgb]{0.5,0.5,0.5}}%
    \else
      \def\colorrgb#1{\color{black}}%
      \def\colorgray#1{\color[gray]{#1}}%
      \expandafter\def\csname LTw\endcsname{\color{white}}%
      \expandafter\def\csname LTb\endcsname{\color{black}}%
      \expandafter\def\csname LTa\endcsname{\color{black}}%
      \expandafter\def\csname LT0\endcsname{\color{black}}%
      \expandafter\def\csname LT1\endcsname{\color{black}}%
      \expandafter\def\csname LT2\endcsname{\color{black}}%
      \expandafter\def\csname LT3\endcsname{\color{black}}%
      \expandafter\def\csname LT4\endcsname{\color{black}}%
      \expandafter\def\csname LT5\endcsname{\color{black}}%
      \expandafter\def\csname LT6\endcsname{\color{black}}%
      \expandafter\def\csname LT7\endcsname{\color{black}}%
      \expandafter\def\csname LT8\endcsname{\color{black}}%
    \fi
  \fi
    \setlength{\unitlength}{0.0500bp}%
    \ifx\gptboxheight\undefined%
      \newlength{\gptboxheight}%
      \newlength{\gptboxwidth}%
      \newsavebox{\gptboxtext}%
    \fi%
    \setlength{\fboxrule}{0.5pt}%
    \setlength{\fboxsep}{1pt}%
\begin{picture}(7200.00,5040.00)%
    \gplgaddtomacro\gplbacktext{%
      \csname LTb\endcsname
      \put(814,704){\makebox(0,0)[r]{\strut{}$100$}}%
      \put(814,1317){\makebox(0,0)[r]{\strut{}$120$}}%
      \put(814,1929){\makebox(0,0)[r]{\strut{}$140$}}%
      \put(814,2542){\makebox(0,0)[r]{\strut{}$160$}}%
      \put(814,3154){\makebox(0,0)[r]{\strut{}$180$}}%
      \put(814,3767){\makebox(0,0)[r]{\strut{}$200$}}%
      \put(814,4379){\makebox(0,0)[r]{\strut{}$220$}}%
      \put(946,484){\makebox(0,0){\strut{}$1$}}%
      \put(1727,484){\makebox(0,0){\strut{}$1.2$}}%
      \put(2508,484){\makebox(0,0){\strut{}$1.4$}}%
      \put(3289,484){\makebox(0,0){\strut{}$1.6$}}%
      \put(4070,484){\makebox(0,0){\strut{}$1.8$}}%
      \put(4851,484){\makebox(0,0){\strut{}$2$}}%
      \put(5632,484){\makebox(0,0){\strut{}$2.2$}}%
      \put(6413,484){\makebox(0,0){\strut{}$2.4$}}%
    }%
    \gplgaddtomacro\gplfronttext{%
      \csname LTb\endcsname
      \put(198,2541){\rotatebox{-270}{\makebox(0,0){\strut{}$\delta_1^1(s)$ [${}^\circ$]}}}%
      \put(3874,154){\makebox(0,0){\strut{}$\sqrt{s}$ [GeV]}}%
      \put(3874,4709){\makebox(0,0){\strut{}Continuations of the $\pi\pi$ $P$-wave phase shift $\delta_1^1$}}%
      \csname LTb\endcsname
      \put(5816,1592){\makebox(0,0)[r]{\strut{}``central''}}%
      \csname LTb\endcsname
      \put(5816,1372){\makebox(0,0)[r]{\strut{}$\sqrt{s_2}=1.5$ GeV}}%
      \csname LTb\endcsname
      \put(5816,1152){\makebox(0,0)[r]{\strut{}$\sqrt{s_2}=2$ GeV}}%
      \csname LTb\endcsname
      \put(5816,932){\makebox(0,0)[r]{\strut{}$\rho'$, $\rho''$}}%
    }%
    \gplbacktext
    \put(0,0){\includegraphics{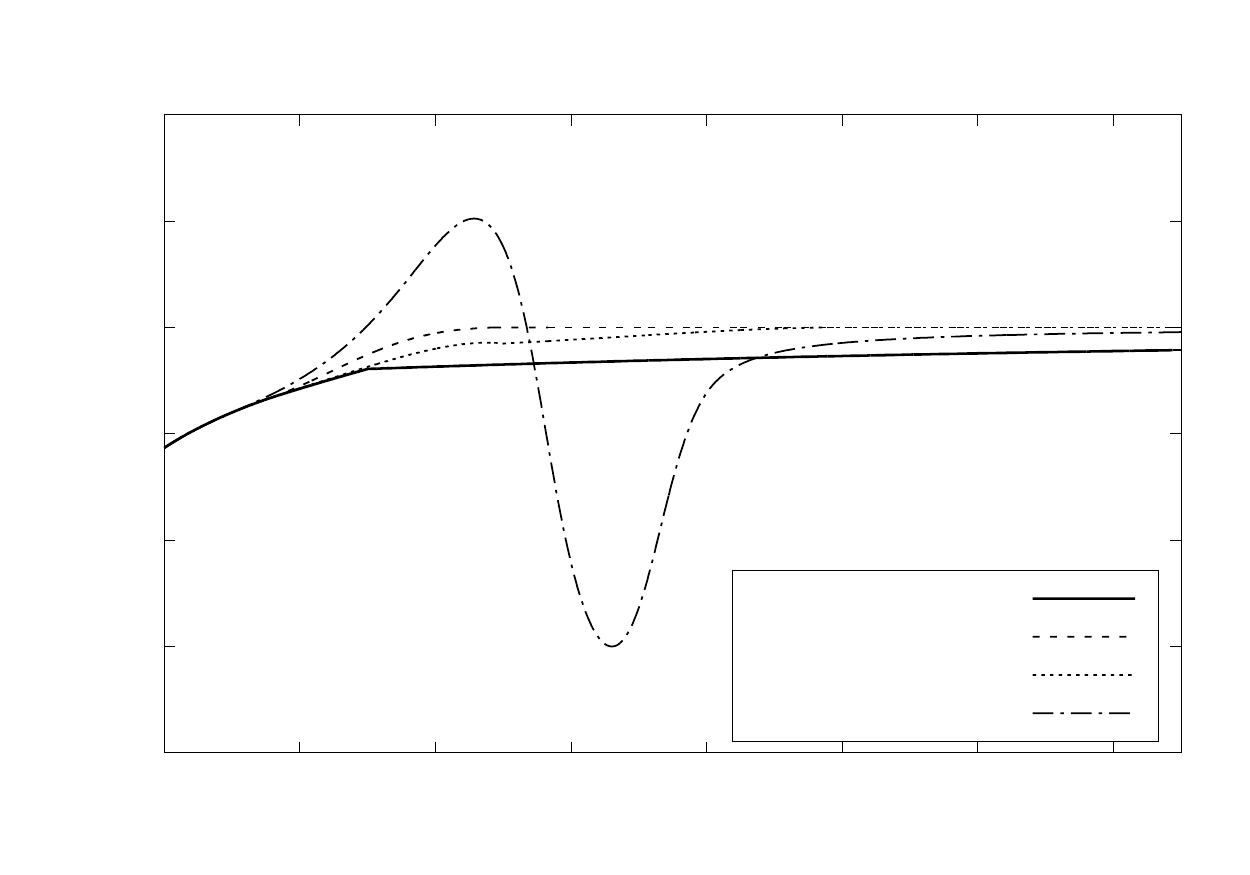}}%
    \gplfronttext
  \end{picture}%
\endgroup

%% file: plots/omega11.tex
\begingroup
  \makeatletter
  \providecommand\color[2][]{%
    \GenericError{(gnuplot) \space\space\space\@spaces}{%
      Package color not loaded in conjunction with
      terminal option `colourtext'%
    }{See the gnuplot documentation for explanation.%
    }{Either use 'blacktext' in gnuplot or load the package
      color.sty in LaTeX.}%
    \renewcommand\color[2][]{}%
  }%
  \providecommand\includegraphics[2][]{%
    \GenericError{(gnuplot) \space\space\space\@spaces}{%
      Package graphicx or graphics not loaded%
    }{See the gnuplot documentation for explanation.%
    }{The gnuplot epslatex terminal needs graphicx.sty or graphics.sty.}%
    \renewcommand\includegraphics[2][]{}%
  }%
  \providecommand\rotatebox[2]{#2}%
  \@ifundefined{ifGPcolor}{%
    \newif\ifGPcolor
    \GPcolorfalse
  }{}%
  \@ifundefined{ifGPblacktext}{%
    \newif\ifGPblacktext
    \GPblacktexttrue
  }{}%
  \let\gplgaddtomacro\g@addto@macro
  \gdef\gplbacktext{}%
  \gdef\gplfronttext{}%
  \makeatother
  \ifGPblacktext
    \def\colorrgb#1{}%
    \def\colorgray#1{}%
  \else
    \ifGPcolor
      \def\colorrgb#1{\color[rgb]{#1}}%
      \def\colorgray#1{\color[gray]{#1}}%
      \expandafter\def\csname LTw\endcsname{\color{white}}%
      \expandafter\def\csname LTb\endcsname{\color{black}}%
      \expandafter\def\csname LTa\endcsname{\color{black}}%
      \expandafter\def\csname LT0\endcsname{\color[rgb]{1,0,0}}%
      \expandafter\def\csname LT1\endcsname{\color[rgb]{0,1,0}}%
      \expandafter\def\csname LT2\endcsname{\color[rgb]{0,0,1}}%
      \expandafter\def\csname LT3\endcsname{\color[rgb]{1,0,1}}%
      \expandafter\def\csname LT4\endcsname{\color[rgb]{0,1,1}}%
      \expandafter\def\csname LT5\endcsname{\color[rgb]{1,1,0}}%
      \expandafter\def\csname LT6\endcsname{\color[rgb]{0,0,0}}%
      \expandafter\def\csname LT7\endcsname{\color[rgb]{1,0.3,0}}%
      \expandafter\def\csname LT8\endcsname{\color[rgb]{0.5,0.5,0.5}}%
    \else
      \def\colorrgb#1{\color{black}}%
      \def\colorgray#1{\color[gray]{#1}}%
      \expandafter\def\csname LTw\endcsname{\color{white}}%
      \expandafter\def\csname LTb\endcsname{\color{black}}%
      \expandafter\def\csname LTa\endcsname{\color{black}}%
      \expandafter\def\csname LT0\endcsname{\color{black}}%
      \expandafter\def\csname LT1\endcsname{\color{black}}%
      \expandafter\def\csname LT2\endcsname{\color{black}}%
      \expandafter\def\csname LT3\endcsname{\color{black}}%
      \expandafter\def\csname LT4\endcsname{\color{black}}%
      \expandafter\def\csname LT5\endcsname{\color{black}}%
      \expandafter\def\csname LT6\endcsname{\color{black}}%
      \expandafter\def\csname LT7\endcsname{\color{black}}%
      \expandafter\def\csname LT8\endcsname{\color{black}}%
    \fi
  \fi
    \setlength{\unitlength}{0.0500bp}%
    \ifx\gptboxheight\undefined%
      \newlength{\gptboxheight}%
      \newlength{\gptboxwidth}%
      \newsavebox{\gptboxtext}%
    \fi%
    \setlength{\fboxrule}{0.5pt}%
    \setlength{\fboxsep}{1pt}%
\begin{picture}(7200.00,5040.00)%
    \gplgaddtomacro\gplbacktext{%
      \csname LTb\endcsname
      \put(682,704){\makebox(0,0)[r]{\strut{}$0$}}%
      \put(682,1072){\makebox(0,0)[r]{\strut{}$5$}}%
      \put(682,1439){\makebox(0,0)[r]{\strut{}$10$}}%
      \put(682,1807){\makebox(0,0)[r]{\strut{}$15$}}%
      \put(682,2174){\makebox(0,0)[r]{\strut{}$20$}}%
      \put(682,2542){\makebox(0,0)[r]{\strut{}$25$}}%
      \put(682,2909){\makebox(0,0)[r]{\strut{}$30$}}%
      \put(682,3277){\makebox(0,0)[r]{\strut{}$35$}}%
      \put(682,3644){\makebox(0,0)[r]{\strut{}$40$}}%
      \put(682,4012){\makebox(0,0)[r]{\strut{}$45$}}%
      \put(682,4379){\makebox(0,0)[r]{\strut{}$50$}}%
      \put(1275,484){\makebox(0,0){\strut{}$-0.2$}}%
      \put(2196,484){\makebox(0,0){\strut{}$0$}}%
      \put(3117,484){\makebox(0,0){\strut{}$0.2$}}%
      \put(4039,484){\makebox(0,0){\strut{}$0.4$}}%
      \put(4960,484){\makebox(0,0){\strut{}$0.6$}}%
      \put(5882,484){\makebox(0,0){\strut{}$0.8$}}%
      \put(6803,484){\makebox(0,0){\strut{}$1$}}%
    }%
    \gplgaddtomacro\gplfronttext{%
      \csname LTb\endcsname
      \put(198,2541){\rotatebox{-270}{\makebox(0,0){\strut{}$|\Omega_1^1(s)|^2$}}}%
      \put(3808,154){\makebox(0,0){\strut{}$s$ [GeV${}^2$]}}%
      \put(3808,4709){\makebox(0,0){\strut{}Omn\`es function from elastic $\pi\pi$ $P$-wave phase shift}}%
    }%
    \gplbacktext
    \put(0,0){\includegraphics{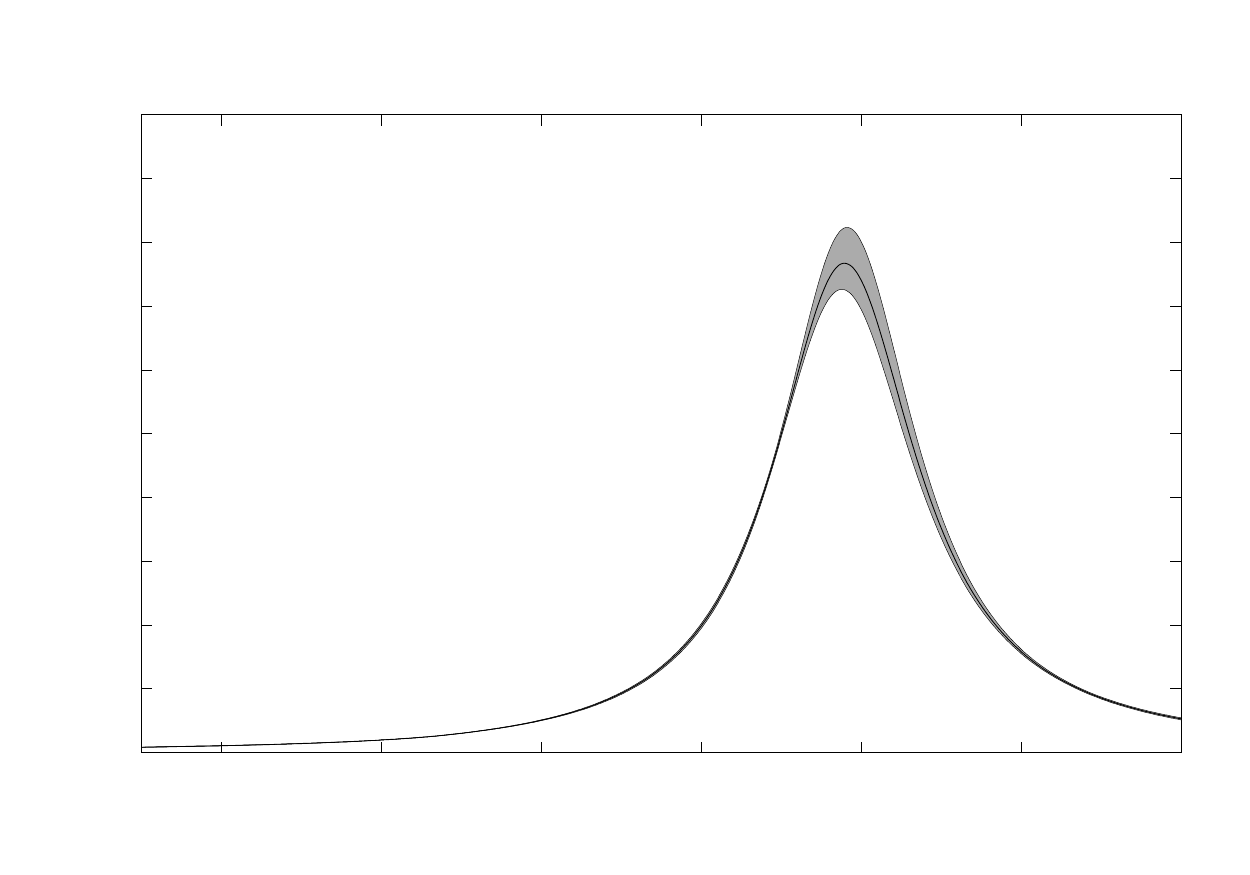}}%
    \gplfronttext
  \end{picture}%
\endgroup

%% file: sections/Fits.tex

\section{\boldmath Fits to $e^+e^-$ data}

In this section, we first describe in Sect.~\ref{sec:FitParameters} the parameters in our representation of the pion VFF. They are either fit to experimental data or treated as sources of systematic uncertainties in the theoretical description. In Sect.~\ref{sec:DataSetsUnbiasedFitting}, we give an overview of the available data sets and describe the procedure that we use to avoid bias in the fit. 
In Sect.~\ref{sec:FitResults}, we present the results of the fits to single experiments and in Sect.~\ref{sec:Combinations}, we perform fits to combinations of the data sets.
In Sect.~\ref{sec:omega}, we compare the fit result for the $\omega$ mass with extractions from other channels and discuss the observed tension.

\subsection{Fit parameters and systematic uncertainties}
\label{sec:FitParameters}

The representation of the pion VFF~\eqref{eq:PionVFF} is given by a product of three functions. Each of them contains parameters that we fit to data from $e^+e^-\to\pi^+\pi^-$ experiments.
First, the Omn\`es function $\Omega_1^1$ contains two free parameters, the values of the elastic $\pi\pi$ scattering $P$-wave phase shift at two points, $\delta_1^1(s_0)$ and $\delta_1^1(s_1)$.
Second, the function $G_\omega$ involves the $\rho$--$\omega$ mixing parameter $\epsilon_\omega$ as a free parameter, while the $\omega$ mass will either be taken as an input or considered a fit parameter as well.
Third, the function $G_\mathrm{in}^N$ describing inelastic contributions contains the $N-1$ fit parameters $c_k$ of the conformal polynomial ($c_1$ is constrained by~\eqref{eq:ConformalParameterPWaveConstraint}).
Finally, we will also consider fit variants in which we allow for an experimental uncertainty in the energy calibration, 
which we will implement by rescaling the energies of the data points constrained by the expected calibration uncertainty
of each experiment in the vicinity of the $\rho$ peak. For a single experiment, this strategy produces a similar effect as fitting the $\omega$ mass,
but for the combined fits it allows us to separate a single $\omega$ mass as determined from the $e^+e^-\to \pi^+\pi^-$ fits from variations among the different experiments
that might be attributed to the energy calibration.

All other parameters in the form factor representation are treated as sources of systematic uncertainties.
First, the 25 additional parameters in the solution of the Roy equations for the phase shift $\delta_1^1$ are varied independently within the ranges estimated in~\cite{Caprini:2011ky}, with the exception of $\iota_1$, which determines the elasticity~\eqref{eq:ElasticityParameter} and does not only appear in the phase shift 
but also in the \EL{} bound~\eqref{eq:ELBound}. This parameter is varied within $\iota_1 \in [0.01,0.09]$, as explained in Sect.~\ref{sec:ELBound}.
A second source of systematic uncertainty concerns the continuation of the phase shift to energies above the validity of the Roy equations as described in Sect.~\ref{sec:PhaseShiftInput}. 
If not fit to $e^+e^-\to\pi^+\pi^-$ data, the omega mass is taken as an input from the PDG~\cite{PDG2018}
 		\beq
			\label{eq:OmegaMass}
			\mw = 782.65(12)\MeV.
		\eeq
Since we do not observe any improvement of the fits by letting the $\omega$ width float, we keep it as an input~\cite{PDG2018}
		\beq
			\Gamma_\omega = 8.49(8)\MeV.
		\eeq
Next, in the conformal polynomial, the point $s_c$ that is mapped to the origin is a free parameter. It should be taken sufficiently far from any branch cuts. We vary it in the range
		\beq
			s_c = - (0.5\ldots2)\GeV^2
		\eeq
		and treat it as another source of systematic uncertainty.
Finally, the order $N$ of the conformal polynomial is varied between $N=2$ and $N=6$.

\subsection{Data sets and unbiased fitting}
\label{sec:DataSetsUnbiasedFitting}

In our fits of the pion VFF, we take into account the high-statistics time-like data sets from the $e^+e^-$ experiments. On the one hand, there are the results from the energy-scan $e^+e^-\to\pi^+\pi^-$ experiments SND~\cite{Achasov:2005rg,Achasov:2006vp} and CMD-2~\cite{Akhmetshin:2001ig,Akhmetshin:2003zn,Akhmetshin:2006wh,Akhmetshin:2006bx} at the VEPP-2M collider in Novosibirsk. On the other hand, the so-called radiative return measurements run at a fixed $e^+e^-$ 
energy and vary the $\pi^+\pi^-$ energy by making use of ISR in the process $e^+e^-\to\pi^+\pi^-\gamma$. These experiments are BaBar~\cite{Aubert:2009ad,Lees:2012cj} at SLAC, KLOE~\cite{Ambrosino:2008aa,Ambrosino:2010bv,Babusci:2012rp,Anastasi:2017eio} at the Frascati $\phi$-factory DA$\Phi$NE, and BESIII~\cite{Ablikim:2015orh} at the BEPCII collider in Beijing.

In addition to these time-like data sets, there is also some experimental information on the space-like form factor available from the scattering of pions off an electron target, performed by the F2 experiment~\cite{Dally:1982zk} at Fermilab and by NA7~\cite{Amendolia:1986wj} at CERN. Although we have checked consistency of the fit with the extraction of the space-like form factor from $e^-p\to e^-\pi^+n$ data by the JLab $F_\pi$ collaboration~\cite{Horn:2006tm,Tadevosyan:2007yd,Blok:2008jy,Huber:2008id}, we do not use these data in our fits because of their remaining model dependence due to the extrapolation to the pion pole.

\begin{table}[t]
	\centering
	\small
	\begin{tabular}{l c c c c}
	\toprule
	Experiment & Region of $s$ [GeV${}^2$] & \# data points & Statistical errors & Systematic errors \\
	\midrule
	F2~\cite{Dally:1982zk}									& $[-0.092, -0.039]$	&	14	&	diagonal	&	1\%		 \\
	NA7~\cite{Amendolia:1986wj}								& $[-0.253, -0.015]$	&	45	&	diagonal	&	0.9\%	 \\
	\midrule
	SND~\cite{Achasov:2005rg,Achasov:2006vp}					& $[0.152,0.941]$	&	45	&	diagonal	&	1.3\% for $s>(420\MeV)^2$ \\
											&		&		&			&	3.2\% for $s<(420\MeV)^2$ \\
	CMD-2~\cite{Akhmetshin:2001ig,Akhmetshin:2003zn,Akhmetshin:2006wh,Akhmetshin:2006bx}
														& $[0.373,0.925]$	&	43	&	diagonal	&	0.6\%	 \\
														& $[0.137,0.270]$	&	10	&	diagonal	&	0.7\%	 \\
														& $[0.360,0.941]$	&	29	&	diagonal	&	0.8\%	 \\
	BaBar~\cite{Aubert:2009ad,Lees:2012cj}						& $[0.093,8.703]$	&	337	&	full covariance	&	full covariance	 \\
														& $[0.093,0.998]$		&	270 \\
	KLOE~\cite{Ambrosino:2008aa,Ambrosino:2010bv,Babusci:2012rp,Anastasi:2017eio}	& $[0.355,0.945]$	&	195	&	full covariance	&	full covariance \\
	BESIII~\cite{Ablikim:2015orh}								& $[0.363,0.806]$	&	60	&	full covariance	&	0.9\%	 \\
	\bottomrule
	\end{tabular}
	\caption{Overview of the data sets that we use for the fits of the pion VFF. In most cases, the systematic uncertainty is an overall normalization uncertainty with 100\% correlation between all data points. For CMD-2, we treat the systematics between~\cite{Akhmetshin:2001ig,Akhmetshin:2003zn}, \cite{Akhmetshin:2006wh}, and~\cite{Akhmetshin:2006bx} as fully correlated, apart from the event separation, which is uncorrelated between the low-energy~\cite{Akhmetshin:2006wh} and high-energy~\cite{Akhmetshin:2001ig,Akhmetshin:2003zn}, \cite{Akhmetshin:2006bx} data sets~\cite{Novosibirsk}. For BaBar and KLOE, the systematic uncertainties have a more complicated covariance structure. From the BaBar data set, we only use the 270 data points below $1\GeV^2$.}
	\label{tab:Experiments}
\end{table}

For all the data sets that we use in the fits, the experimental uncertainties are split into statistical and systematic errors, see Table~\ref{tab:Experiments} for an overview. In the case of the space-like data sets and the energy-scan $e^+e^-$ experiments, the statistical uncertainties are assumed to be uncorrelated between the data points.
The systematic errors in general are multiplicative uncertainties similar to overall normalization errors. If fits to data with this type of uncertainties are performed by minimizing a $\chi^2$ function that is constructed with the naive covariance matrix
\beq
	\chi^2 = \sum_{i,j} ( f(x_i) - y_i ) \mathrm{Cov}(i,j)^{-1} ( f(x_j) - y_j ),
\eeq
one usually introduces a bias, as first observed by D'Agostini~\cite{DAgostini:1993arp}. The bias can be severe especially when combining different data sets with normalization uncertainties. 
We use an iterative method to avoid this bias as proposed by the NNPDF collaboration~\cite{Ball:2009qv}. To this end, we define a systematic covariance matrix for relative values
\beq
	\mathrm{Cov}_\mathrm{rel}^\mathrm{syst}(i,j) = \frac{\mathrm{Cov}^\mathrm{syst}(i,j)}{y_i y_j}
\eeq
and use the following covariance matrix in the $\chi^2$ function:
\beq
	\mathrm{Cov}(i,j) = \mathrm{Cov}^\mathrm{stat}(i,j) + f(x_i) f(x_j) \mathrm{Cov}_\mathrm{rel}^\mathrm{syst}(i,j),
\eeq
i.e.\ the relative systematic covariance is weighted by the values of the fit model and not the data. We assume some initial value for the model parameters and iterate the fit with a new covariance matrix constructed using the model function of the previous fit iteration. The iterative fit converges rapidly, typically after only a couple of iteration steps.

In the case of the space-like data sets, our fit function $f(x_i)$ is the squared modulus of the form factor $|F_\pi^V(s_i)|^2$ at the center-of-mass squared energies $s_i$ of the data points. For all time-like data sets, we use the bare cross sections, which are already undressed of VP effects, and we correct for FSR effects as explained in Sect.~\ref{sec:HVPRadiativeCorrections} to relate the bare cross section to the form factor in pure QCD.
In contrast, the VFF data directly provided by experiment still contain VP effects and therefore cannot be consistently fit with our QCD-only parametrization.
Hence, in the case of the energy-scan experiments SND and CMD-2, the fit function $f(x_i)$ is the FSR-inclusive bare cross section at the given center-of-mass squared energies $s_i$ of the data points,
with the fit function being derived from the QCD VFF accounting for the kinematic factors from~\eqref{eq:BareCrossSectionVSFormFactor} and 
FSR by means of~\eqref{FSR}.
In the case of the radiative return experiments, the provided cross-section measurements should be considered as an average value integrated over each energy bin~\cite{KLOE,BaBar}. Since the experiments do not provide a bin center weighted by the experimental distribution within the bin, we take as the fit function the theoretical bare cross section including FSR integrated over the energy bins
\beq
  f(x_i) := \frac{1}{s_i^\mathrm{max}-s_i^\mathrm{min}} \int_{s_i^\mathrm{min}}^{s_i^\mathrm{max}} ds \left[ 1 + \frac{\alpha}{\pi} \eta(s) \right] \sigma^{(0)}(s),
\eeq
since this prescription should be closest to a reweighting based on the experimental data themselves. 
The overall effect, equivalent to shifting the bin center according to the theoretical distribution, is small, but becomes relevant in the vicinity of the $\rho$--$\omega$ interference where the VFF is changing rapidly.

Finally, we implement the \EL{} bound by adding a penalty term to the $\chi^2$ function that only contributes if the difference between elastic $\pi\pi$ phase and form factor phase is larger than the central value of the bound:
\begin{align}
	\chi^2_\text{E\L{}} = \sum_i \frac{\big(\Delta(s_i) - \Delta_i^\mathrm{max}\big)^2}{\sigma_{\Delta_i^\mathrm{max}}^2} \theta\big( \Delta(s_i) - \Delta_i^\mathrm{max} \big), \qquad \Delta(s) = | \psi(s) - \delta_1^1(s) |^2 ,
\end{align}
where $\theta$ is the Heaviside step function. For $\sigma_{\Delta_i^\mathrm{max}}$, we use the uncertainty on the bound due to the cross section ratio $r$. The variation of the elasticity parameter is treated as a systematic uncertainty.

Since the data compilation~\cite{Eidelman:2003uh} only considers the contribution to the cross-section ratio $r$ from $I=1$ channels, we do not include the isospin-breaking factor $G_\omega(s)$ in the bound, i.e.\ we only constrain the phase of the inelasticity factor by identifying
\begin{align}
	\Delta(s) := \Big( \mathrm{arg} \left( G_\mathrm{in}^N(s) \right) \Big)^2,
\end{align}
but in any case away from the $\omega$ resonance the phase of $G_\omega$ is tiny. In the fit results, we do not include the data points of the \EL{} bound in the counting of the degrees of freedom, otherwise one might encounter a situation where small shifts in the model function change the number of degrees of freedom. 
This treatment is further justified by the fact that the contribution of $\chi^2_\text{E\L{}}$ to the total $\chi^2$ is typically very small.

\subsection{Fit results}
\label{sec:FitResults}

In the following, we discuss different fit strategies by comparing the goodness of the fits to single time-like data sets. We also perform simultaneous fits to a single time-like data set and the space-like data sets. These studies allow us to define an optimal fit strategy that we will use in Sect.~\ref{sec:Combinations} for fits to a combination of time-like (and space-like) data sets.

\begin{table}[t]
	\centering
	\scriptsize
	\begin{tabular}{l r l l l l l l}
	\toprule
	 & $\chi^2/$dof$\qquad$ & $p$-value & $\delta_1^1(s_0)$ [$^\circ$] & $\delta_1^1(s_1)$ [$^\circ$] & $10^3 \times \epsilon_\omega$ & $c_2$ & \hspace{-0.5cm} $10^{10} \times a_\mu^{\pi\pi}|_{[0.6,0.9]}$ \\
	\midrule
	SND		&	$243.9 / 41 = 5.95$		&	$1.1\times10^{-30}$	&	$110.3(2)$	&	$165.3(2)$	&	$1.94(4)$		&	$-0.109(8)$	&	$370.5(2.8)$	\\
	CMD-2	&	$178.0 / 78 = 2.28$		&	$8.8\times10^{-10}$	&	$110.0(3)$	&	$165.7(2)$	&	$1.75(5)$		&	$-0.099(7)$	&	$370.8(2.3)$	\\
	BaBar	&	$425.9 / 266 = 1.60$		&	$1.6\times10^{-9}$	&	$110.6(2)$	&	$165.9(2)$	&	$2.09(3)$		&	$-0.106(7)$	&	$376.4(1.9)$	\\
	KLOE	&	$345.0 / 191 = 1.81$		&	$6.0\times10^{-11}$	&	$110.6(1)$	&	$165.8(1)$	&	$1.80(3)$		&	$-0.077(3)$	&	$367.1(1.1)$	\\
	\bottomrule
	\end{tabular}
	\caption{Fit results for fixed $\omega$ parameters. The uncertainties are the fit errors only. The value for $a_\mu^{\pi\pi}$ denotes the $\pi\pi$ contribution from the energy region $\sqrt{s}\in[0.6,0.9]\GeV$.}
	\label{tab:FitsNoOmega}
\end{table}

\subsubsection[Fixed $\omega$ mass]{\boldmath Fixed $\omega$ mass}

In a first step, we fix the mass and width of the $\omega$ at the PDG values~\cite{PDG2018}. For simplicity, we use only one free parameter in the conformal polynomial, i.e.\ $N=2$. Therefore, in total we have four fit parameters: the two values of the phase shift, the $\rho$--$\omega$ mixing parameter $\epsilon_\omega$, and one parameter $c_2$ in the conformal polynomial.
In Table~\ref{tab:FitsNoOmega}, we show the results of the fits to single time-like data sets. Apart from the fit parameters, we show the value for the two-pion HVP contribution to $a_\mu$ from the energy region $\sqrt{s}\in[0.6,0.9]\GeV$ (including the FSR contribution according to~\eqref{FSR}). Although the values for $a_\mu$ are reasonable, the fit quality in general is very poor. The $p$-values clearly show that these fits are unacceptable.

This conclusion is most severe for the BESIII data set, for which we find a reduced $\chi^2$ of the order of $10$. This behavior can be traced back to the statistical covariance matrix, e.g.\ the exact same difficulties arise for any kind of global fit function. For instance, the Gounaris--Sakurai (GS)~\cite{Gounaris:1968mw} fits presented in~\cite{Ablikim:2015orh} are performed using the diagonal errors only, 
while a fit using the full covariance matrix breaks down in the same way as a fit using our dispersive representation. Moreover, this observation stands in marked contrast say to the BaBar data, 
for which a GS fit was performed including both systematic and off-diagonal statistical uncertainties as well, leading to a $\chi^2$ in a similar range as ours~\cite{Lees:2012cj}.
We conclude that with the statistical covariance matrix as provided together with~\cite{Ablikim:2015orh} no statistically meaningful description of the data is possible, and will therefore not consider 
the BESIII data set in the following.\footnote{The covariance matrix is currently being revisited by the BESIII collaboration~\cite{BESIII-private}.}

\begin{table}[t]
	\centering
	\scriptsize
	\begin{tabular}{l r c l l l l l l}
	\toprule
	 & $\chi^2/$dof$\qquad$ & $p$-value & $\mw$ [MeV]	&	$\delta_1^1(s_0)$ [$^\circ$] & $\delta_1^1(s_1)$ [$^\circ$] & $10^3 \times \epsilon_\omega$ & $c_2$ & \hspace{-0.5cm} $10^{10} \times a_\mu^{\pi\pi}|_{[0.6,0.9]}$ \\
	\midrule
	SND		&	$58.8 / 40 = 1.47$		&	$2.8\%$			&	$781.47(8)$	&	$110.1(2)$	&	$165.7(2)$	&	$2.03(4)$		&	$-0.099(8)$	&	$371.5(2.8)$	\\
	CMD-2	&	$92.0 / 77 = 1.19$		&	$12\%$			&	$781.97(7)$	&	$110.0(3)$	&	$166.2(2)$	&	$1.89(5)$		&	$-0.085(9)$	&	$373.0(2.6)$	\\
	BaBar	&	$305.4 / 265 = 1.15$		&	$4.5\%$			&	$781.85(8)$	&	$110.2(2)$	&	$165.8(2)$	&	$2.04(3)$		&	$-0.105(7)$	&	$374.2(1.9)$	\\
	KLOE	&	$289.9 / 190 = 1.53$		&	$4.1\times10^{-6}$	&	$781.62(11)$	&	$110.4(1)$	&	$165.7(1)$	&	$1.97(4)$		&	$-0.075(3)$	&	$366.1(1.1)$	\\
	\bottomrule
	\end{tabular}
	\caption{The same as Table~\ref{tab:FitsNoOmega}, but with the $\omega$ mass as a free fit parameter.}
	\label{tab:FitsOmegaMass}
\end{table}

\begin{table}[t]
	\centering
	\scriptsize
	\begin{tabular}{l r c c l l l l l}
	\toprule
	 & $\chi^2/$dof$\qquad$ & $p$-value  & $10^3 \times \xi_j \hspace{-0.1cm}$	&	$\delta_1^1(s_0)$ [$^\circ$] & $\delta_1^1(s_1)$ [$^\circ$] & $10^3 \times \epsilon_\omega$ & $c_2$ & \hspace{-0.5cm} $10^{10} \times a_\mu^{\pi\pi}|_{[0.6,0.9]}$ \\
	\midrule
	SND		&	$58.3 / 40 = 1.46$		&	$3.1\%$			&	$2.4(2)$		&	$109.4(3)$	&	$165.8(2)$	&	$2.03(4)$		&	$-0.102(8)$	&	$372.3(2.9)$	\\
	CMD-2	&	$92.0 / 77 = 1.19$		&	$12\%$			&	$1.4(1)$		&	$109.6(3)$	&	$166.2(2)$	&	$1.89(5)$		&	$-0.086(9)$	&	$373.2(2.6)$	\\
	BaBar	&	$304.6 / 265 = 1.15$		&	$4.8\%$			&	$1.6(2)$		&	$109.8(2)$	&	$165.8(2)$	&	$2.04(3)$		&	$-0.107(7)$	&	$374.6(1.9)$	\\
	KLOE	&	$290.4 / 190 = 1.53$		&	$3.7\times10^{-6}$	&	$2.0(2)$		&	$109.8(1)$	&	$165.8(1)$	&	$1.97(4)$		&	$-0.076(3)$	&	$366.5(1.1)$	\\
	\bottomrule
	\end{tabular}
	\caption{The same as Table~\ref{tab:FitsNoOmega}, but with an energy rescaling for each experiment according to~\eqref{eq:EnergyRescaling}.}
	\label{tab:FitsEnergyRescaling}
\end{table}

\begin{table}[t]
	\centering
	\scalebox{0.82}{
	\small
	\begin{tabular}{l r c l c l l l l l}
	\toprule
	 & $\chi^2/$dof$\qquad$ & $p$-value & $\mw$ [MeV] & $10^3 \times \xi_j \hspace{-0.1cm}$	&	$\delta_1^1(s_0)$ [$^\circ$] & $\delta_1^1(s_1)$ [$^\circ$] & $10^3 \times \epsilon_\omega$ & $c_2$ & \hspace{-0.5cm} $10^{10} \times a_\mu^{\pi\pi}|_{[0.6,0.9]}$ \\
	\midrule
	SND		&	$58.8 / 40 = 1.47$		&	$2.8\%$	&			$781.49(27)$	&	$0.0(5)$		&	$110.1(3)$	&	$165.7(2)$	&	$2.03(4)$		&	$-0.099(8)$	&	$371.6(2.8)$	\\
	CMD-2	&	$92.0 / 77 = 1.19$		&	$12\%$	&			$781.97(27)$	&	$0.0(5)$		&	$110.0(3)$	&	$166.2(2)$	&	$1.89(5)$		&	$-0.085(9)$	&	$373.0(2.6)$	\\
	BaBar	&	$305.4 / 265 = 1.15$		&	$4.5\%$	&			$781.86(13)$	&	$0.0(2)$		&	$110.2(2)$	&	$165.8(2)$	&	$2.04(3)$		&	$-0.105(7)$	&	$374.2(1.9)$	\\
	KLOE	&	$289.9 / 190 = 1.53$		&	$4.1\times10^{-6}$	&	$781.62(17)$	&	$0.0(3)$		&	$110.4(1)$	&	$165.7(1)$	&	$1.97(4)$		&	$-0.075(3)$	&	$366.1(1.1)$	\\
	\bottomrule
	\end{tabular}
	}
	\caption{The same as Table~\ref{tab:FitsNoOmega}, but with both free $\omega$ mass and energy rescaling.}
	\label{tab:FitsEnergyRescalingOmegaMass}
\end{table}

\subsubsection[Fitting the $\omega$ mass]{\boldmath Fitting the $\omega$ mass}

In a next step, we use the $\omega$ mass as a free fit parameter and disregard the input from the PDG. The results of the fits to single $e^+e^-$ data sets are shown in Table~\ref{tab:FitsOmegaMass}. 
The fits to the energy-scan experiments and BaBar are now of good quality.
Unfortunately, the fit to KLOE is only improved slightly, and fitting the $\omega$ width as well does not improve the fit either.

However, the fit result for the $\omega$ mass is not in agreement with the value~\eqref{eq:OmegaMass} from the PDG~\cite{PDG2018}, which is dominated by $e^+e^-\to3\pi$ and $e^+e^-\to\pi^0\gamma$ experiments at SND and CMD-2~\cite{Achasov:2003ir,Akhmetshin:2003zn,Akhmetshin:2004gw} as well as from $\bar p p\to\omega\pi^0\pi^0$~\cite{Amsler:1993pr}. Due to the fact that in the two-pion channel the $\omega$ resonance appears very close to the broader $\rho$ resonance and only due to an isospin-violating effect, it seems unlikely that the extraction of the $\omega$ mass from the two-pion channel should be more reliable than 
from these channels. Therefore, one might suspect that consistency among the different channels could require allowing for another source of uncertainty 
related to the energy calibration in the respective experiments. This possible explanation is further pursued below in terms of fits that implement precisely such an energy rescaling.

Finally, we note that the fit values of the phase shift are in all fits perfectly compatible with the values~\eqref{eq:Delta11ParametersRoy} used in the Roy-equation analysis~\cite{Caprini:2011ky}, and, even more importantly, consistent among the different data sets at a level well below the uncertainties quoted in~\eqref{eq:Delta11ParametersRoy}. This shows the potential in further improving the $\pi\pi$ $P$-wave phase shift from the present VFF fits.

\begin{table}[t]
	\centering
	\scalebox{0.82}{
	\small
	\arraycolsep=-0.3cm
	\begin{tabular}{l r c l c l l l l l}
	\toprule
	 & $\chi^2/$dof$\qquad$ & $p$-value & $\mw$ [MeV] & $10^3 \times \xi_j \hspace{-0.1cm}$	&	$\delta_1^1(s_0)$ [$^\circ$] & $\delta_1^1(s_1)$ [$^\circ$] & $10^3 \times \epsilon_\omega$ & $c_2$ & \hspace{-0.5cm} $10^{10} \times a_\mu^{\pi\pi}|_{[0.6,0.9]}$ \\
	\midrule
	$\;\begin{array}{l} \text{KLOE08} \\ \text{KLOE10} \\ \text{KLOE12} \end{array}$
			&	$268.5 / 190 = 1.41$	&	$1.5\times10^{-4}$	&	$781.78(14)$	&	$\begin{matrix*}[r]	0.6(2)	\\	-0.3(2)	\\	-0.2(2) \end{matrix*}$	&	$110.2(1)$	&	$165.7(1)$	&	$1.98(4)$	&	$-0.073(3)$	&	$365.3(1.1)$	\\
	\midrule
	$\;\begin{array}{l} \text{KLOE08}'' \\ \text{KLOE10} \\ \text{KLOE12} \end{array}$
			&	$235.2 / 188 = 1.25$	&	$1.1\%$			&	$781.78(14)$	&	$\begin{matrix*}[r]	0.5(2)	\\	-0.3(2)	\\	-0.2(2) \end{matrix*}$	&	$110.2(1)$	&	$165.7(1)$	&	$1.98(4)$	&	$-0.072(3)$	&	$365.0(1.1)$	\\
	\bottomrule
	\end{tabular}
	}
	\caption{The same as Table~\ref{tab:FitsEnergyRescalingOmegaMass} with individual energy rescalings for the KLOE experiments, but a common $\omega$ mass. In the set KLOE08$''$, two outliers have been deleted, see Sect.~\ref{sec:Outliers}.}
	\label{tab:FitsEnergyRescalingKLOE}
\end{table}

\subsubsection{Energy rescaling}

Instead of fitting the $\omega$ mass, proper alignment with its PDG value could be ensured by rescaling the energies of the time-like data points $i$ by
\beq
	\label{eq:EnergyRescaling}
	\sqrt{s_i} \mapsto \sqrt{s_i} + \xi_j ( \sqrt{s_i} - 2M_\pi ),
\eeq
where $\xi_j$ is a small rescaling factor for each experiment $j$ and we have chosen this mapping to leave the two-pion threshold invariant. The rescaling of the energy affects the relation~\eqref{eq:BareCrossSectionVSFormFactor} between the form factor and the bare cross section (we neglect the rescaling in the FSR correction). The effect can be described by
\beq
	|F_\pi^V|_i^2 \mapsto |F_\pi^V|_i^2 \Big( 1 + \xi_j A(s_i) + \O(\xi_i^2) + \O(m_e^2) \Big) \, , \quad A(s) = \frac{2(s-10M_\pi^2)}{s+2\sqrt{s}M_\pi},
\eeq
where $A(s) \in [-1.5,2]$ for $s\ge 4M_\pi^2$.

The results of this fit are shown in Table~\ref{tab:FitsEnergyRescaling}. They are almost indistinguishable from the ones where the $\omega$ mass is fit. We also note that the exact form of the rescaling~\eqref{eq:EnergyRescaling} proves immaterial, given that a simpler rescaling $s_i \mapsto \xi_j^2 s_i$ or a small energy shift are possible as well and lead to almost identical results. 
As the energy rescaling is at the permille level, the effect on the integrated $a_\mu$ is very small, while the improvement in the $\chi^2$ compared to the fit with fixed $\omega$ mass and no energy rescaling is critical to obtain acceptable fits. In the end, it appears to be simply related to the correct alignment of the $\omega$ resonance, which, when insisting on the 
PDG value~\eqref{eq:OmegaMass}, necessitates some rescaling as in~\eqref{eq:EnergyRescaling}. However, the implied energy calibration uncertainties as large as $1\MeV$ in the $\rho$ peak 
are significantly larger than estimated in the respective experiments~\cite{Novosibirsk,KLOE,BaBar}, pointing to a corresponding tension in the $\omega$ mass among different channels. 
To separate these issues, in particular in the combined fits, we will from now on keep both a global $\omega$ mass and individual rescalings for each experiment as free fit parameters, but
constrain each rescaling by an additional penalty $\Delta\chi^2_j=(\xi_j M_\rho/\Delta E_j)^2$, where the calibration uncertainty in the $\rho$ peak is estimated as $\Delta E=0.4\MeV$ for the Novosibirsk data sets~\cite{Novosibirsk}, $0.16\MeV$ for BaBar~\cite{Lees:2012cj}, and $0.2\MeV$ for KLOE~\cite{KLOE}. In contrast to the \EL{} bound, we will count these terms as additional data points in the number of degrees of freedom. The results shown in Table~\ref{tab:FitsEnergyRescalingOmegaMass} illustrate the fact that a free $\omega$ mass and an energy rescaling are all but equivalent, with 
the corresponding flat direction broken by the requirement that the energy rescaling not be larger than acceptable given the estimate of the experimental calibration uncertainty.  

In the case of KLOE, we have used the combination of the KLOE08, KLOE10, and KLOE12 results~\cite{Anastasi:2017eio}, but in all fit variants considered so far the fit quality is significantly worse than for the other experiments. However, as shown in Table~\ref{tab:FitsEnergyRescalingKLOE}, we observe a significant improvement of the $\chi^2$
if we allow for different energy rescalings $\xi_j$ for each of the three KLOE experiments. Since it may indeed be that energy calibration uncertainties differ among the three KLOE data sets,
we will allow for three individual rescalings in the following, each constrained by an uncertainty of $0.2\MeV$ at the $\rho$ peak.

\subsubsection{Possible outliers}
\label{sec:Outliers}

\begin{figure}[t]
	\centering
	\begin{subfigure}{8cm}
		\hspace{-0.5cm}
		\input{plots/chi2-KLOE}
		\caption{}
		\label{img:chi2KLOE}
	\end{subfigure}
	\begin{subfigure}{8cm}
		\hspace{-0.5cm}
		\input{plots/chi2-BESIII}
		\caption{}
		\label{img:chi2BESIII}
	\end{subfigure}
	\caption{{\bf (a)} Individual bin contributions to the $\chi^2$ in the KLOE fit, restricted to the 60 bins of KLOE08. Bins \#15 and \#22 appear as outliers and are marked by red circles. {\bf (b)} Individual bin contributions to the $\chi^2$ in the BESIII fit. There are huge positive and negative contributions and it is not possible to identify single outliers.}
	\label{img:chi2}
\end{figure}
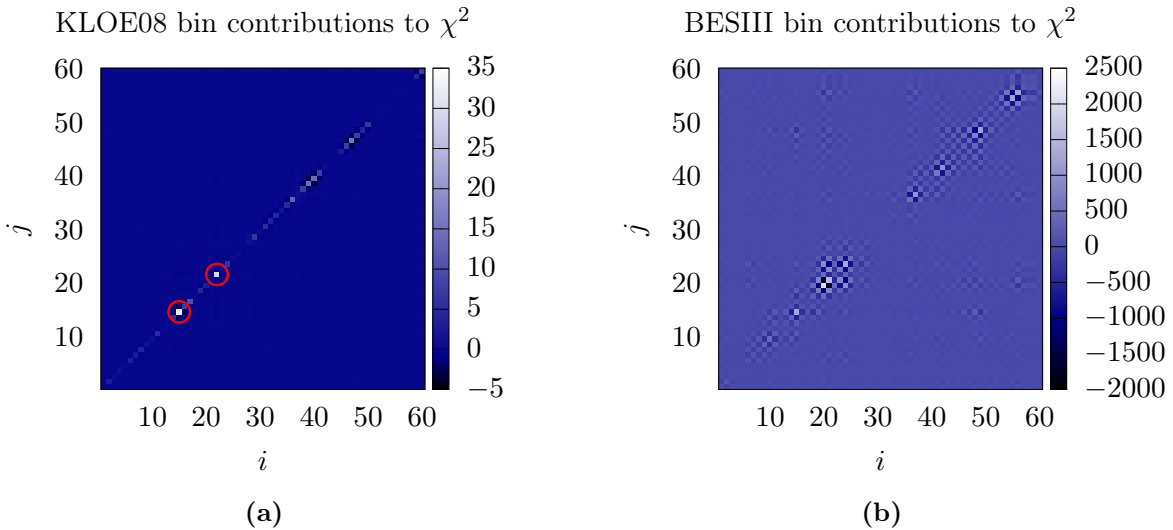

Let us scrutinize the fit results to KLOE and BESIII. By considering the individual contributions to the $\chi^2$ from each energy bin, we were able to identify in the KLOE set two bins with
wildly disproportionate contributions to the $\chi^2$: if we remove bins \#15 and \#22 from the KLOE08 set, the total $\chi^2$ reduces by more than 30 units, as shown in Table~\ref{tab:FitsEnergyRescalingKLOE} (the set with deleted outliers is marked as KLOE08$''$). In Fig.~\ref{img:chi2KLOE}, we show the individual bin contributions,
\beq
	\chi^2_{ij} = ( f(x_i) - y_i ) \mathrm{Cov}(i,j)^{-1} ( f(x_j) - y_j ),
\eeq
to the $\chi^2$ in the KLOE fit, restricted to the 60 bins of KLOE08.
The bins \#15 and \#22 are marked by red circles. They correspond to $703.56\MeV$ and $751.66\MeV$, where the form factor is expected to show no particular structure. In fact, even in the vicinity of the $\rho$--$\omega$ interference, where the VFF varies much more rapidly, no conspicuous contributions to the $\chi^2$ arise, see Fig.~\ref{img:chi2KLOE}, while bins \#15 and \#22   
are clearly visible. This suggests to discard them as obvious outliers. We will denote the corresponding data set by KLOE$''$ in the remainder of the paper, but also show results for the full KLOE data set. In the end, the main impact is restricted to the goodness of the fit, the results for the HVP contribution to $a_\mu$ or the pion radius are hardly affected.  

Unfortunately, in the case of BESIII we were not able to identify similar outliers. In Fig.~\ref{img:chi2BESIII}, we show the individual bin contributions to the $\chi^2$ in the BESIII fit. We observe fluctuations between huge positive and negative values. If we only take into account the diagonal elements of the covariance matrix, a perfect fit (with a reduced $\chi^2$ around 1) is possible. 
As mentioned above, this suggests that there might be a problem with the
BESIII covariance matrix. We remark in addition that the diagonal elements
of the statistical covariance matrix in the supplementary material
of~\cite{Ablikim:2015orh} do not agree with the diagonal errors published
in the same reference.

\subsubsection{Including space-like data sets}

We now perform fits to a combination of one time-like data set and the space-like data from NA7 (we also tried including F2, in addition, but the gain in statistics is entirely negligible and we drop the corresponding data set for simplicity). Although for $(g-2)_\mu$ we only integrate over the time-like region above the threshold, $s\ge 4M_\pi^2$, analyticity provides the connection between the time-like and space-like region, so that we can use experimental input from both regions to constrain the form factor.
In principle, the same discussion of radiative corrections as in Sect.~\ref{sec:HVPRadiativeCorrections} arises, but fortunately the applied radiative corrections in the space-like data sets
include VP~\cite{Kahane:1964zz,Adylov:1977kj}, so that the provided data for the form factor can be used without further adjustments.

In Table~\ref{tab:FitsEnergyRescalingSpacelike} we show the results of the combined fits including NA7. The NA7 data are perfectly compatible with the fits to all time-like experiments. Since they have much larger uncertainties than the $e^+e^-$ experiments, their influence on the fit result is minor, mainly leading to a smaller $\chi^2/\mathrm{dof}$. This is even more so in the case of the F2 data, which do not have any observable influence on the fit.

\begin{table}[t]
	\centering
	\scalebox{0.80}{
	\small
	\begin{tabular}{l r c l r l l l l l}
	\toprule
	 & $\chi^2/$dof$\qquad$ & $p$-value & $\mw$ [MeV] & $10^3 \times \xi_j \hspace{-0.1cm}$	&	$\delta_1^1(s_0)$ [$^\circ$] & $\delta_1^1(s_1)$ [$^\circ$] & $10^3 \times \epsilon_\omega$ & $c_2$ & \hspace{-1.0cm} $10^{10} \times a_\mu^{\pi\pi}|_{[0.6,0.9]}$ \\
	\midrule
	SND, NA7		&	$101.7 / 85 = 1.20$	&	$10\%$			&	$781.49(27)$	&	$0.0(5)$	&	$110.1(3)$	&	$165.7(2)$	&	$2.03(4)$	&	$-0.100(8)$	&	$371.9(2.8)$	\\
	CMD-2, NA7	&	$135.3 / 122 = 1.11$	&	$19\%$			&	$781.97(27)$	&	$0.0(5)$	&	$110.0(3)$	&	$166.2(2)$	&	$1.89(5)$	&	$-0.086(9)$	&	$373.3(2.5)$	\\
	BaBar, NA7	&	$348.1 / 310 = 1.12$	&	$6.7\%$			&	$781.86(13)$	&	$0.0(2)$	&	$110.2(2)$	&	$165.8(2)$	&	$2.04(3)$	&	$-0.106(7)$	&	$374.3(1.9)$	\\
	KLOE, NA7	&	$312.1 / 235 = 1.33$	&	$5.6\times10^{-4}$	&	$781.78(14)$	&	$\!\!\!\Bigg\{\begin{matrix*}[r] 0.6(2) \\ -0.3(2) \\ -0.2(2) \end{matrix*}$	&	$110.2(1)$	&	$165.7(1)$	&	$1.98(4)$	&	$-0.073(3)$	&	$365.4(1.1)$	\\
	KLOE$''$, NA7	&	$278.8 / 233 = 1.20$	&	$2.1\%$			&	$781.78(14)$	&	$\!\!\!\Bigg\{\begin{matrix*}[r] 0.5(2) \\ -0.3(2) \\ -0.2(2) \end{matrix*}$	&	$110.2(1)$	&	$165.7(1)$	&	$1.98(4)$	&	$-0.072(3)$	&	$365.0(1.1)$	\\
	\bottomrule
	\end{tabular}
	}
	\caption{Combined fits to one time-like experiment and the space-like NA7 data set. In the KLOE$''$ set, the two outliers in KLOE08 have been removed. No rescaling of $s$ has been applied to the space-like data.}
	\label{tab:FitsEnergyRescalingSpacelike}
\end{table}

\subsubsection{Varying the order of the conformal polynomial}

\begin{figure}[H]
	\centering
	\scalebox{0.95}{
	\input{plots/phasediff}}%
	\caption{Phase of the inelastic contribution $G_\mathrm{in}^N(s)$ for $N-1=4$ free parameters, shown together with the \EL{} bound ($\iota_1=0.05$).}
	\label{img:phasediff}
\end{figure}
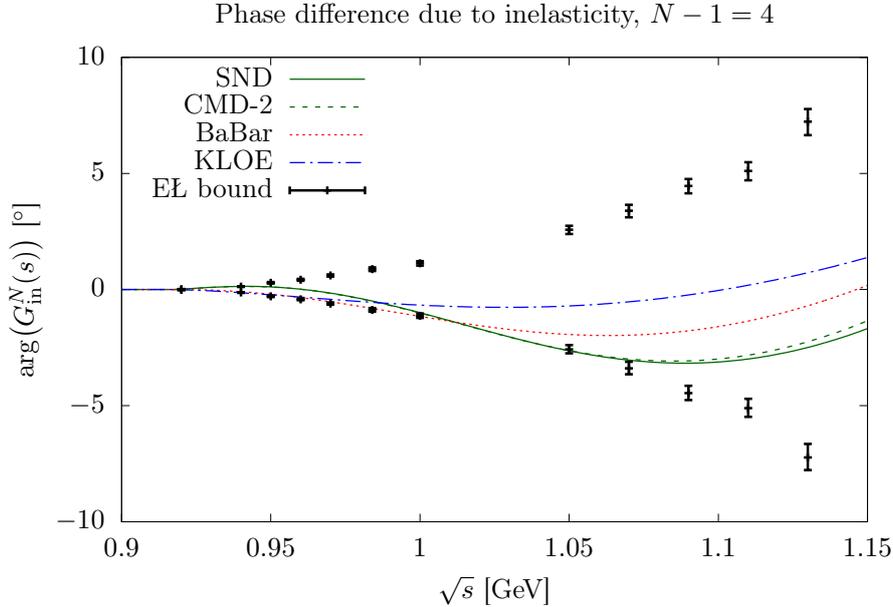

In a final step, we vary the order $N$ of the conformal polynomial used to describe inelasticity effects. Due to the $P$-wave constraint~\eqref{eq:ConformalParameterPWaveConstraint}, the number of free parameters in the conformal polynomial is $N-1$. The fit results for $N-1=1\ldots5$ are shown in Table~\ref{tab:FitsConfPoly}. The fit quality is good in all cases, provided that we remove the two outliers from the KLOE08 set. For small $N$, the \EL{} bound is fulfilled either automatically or imposed at only one point, while for larger $N$, the number of points where the bound is activated increases. We have performed fits with up to $N-1=7$ free parameters in the conformal polynomial. In the case of BaBar and KLOE, the $\chi^2$ does not improve any more for $N-1>4$, 
while for SND and CMD-2 some further improvement might be inferred, but due to the large number of parameters their fit values become unnaturally large and highly correlated. 
In all cases, the results for $a_\mu^{\pi\pi}$ remain stable for larger values of $N$, with the main effect that the parameters of the conformal polynomial receive large uncertainties and the \EL{} bound becomes an increasingly important constraint. 
Moreover, the phase of the inelasticity contribution $G_\mathrm{in}^N$ starts to oscillate for higher values of $N$, indicating further that very large values of $N$ do not correspond to a physically acceptable solution.
Therefore, we choose $N-1=4$ free parameters as the central fit configuration and take the effects due to the variation of $N-1=1\ldots5$ as a systematic uncertainty. 
The inelastic phase for $N-1=4$ is shown in Fig.~\ref{img:phasediff} together with the \EL{} bound.

\begin{table}[H]
	\centering
	\scalebox{0.55}{
	\small
	\begin{tabular}{l c r c l r l l l l l l l l l l l}
	\toprule
		& $N-1$ & $\chi^2/$dof$\qquad$ & $p$-value & $\mw$ [MeV] & $10^3 \times \xi_j \hspace{-0.1cm}$	&	$\delta_1^1(s_0)$ [$^\circ$] & $\delta_1^1(s_1)$ [$^\circ$] & $10^3 \times \epsilon_\omega$ & $c_2$ & $c_3$ & $c_4$ & $c_5$ & $c_6$
		& \hspace{-1.0cm} $10^{10} \times a_\mu^{\pi\pi}|_{[0.6,0.9]}$ \\
	\midrule
	SND		&	1	&	$58.8 / 40 = 1.47$		&	$2.8\%$			&	$781.49(27)$	&	$0.0(5)$	&	$110.1(3)$	&	$165.7(2)$	&	$2.03(4)$	&	$-0.099(8)$	&	&	&	&	&	$371.6(2.8)$	\\
	CMD-2	&	1	&	$92.0 / 77 = 1.19$		&	$12\%$			&	$781.97(27)$	&	$0.0(5)$	&	$110.0(3)$	&	$166.2(2)$	&	$1.89(5)$	&	$-0.085(9)$	&	&	&	&	&	$373.0(2.6)$	\\
	BaBar	&	1	&	$305.4 / 265 = 1.15$		&	$4.5\%$			&	$781.86(13)$	&	$0.0(2)$	&	$110.2(2)$	&	$165.8(2)$	&	$2.04(3)$	&	$-0.105(7)$	&	&	&	&	&	$374.2(1.9)$	\\
	KLOE	&	1	&	$268.5 / 190 = 1.41$		&	$1.5\times10^{-4}$	&	$781.78(14)$	&	$\Bigg\{\begin{matrix*}[r] 0.6(2) \\ -0.3(2) \\ -0.2(2) \end{matrix*}$	&	$110.2(1)$	&	$165.7(1)$	&	$1.98(4)$	&	$-0.073(3)$	&	&	&	&	&	$365.3(1.1)$	\\
	KLOE$''$	&	1	&	$235.2 / 188 = 1.25$		&	$1.1\%$			&	$781.78(14)$	&	$\Bigg\{\begin{matrix*}[r] 0.5(2) \\ -0.3(2) \\ -0.2(2) \end{matrix*}$	&	$110.2(1)$	&	$165.7(1)$	&	$1.98(4)$	&	$-0.072(3)$	&	&	&	&	&	$365.0(1.1)$	\\
	\midrule
	SND		&	2	&	$57.5 / 39 = 1.47$		&	$2.8\%$			&	$781.48(27)$	&	$0.0(5)$	&	$110.1(3)$	&	$165.6(3)$	&	$2.04(4)$	&	$-0.18(13)$	&	$\phantom{+}0.03(6)$	&	&	&	&	$375.1(4.6)$	\\
	CMD-2	&	2	&	$92.0 / 76 = 1.21$		&	$10\%$			&	$781.98(27)$	&	$0.0(5)$	&	$109.9(4)$	&	$166.2(3)$	&	$1.89(5)$	&	$-0.10(13)$	&	$\phantom{+}0.01(6)$	&	&	&	&		$373.1(2.9)$	\\
	BaBar	&	2	&	$302.5 / 264 = 1.15$		&	$5.1\%$			&	$781.85(13)$	&	$0.0(2)$	&	$110.2(2)$	&	$165.6(2)$	&	$2.06(3)$	&	$-0.20(9)$		&	$\phantom{+}0.04(4)$	&	&	&	&		$376.5(2.5)$	\\
	KLOE	&	2	&	$259.8 / 189 = 1.37$		&	$4.9\times10^{-4}$	&	$781.81(14)$	&	$\Bigg\{\begin{matrix*}[r] 0.6(2) \\ -0.3(2) \\ -0.2(2) \end{matrix*}$	&	$110.2(1)$	&	$165.5(1)$	&	$1.99(4)$	&	$-0.19(4)$	&	$\phantom{+}0.05(2)$	&	&	&	&		$368.0(1.4)$	\\
	KLOE$''$	&	2	&	$227.0 / 187 = 1.21$		&	$2.4\%$			&	$781.80(14)$	&	$\Bigg\{\begin{matrix*}[r] 0.5(2) \\ -0.3(2) \\ -0.2(2) \end{matrix*}$	&	$110.1(1)$	&	$165.5(1)$	&	$1.99(4)$	&	$-0.18(4)$	&	$\phantom{+}0.05(2)$	&	&	&	&		$367.5(1.4)$	\\
	\midrule
	SND		&	3	&	$57.1 / 38 = 1.50$		&	$2.4\%$			&	$781.48(27)$	&	$0.0(5)$	&	$110.0(4)$	&	$165.5(3)$	&	$2.04(4)$	&	$-0.33(29)$	&	$\phantom{+}0.17(23)$	&	$-0.04(6)$	&	&	&		$376.1(4.7)$	\\
	CMD-2	&	3	&	$92.0 / 75 = 1.23$		&	$8.9\%$			&	$781.98(27)$	&	$0.0(5)$	&	$109.9(4)$	&	$166.2(3)$	&	$1.89(5)$	&	$-0.15(29)$	&	$\phantom{+}0.05(23)$	&	$-0.02(7)$	&	&	&		$373.2(2.9)$	\\
	BaBar	&	3	&	$301.4 / 263 = 1.15$		&	$5.2\%$			&	$781.85(13)$	&	$0.0(2)$	&	$110.2(2)$	&	$165.6(2)$	&	$2.05(3)$	&	$\phantom{+}0.02(17)$	&	$-0.20(13)$	&	$\phantom{+}0.09(4)$	&	&	&		$376.1(2.5)$	\\
	KLOE	&	3	&	$255.9 / 188 = 1.36$		&	$7.3\times10^{-4}$	&	$781.80(14)$	&	$\Bigg\{\begin{matrix*}[r] 0.6(2) \\ -0.3(2) \\ -0.2(2) \end{matrix*}$	&	$110.2(1)$	&	$165.5(1)$	&	$1.99(4)$	&	$\phantom{+}0.00(8)$	&	$-0.15(7)$	&	$\phantom{+}0.07(2)$	&	&	&		$367.4(1.4)$	\\
	KLOE$''$	&	3	&	$223.1 / 186 = 1.20$		&	$3.3\%$			&	$781.80(14)$	&	$\Bigg\{\begin{matrix*}[r] 0.5(2) \\ -0.3(2) \\ -0.2(2) \end{matrix*}$	&	$110.1(1)$	&	$165.5(1)$	&	$1.99(4)$	&	$\phantom{+}0.01(8)$	&	$-0.15(7)$	&	$\phantom{+}0.07(2)$	&	&	&		$367.0(1.4)$	\\
	\midrule
	SND		&	4	&	$51.9 / 37 = 1.40$		&	$5.3\%$			&	$781.49(27)$	&	$0.0(5)$	&	$110.5(4)$	&	$165.7(3)$	&	$2.03(4)$	&	$\phantom{+}1.76(67)$	&	$-2.69(78)$	&	$\phantom{+}1.81(43)$	&	$-0.47(10)$	&	&		$373.6(4.7)$	\\
	CMD-2	&	4	&	$87.4 / 74 = 1.18$		&	$14\%$			&	$781.98(27)$	&	$0.0(5)$	&	$110.5(5)$	&	$166.4(3)$	&	$1.88(5)$	&	$\phantom{+}2.02(70)$	&	$-2.93(82)$	&	$\phantom{+}1.92(46)$	&	$-0.49(10)$	&	&	$372.2(2.9)$	\\
	BaBar	&	4	&	$299.1 / 262 = 1.14$		&	$5.7\%$			&	$781.86(13)$	&	$0.0(2)$	&	$110.4(3)$	&	$165.7(2)$	&	$2.04(3)$	&	$\phantom{+}1.18(63)$	&	$-1.86(88)$	&	$\phantom{+}1.20(61)$	&	$-0.29(17)$	&	&		$375.3(2.5)$	\\
	KLOE	&	4	&	$254.5 / 187 = 1.36$		&	$7.4\times10^{-4}$	&	$781.82(14)$	&	$\Bigg\{\begin{matrix*}[r] 0.6(2) \\ -0.3(2) \\ -0.2(2) \end{matrix*}$	&	$110.4(2)$	&	$165.6(1)$	&	$1.97(4)$	&	$\phantom{+}0.62(53)$	&	$-0.98(72)$	&	$\phantom{+}0.60(46)$	&	$-0.13(12)$	&	&		$366.8(1.5)$	\\
	KLOE$''$	&	4	&	$222.5 / 185 = 1.20$		&	$3.1\%$			&	$781.81(14)$	&	$\Bigg\{\begin{matrix*}[r] 0.5(2) \\ -0.3(2) \\ -0.2(2) \end{matrix*}$	&	$110.3(2)$	&	$165.6(1)$	&	$1.98(4)$	&	$\phantom{+}0.45(54)$	&	$-0.75(72)$	&	$\phantom{+}0.46(46)$	&	$-0.10(12)$	&	&		$366.5(1.5)$	\\
	\midrule
	SND		&	5	&	$51.5 / 36 = 1.43$		&	$4.6\%$			&	$781.49(27)$	&	$0.0(5)$	&	$110.5(4)$	&	$165.7(3)$	&	$2.03(4)$	&	$\phantom{+}2.47(97)$	&	$-4.16(1.51)$	&	$\phantom{+}3.44(1.40)$	&	$-1.39(72)$	&	$\phantom{+}0.21(16)$	&	$373.4(4.7)$	\\
	CMD-2	&	5	&	$87.3 / 73 = 1.20$		&	$12\%$			&	$781.98(27)$	&	$0.0(5)$	&	$110.5(5)$	&	$166.4(3)$	&	$1.88(5)$	&	$\phantom{+}1.74(93)$	&	$-2.33(1.34)$	&	$\phantom{+}1.26(1.09)$	&	$-0.12(49)$	&	$-0.09(10)$	&	$372.3(2.9)$	\\
	BaBar	&	5	&	$298.9 / 261 = 1.15$		&	$5.4\%$			&	$781.86(13)$	&	$0.0(2)$	&	$110.4(3)$	&	$165.6(2)$	&	$2.04(3)$	&	$\phantom{+}1.97(1.75)$	&	$-3.34(3.18)$	&	$\phantom{+}2.71(3.12)$	&	$-1.09(1.59)$	&	$\phantom{+}0.18(33)$	&	$375.1(2.6)$	\\
	KLOE	&	5	&	$254.1 / 186 = 1.37$		&	$6.7\times10^{-4}$	&	$781.82(14)$	&	$\Bigg\{\begin{matrix*}[r] 0.6(2) \\ -0.3(2) \\ -0.2(2) \end{matrix*}$	&	$110.3(2)$	&	$165.6(1)$	&	$1.98(4)$	&	$-0.10(1.05)$	&	$\phantom{+}0.48(1.84)$	&	$-0.97(1.76)$	&	$\phantom{+}0.75(88)$	&	$-0.20(18)$	&	$366.9(1.5)$	\\
	KLOE$''$	&	5	&	$221.8 / 184 = 1.21$		&	$3.0\%$			&	$781.80(14)$	&	$\Bigg\{\begin{matrix*}[r] 0.5(2) \\ -0.3(2) \\ -0.2(2) \end{matrix*}$	&	$110.2(2)$	&	$165.6(1)$	&	$1.98(4)$	&	$-0.45(1.05)$	&	$\phantom{+}1.08(1.85)$	&	$-1.53(1.76)$	&	$\phantom{+}1.02(88)$	&	$-0.25(18)$	&	$366.7(1.5)$	\\
	\bottomrule
	\end{tabular}
	}
	\caption{Fits with various values for the order $N$ of the conformal polynomial that describes the inelasticities.}
	\label{tab:FitsConfPoly}
\end{table}

\subsection{Combining data sets}
\label{sec:Combinations}

We now present the results of our final fit configuration. We use $N-1=4$ free parameters in the conformal polynomial, let the $\omega$ mass free, and use an energy rescaling for each of the time-like data sets constrained by the expected energy calibration uncertainty, in the case of KLOE three separate rescaling parameters for KLOE08/10/12. From the KLOE08 data set, we remove the two outliers (``KLOE$''$''), but also show 
the fits to the full set to demonstrate that while the improvement in the $\chi^2$ is significant, the impact on HVP is very minor. All sources of systematic uncertainties described in Sect.~\ref{sec:FitParameters} are considered, leading to the fit results for the parameters $\mw$, $\xi_j$, $\epsilon_\omega$, and the values of the phase shift at $s_0$ and $s_1$ as shown in Table~\ref{tab:FinalFitsSingleExperiments}. The fit errors are inflated by a scale factor
\beq
	\label{eq:Chi2ScaleFactor}
	S = \sqrt{\chi^2 / \mathrm{dof} },
\eeq
according to the PDG averaging prescription~\cite{PDG2018}.

Next, we perform simultaneous fits to combinations of the data sets. As the fit quality is equally good in all fits to single experiments, we do not introduce any weighting factors, but only apply the inflation factor $S$~\eqref{eq:Chi2ScaleFactor}, which increases the fit errors by $12\%$ to $19\%$.\footnote{We remark that in particular for high statistics the prescription~\eqref{eq:Chi2ScaleFactor} does not fully account for a situation where the systematic uncertainties in the experiments were underestimated.}  The results of these fits are given in Table~\ref{tab:FinalFitsCombinations}.

\begin{figure}[t!]
	\centering
	\input{plots/vff-spacelike}
	\caption{Fit result for the pion VFF in the space-like region, together with the NA7 data.}
	\label{img:VFFfitSpacelike}
\end{figure}
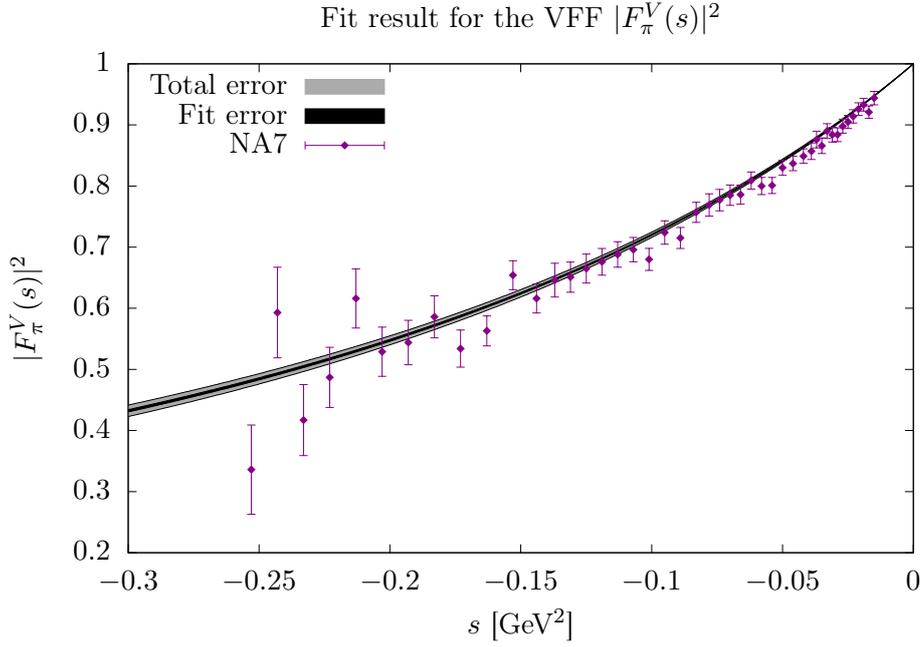

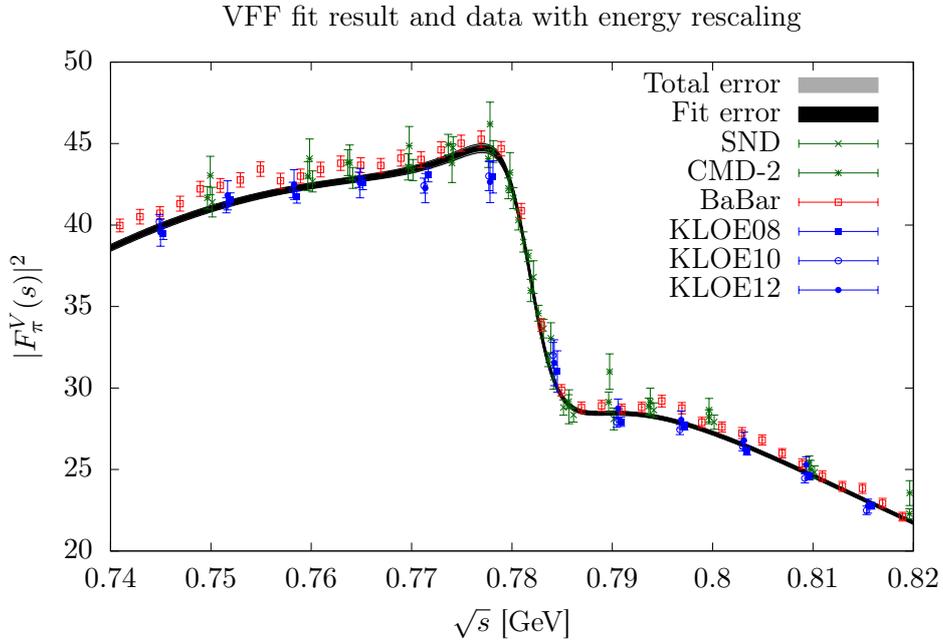
\begin{figure}[t]
	\centering
	\input{plots/vff-zoom-Xi}
	\caption{Fit result for the pion VFF in the $\rho$--$\omega$ interference region, together with the $e^+e^-$ data sets. The data points are shown with the energy rescaling~\eqref{eq:EnergyRescaling} and the curve is the fit result with~\eqref{omega_fit} for the $\omega$ mass.}
	\label{img:VFFfitZoomXi}
\end{figure}

\begin{figure}[t]
	\centering
	\input{plots/vff-rel-Xi}
	\caption{Relative difference between the data points (including the energy rescaling~\eqref{eq:EnergyRescaling}) and the fit result for the VFF, normalized to the fit result for $|F_\pi^V(s)|^2$.
	As in all plots, we show fit errors and total uncertainties as two separate error bands. The total uncertainty is given by the fit error and the systematic uncertainty, added in quadrature.}
	\label{img:VFFfitRelativeXi}
\end{figure}
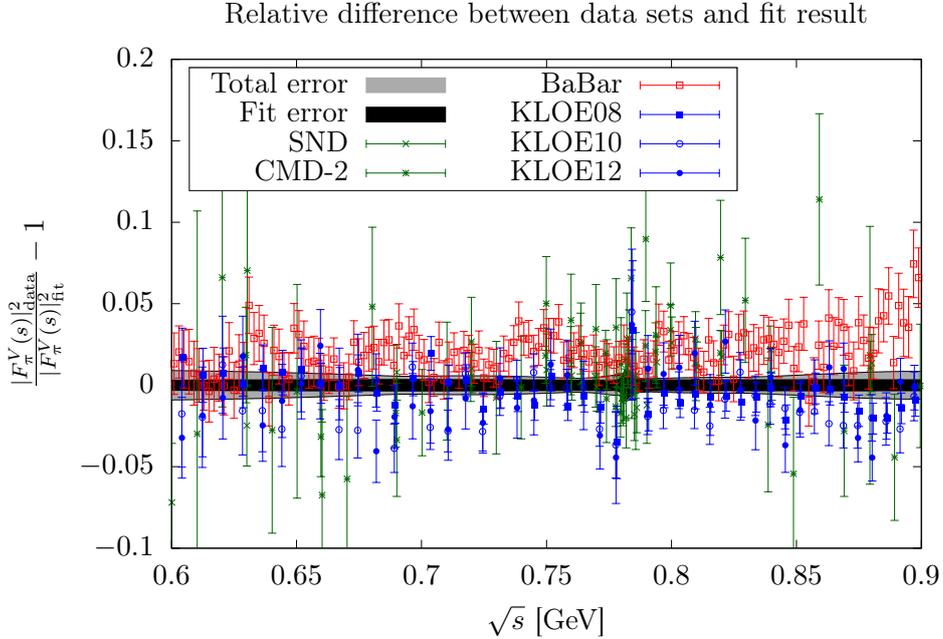

In Fig.~\ref{img:VFFfit}, we show the fit result for the VFF both in the time- and space-like region together with all the data sets used in the fit. At this scale, the uncertainties of the fit result are barely visible.
In Fig.~\ref{img:VFFfitSpacelike}, we show the space-like region of the VFF together with the NA7 data.
In Fig.~\ref{img:VFFfitZoomXi}, we focus on the $\rho$--$\omega$ interference region, in order to make it possible to distinguish between the dense time-like data sets. For comparison, we show in Fig.~\ref{img:VFFfitZoom} the same plot without the energy rescaling~\eqref{eq:EnergyRescaling} and for PDG $\omega$ mass, so that the effect of the exact alignment of the $\omega$ resonance position becomes apparent.
In Fig.~\ref{img:VFFfitRelativeXi}, we show for the region $[0.6,0.9]\GeV$
the relative deviation of the data points from the fit result, normalized
to the fit value of $|F_\pi^V(s)|^2$.  
In this variant, one can clearly observe the well-known tension between the
BaBar and KLOE data sets~\cite{Davier:2017zfy,Keshavarzi:2018mgv}: the
BaBar data lie systematically above the KLOE results, and the fit 
finds the average as dictated by the experimental covariance matrices.

\begin{table}[t]
	\centering
	\small
	\begin{tabular}{l r r r r r r }
	\toprule
		& $\chi^2/$dof$\qquad$ & $\mw$ [MeV] & $10^3 \times \xi_j \hspace{-0.1cm}$	&	$\delta_1^1(s_0)$ [$^\circ$] & $\delta_1^1(s_1)$ [$^\circ$] & $10^3 \times \epsilon_\omega$  \\
	\midrule
	SND		&	$51.9 / 37 = 1.40$		&	$781.49(32)(2)$	&	$0.0(6)(0)$	&	$110.5(5)(8)$	&	$165.7(0.3)(2.4)$	&	$2.03(5)(2)$	\\
	CMD-2	&	$87.4 / 74 = 1.18$		&	$781.98(29)(1)$	&	$0.0(6)(0)$	&	$110.5(5)(8)$	&	$166.4(0.4)(2.4)$	&	$1.88(6)(2)$	\\
	BaBar	&	$299.1 / 262 = 1.14$		&	$781.86(14)(1)$	&	$0.0(2)(0)$	&	$110.4(3)(7)$	&	$165.7(0.2)(2.5)$	&	$2.04(3)(2)$	\\
	KLOE	&	$254.5 / 187 = 1.36$		&	$781.82(17)(4)$	&	$\Bigg\{\begin{matrix*}[r] 0.6(2)(0) \\ -0.3(2)(0) \\ -0.2(3)(0) \end{matrix*}$	&	$110.4(2)(6)$	&	$165.6(0.1)(2.4)$	&	$1.97(4)(2)$	\\
	KLOE$''$	&	$222.5 / 185 = 1.20$		&	$781.81(16)(3)$	&	$\Bigg\{\begin{matrix*}[r] 0.5(2)(0) \\ -0.3(2)(0) \\ -0.2(3)(0) \end{matrix*}$	&	$110.3(2)(6)$	&	$165.6(0.1)(2.4)$	&	$1.98(4)(1)$	\\
	\bottomrule
	\end{tabular}
	\caption{Final fits to single $e^+e^-$ experiments with $N-1=4$ free parameters in the conformal polynomial. The first error is the fit uncertainty, inflated by $\sqrt{\chi^2/\mathrm{dof}}$, the second error is the combination of all systematic uncertainties.}
	\label{tab:FinalFitsSingleExperiments}
\end{table}

\begin{table}[t]
	\centering
	\small
	\begin{tabularx}{\textwidth}{ l @{\extracolsep{\fill}}c r c c c l}
	\toprule
		&& $\chi^2/$dof$\qquad$ & $\delta_1^1(s_0)$ [$^\circ$] & $\delta_1^1(s_1)$ [$^\circ$] & $10^3 \times \epsilon_\omega$  & $\mw$ [MeV] \\
	\midrule
	Energy scan				&&	$152.5 / 119 = 1.28$		&	$110.4(3)(8)$	&	$166.0(0.2)(2.4)$	&	$1.97(4)(2)$	&	$781.75(22)(1)$ \\
	All $e^+e^-$				&&	$764.5 / 584 = 1.31$		&	$110.5(1)(7)$	&	$165.8(0.1)(2.4)$	&	$2.02(2)(3)$	&	$781.68(9)(4)$	\\
	All $e^+e^-$, NA7			&&	$809.0 / 629 = 1.29$		&	$110.4(1)(7)$	&	$165.8(0.1)(2.4)$	&	$2.02(2)(3)$	&	$781.68(9)(3)$	\\
	All $e^+e^-$ (KLOE$''$)		&&	$731.6 / 582 = 1.26$		&	$110.4(1)(7)$	&	$165.8(0.1)(2.4)$	&	$2.02(2)(3)$	&	$781.68(9)(4)$	\\
	All $e^+e^-$ (KLOE$''$), NA7	&&	$776.2 / 627 = 1.24$		&	$110.4(1)(7)$	&	$165.7(0.1)(2.4)$	&	$2.02(2)(3)$	&	$781.68(9)(3)$	\\
	\end{tabularx}
	\begin{tabularx}{\textwidth}{l @{\extracolsep{\fill}}c c c c c c}
	\midrule
		&& $c_2$ & $c_3$ & $c_4$ & $c_5$ \\
	\midrule
	Energy scan				&&	$1.79(53)(80)$		&	$-2.70(0.62)(1.14)$	&	$1.80(35)(77)$	&	$-0.46(8)(20)$		\\
	All $e^+e^-$				&&	$1.08(21)(63)$		&	$-1.63(25)(83)$		&	$1.03(15)(50)$	&	$-0.24(4)(12)$		\\
	All $e^+e^-$, NA7			&&	$1.03(20)(61)$		&	$-1.58(25)(80)$		&	$1.00(15)(49)$	&	$-0.23(4)(11)$		\\
	All $e^+e^-$ (KLOE$''$)		&&	$1.07(20)(62)$		&	$-1.62(25)(82)$		&	$1.02(15)(50)$	&	$-0.24(4)(12)$		\\
	All $e^+e^-$ (KLOE$''$), NA7	&&	$1.03(20)(60)$		&	$-1.57(25)(80)$		&	$0.99(15)(49)$	&	$-0.23(4)(11)$		\\
	\end{tabularx}
	\begin{tabularx}{\textwidth}{l @{\extracolsep{\fill}}c l l l l l l}
	\midrule
	$10^3 \times \xi_j$	&& SND & CMD-2 & BaBar & KLOE08 & KLOE10 & KLOE12 \\
	\midrule
	Energy scan				&&	$0.5(4)(0)$	&	$-0.5(4)(0)$ \\
	All $e^+e^-$				&&	$0.3(2)(0)$	&	$-0.7(2)(0)$	&	$-0.2(2)(0)$	&	$0.7(2)(0)$	&	$-0.3(2)(0)$	&	$-0.0(3)(0)$	\\
	All $e^+e^-$, NA7			&&	$0.3(2)(0)$	&	$-0.7(2)(0)$	&	$-0.2(2)(0)$	&	$0.7(2)(0)$	&	$-0.3(2)(0)$	&	$-0.0(2)(0)$	\\
	All $e^+e^-$ (KLOE$''$)		&&	$0.3(2)(0)$	&	$-0.7(2)(0)$	&	$-0.1(2)(0)$	&	$0.6(2)(0)$	&	$-0.3(2)(0)$	&	$-0.0(2)(0)$	\\
	All $e^+e^-$ (KLOE$''$), NA7	&&	$0.3(2)(0)$	&	$-0.7(2)(0)$	&	$-0.1(2)(0)$	&	$0.6(2)(0)$	&	$-0.3(2)(0)$	&	$-0.0(2)(0)$	\\
	\bottomrule
	\end{tabularx}
	\caption{Final fits to combinations of data sets with $N-1=4$ free parameters in the conformal polynomial. The first error is the fit uncertainty, inflated by $\sqrt{\chi^2/\mathrm{dof}}$, the second error is the combination of all systematic uncertainties, apart from the case of the parameters $c_i$ in the second panel, where it includes only the uncertainties due to the phase input and $\Gamma_\omega$, but not the variation in $s_c$ and $N$.
	The third panel gives the energy rescalings that belong to the respective fits.}
	\label{tab:FinalFitsCombinations}
\end{table}


\begin{figure}[t!]
	\centering
	\input{plots/vff}
	\caption{Fit result for the pion VFF $|F_\pi^V(s)|^2$ including (barely visible) uncertainties, together with all the data sets used in the fit. For the $e^+e^-$ experiments, the experimental values of the form factor are obtained from the published cross section according to Sect.~\ref{sec:HVPRadiativeCorrections}.}
	\label{img:VFFfit}
\end{figure}
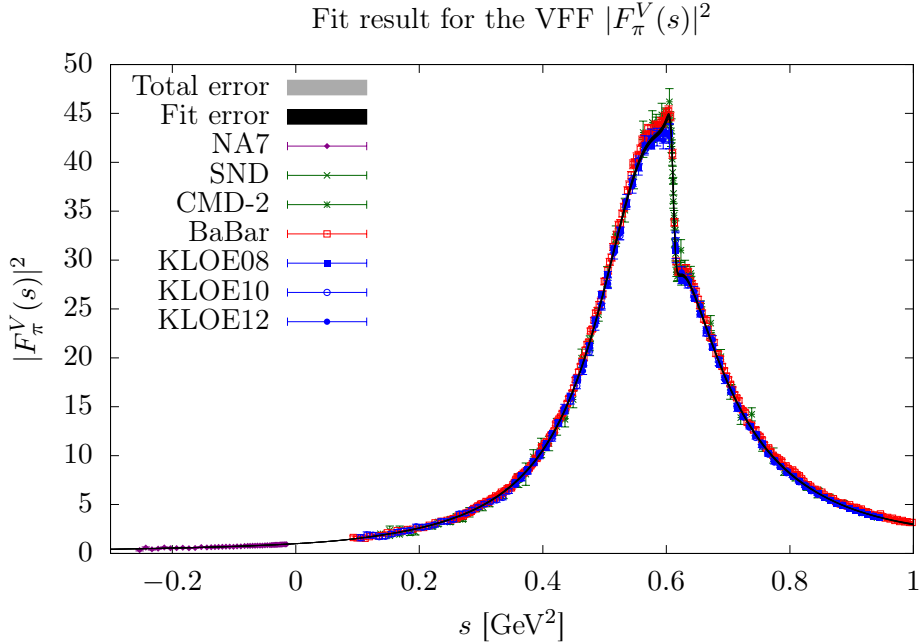

\begin{figure}[t]
	\centering
	\input{plots/vff-zoom}
	\caption{Fit result for the pion VFF in the $\rho$--$\omega$ interference region, together with the $e^+e^-$ data sets. The curve is the result of the VFF fit to the data points including energy rescaling as shown in Fig.~\ref{img:VFFfitZoomXi}, but with an $\omega$ mass reset to the PDG value and compared to the original data points without energy rescaling.}
	\label{img:VFFfitZoom}
\end{figure}
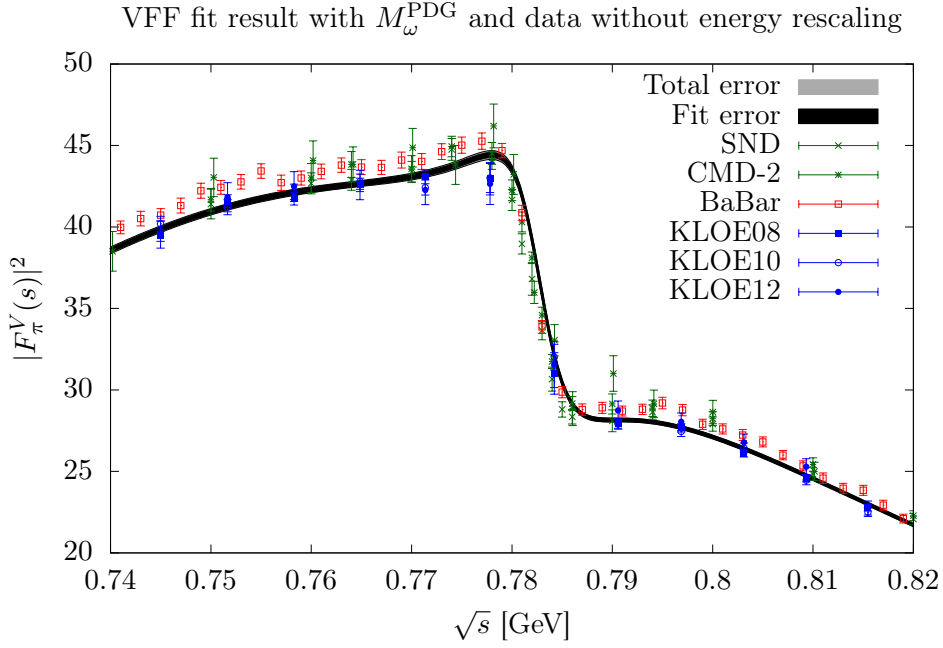

\clearpage

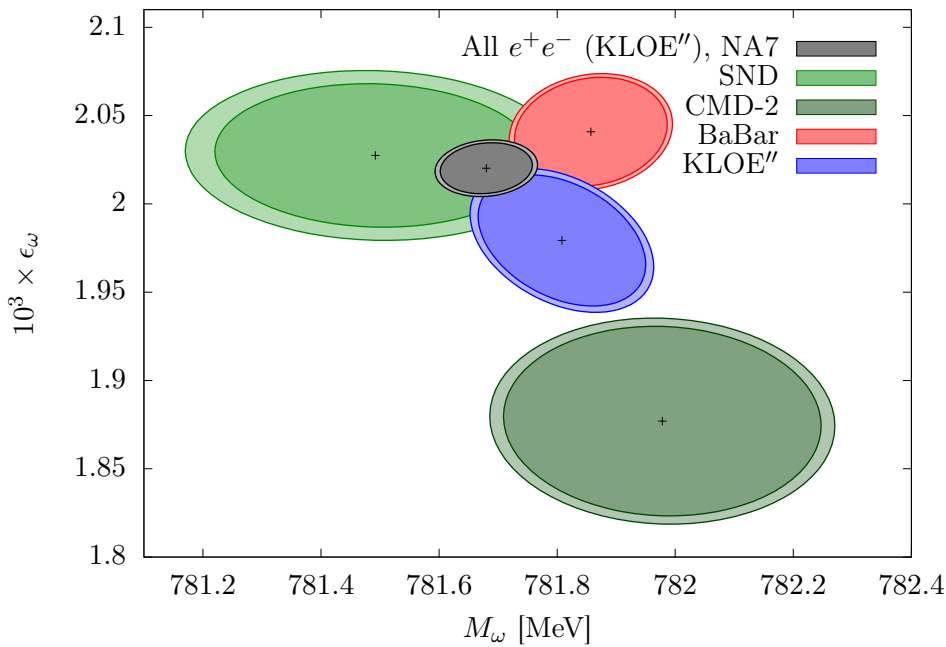
\begin{figure}[t]
	\centering
	\input{plots/ellipses}
	\caption{Error ellipses for the parameters $\epsilon_\omega$ and $M_\omega$ resulting from fits to single experiments and the fit to the combination of all experiments. The smaller ellipses are standard error ellipses that correspond to $\Delta\chi^2 = 1$, the larger ellipses are inflated by the scale factor~\eqref{eq:Chi2ScaleFactor}.}
	\label{img:CorrelationsOmegaPars}
\end{figure}

\subsection[Extraction of the $\omega$ mass]{\boldmath Extraction of the $\omega$ mass}
\label{sec:omega}

At first sight, it is surprising that our final number for the $\omega$ mass resulting from the fit to the combination of all experiments
\beq
\label{omega_fit}
\mw=781.68(9)(3)\MeV, 
\eeq 
see Table~\ref{tab:FinalFitsCombinations}, is slightly below a naive weighted average of the results from the fits to single experiments in Table~\ref{tab:FinalFitsSingleExperiments} and even below the fit results to BaBar or KLOE alone. However, most of this effect is explained by the correlations with the other fit parameters. If one performs a multivariate weighted average of the fit parameters
\begin{align}
	\vec p_\text{aver} = \sigma \sum_j \sigma_j^{-1} \vec p_j ,
	\quad \sigma = \Big( \sum_j \sigma_j^{-1} \Big)^{-1} ,
\end{align}
where $\vec p_j$ is the vector of parameters fit to experiment $j$ and $\sigma_j$ denotes the covariance matrix of the fit parameters $\vec p_j$, one obtains a result very close to the outcome of a fit to the combination of all experiments, in particular, one finds $M_\omega^\text{aver} = 781.67(9)\MeV$. A rather large correlation of $-39\%$ is present between $M_\omega$ and $\epsilon_\omega$ in the fit to KLOE. This correlation partly explains why the combined value~\eqref{omega_fit} lies below the fit results to either BaBar or KLOE, as illustrated in the plot of the error ellipses in Fig.~\ref{img:CorrelationsOmegaPars}. Small differences between the multivariate weighted average and the result of the combined fit can be observed for the other parameters, which is a sign of the non-linear dependence of the fit function on the parameters.

Both our final result~\eqref{omega_fit} for the $\omega$ mass and the results from fits to single experiments in Table~\ref{tab:FinalFitsSingleExperiments} disagree with the PDG
average~\eqref{eq:OmegaMass}.
Since this discrepancy is not driven by a single
experiment, it seems to indicate that $e^+e^-\to\pi^+\pi^-$ data
indeed unanimously favor an $\omega$ mass substantially lower than the
current PDG average. The latter is mainly based on
$\mw=782.79(8)(9)$~\cite{Achasov:2003ir}, 
$\mw=782.68(9)(4)$~\cite{Akhmetshin:2003zn} from $e^+e^-\to 3\pi$, but also
includes $\mw=783.20(13)(16)$~\cite{Akhmetshin:2004gw} from $e^+e^-\to \pi^0\gamma$
and $\mw=781.96(13)(17)\MeV$ from $\bar pp\to\omega\pi^0\pi^0$~\cite{Amsler:1993pr},
both of which do not influence the average much (or at least cancel each other effectively). 
Our determination is thus closer to the $\bar p p$ reaction, but in conflict with the $3\pi$ data.

This observation is not new, see e.g.~\cite{Hanhart:2016pcd}, and the
mismatch has already been pointed out by the BaBar
collaboration~\cite{Lees:2012cj}. The latter analysis is worthwhile
reviewing in some detail here because it may offer an indication which 
effects could be contributing to the puzzle. In~\cite{Lees:2012cj},
the data are analyzed with the
help of a sum of GS parametrizations for all the relevant
resonances in their energy range, including relative phases between
the various GS terms, which are allowed to float and determined by a fit to the
data. In the unconstrained fit they obtain $\mw=781.91(18)(16)\MeV$ and a
relative phase $\phi_{\rho \omega}=-0.6(2.1)^\circ$, compatible with
zero. This is in very good agreement with~\eqref{omega_fit}, but not with
the PDG. They then observe that there is a strong correlation between
$\phi_{\rho \omega}$ and $\mw$ and that if one fixes the $\omega$ mass at
the PDG value one can still obtain a good fit: in this case, however, the
value of $\phi_{\rho \omega}$ changes to $7.8(1.3)^\circ$, which
is compatible with the phase obtained by CMD-2 in their fit
to the $e^+ e^- \to \pi^+ \pi^-$ data~\cite{Akhmetshin:2006bx}: $\phi_{\rho
  \omega}=10.4(3.8)^\circ$. Conversely, if they constrain this phase in the fit to the BaBar
  data to the CMD-2 value, the $\omega$ mass becomes $\mw=782.68(12)(27)\MeV$, perfectly compatible with the PDG.

Assuming that the fit quality has not decreased significantly in the latter
fit, we can conclude that if CMD-2 and BaBar use similar parametrizations to
fit their respective data, the values of the parameters they obtain are
compatible with each other (taking into account correlations). Still, the
first unconstrained fit shows that the BaBar data prefer a value of the
$\rho$--$\omega$ phase compatible with zero, which is a very good
sign because the presence of such a phase would violate analyticity and
unitarity (as it would give a complex form factor even below the $\pi \pi$
threshold). In our framework such a phase is strictly forbidden and the
agreement with the unconstrained BaBar fit both on the absence of this
phase as well as on the value of the omega mass is very reassuring.

This raises the question whether also other determinations of the omega
mass have used unphysical parametrizations, and the answer is unfortunately
positive: the determination based on $e^+e^-\to
\pi^0\gamma$~\cite{Akhmetshin:2004gw} includes such a relative phase 
of $13.3^\circ$, which might explain the resulting value for $\mw$ even higher than from the $3\pi$ channel. 
In contrast, the extractions based on $e^+e^-\to 3\pi$ do not include a Breit--Wigner 
$\rho$-resonance in their parametrization, only the $\omega$ and $\phi$ resonances together with a smooth background, 
but the complex phases between $\omega$, $\phi$, and background could still distort the extracted $\omega$ mass.
Both for $e^+e^-\to3\pi$ and $e^+e^-\to\pi^0\gamma$ representations exist that do not suffer from 
these shortcomings~\cite{Hoferichter:2014vra,Hoferichter:2018kwz}. In these papers,
good fits were obtained while using the PDG $\omega$ parameters as input, which indicates
a substantial model dependence in the extraction from $e^+e^-\to
\pi^0\gamma$~\cite{Akhmetshin:2004gw}, but likely only a small effect in $e^+e^-\to 3\pi$~\cite{Achasov:2003ir,Akhmetshin:2003zn}.
For a firm conclusion more thorough fits to $3\pi$ and $\pi^0\gamma$ data 
including the respective uncertainties would be necessary.

For these reasons, the high significance of the discrepancy, more than $5\sigma$ if taken at face value,
is puzzling, in particular given that the extraction from the
isospin-conserving $3\pi$ channel should, in principle, be more reliable
than the isospin-breaking effect in $e^+e^-\to \pi^+\pi^-$. Another
potential subtlety could concern the definition of the $\omega$ mass in
view of electromagnetic corrections, but estimates of the corresponding
effect~\cite{Hanhart:inprep} 
\beq
\Delta \mw = \frac{e^2}{2g_{\omega\gamma}^2}\mw=0.13\MeV,
\eeq
with $g_{\omega\gamma}=16.7(2)$~\cite{Hoferichter:2017ftn}, are well below
the observed tension, albeit potentially relevant at the level of the
uncertainty quoted in the PDG average. Similarly, the $\omega$ parameters
in our parameterization~\eqref{Gomega} do not strictly correspond to the pole parameters yet,
with corrections that scale with the $\omega$ width, but those effects seem to be in line with the naive estimate $\Gamma_\omega^2/\mw\sim 0.1\MeV$, 
e.g.\ the difference between~\eqref{Gomega} and~\eqref{gomega} is below this threshold.   

In this paper, we aim to derive the HVP contribution to $a_\mu$ based on
the available experimental information on $e^+e^-\to \pi^+\pi^-$ subject to
a comprehensive analysis of the constraints from analyticity and
unitarity. From this point of view, there is no indication to assume a
common systematic effect in all experiments that would restore agreement
with the $3\pi$ channel, we will therefore pursue the analysis of the HVP
contribution based on the fits in the preceding subsection. However, in
addition to the known tension between the KLOE and BaBar data, this
discrepancy in the $\omega$ mass extracted from the $2\pi$ and $3\pi$
channels deserves further attention.

%% file: plots/chi2-KLOE.tex
\begingroup
  \makeatletter
  \providecommand\color[2][]{%
    \GenericError{(gnuplot) \space\space\space\@spaces}{%
      Package color not loaded in conjunction with
      terminal option `colourtext'%
    }{See the gnuplot documentation for explanation.%
    }{Either use 'blacktext' in gnuplot or load the package
      color.sty in LaTeX.}%
    \renewcommand\color[2][]{}%
  }%
  \providecommand\includegraphics[2][]{%
    \GenericError{(gnuplot) \space\space\space\@spaces}{%
      Package graphicx or graphics not loaded%
    }{See the gnuplot documentation for explanation.%
    }{The gnuplot epslatex terminal needs graphicx.sty or graphics.sty.}%
    \renewcommand\includegraphics[2][]{}%
  }%
  \providecommand\rotatebox[2]{#2}%
  \@ifundefined{ifGPcolor}{%
    \newif\ifGPcolor
    \GPcolorfalse
  }{}%
  \@ifundefined{ifGPblacktext}{%
    \newif\ifGPblacktext
    \GPblacktexttrue
  }{}%
  \let\gplgaddtomacro\g@addto@macro
  \gdef\gplbacktext{}%
  \gdef\gplfronttext{}%
  \makeatother
  \ifGPblacktext
    \def\colorrgb#1{}%
    \def\colorgray#1{}%
  \else
    \ifGPcolor
      \def\colorrgb#1{\color[rgb]{#1}}%
      \def\colorgray#1{\color[gray]{#1}}%
      \expandafter\def\csname LTw\endcsname{\color{white}}%
      \expandafter\def\csname LTb\endcsname{\color{black}}%
      \expandafter\def\csname LTa\endcsname{\color{black}}%
      \expandafter\def\csname LT0\endcsname{\color[rgb]{1,0,0}}%
      \expandafter\def\csname LT1\endcsname{\color[rgb]{0,1,0}}%
      \expandafter\def\csname LT2\endcsname{\color[rgb]{0,0,1}}%
      \expandafter\def\csname LT3\endcsname{\color[rgb]{1,0,1}}%
      \expandafter\def\csname LT4\endcsname{\color[rgb]{0,1,1}}%
      \expandafter\def\csname LT5\endcsname{\color[rgb]{1,1,0}}%
      \expandafter\def\csname LT6\endcsname{\color[rgb]{0,0,0}}%
      \expandafter\def\csname LT7\endcsname{\color[rgb]{1,0.3,0}}%
      \expandafter\def\csname LT8\endcsname{\color[rgb]{0.5,0.5,0.5}}%
    \else
      \def\colorrgb#1{\color{black}}%
      \def\colorgray#1{\color[gray]{#1}}%
      \expandafter\def\csname LTw\endcsname{\color{white}}%
      \expandafter\def\csname LTb\endcsname{\color{black}}%
      \expandafter\def\csname LTa\endcsname{\color{black}}%
      \expandafter\def\csname LT0\endcsname{\color{black}}%
      \expandafter\def\csname LT1\endcsname{\color{black}}%
      \expandafter\def\csname LT2\endcsname{\color{black}}%
      \expandafter\def\csname LT3\endcsname{\color{black}}%
      \expandafter\def\csname LT4\endcsname{\color{black}}%
      \expandafter\def\csname LT5\endcsname{\color{black}}%
      \expandafter\def\csname LT6\endcsname{\color{black}}%
      \expandafter\def\csname LT7\endcsname{\color{black}}%
      \expandafter\def\csname LT8\endcsname{\color{black}}%
    \fi
  \fi
    \setlength{\unitlength}{0.0500bp}%
    \ifx\gptboxheight\undefined%
      \newlength{\gptboxheight}%
      \newlength{\gptboxwidth}%
      \newsavebox{\gptboxtext}%
    \fi%
    \setlength{\fboxrule}{0.5pt}%
    \setlength{\fboxsep}{1pt}%
\begin{picture}(5400.00,3780.00)%
    \gplgaddtomacro\gplbacktext{%
      \csname LTb\endcsname
      \put(1154,1086){\makebox(0,0)[r]{\strut{}$10$}}%
      \put(1154,1489){\makebox(0,0)[r]{\strut{}$20$}}%
      \put(1154,1891){\makebox(0,0)[r]{\strut{}$30$}}%
      \put(1154,2294){\makebox(0,0)[r]{\strut{}$40$}}%
      \put(1154,2696){\makebox(0,0)[r]{\strut{}$50$}}%
      \put(1154,3099){\makebox(0,0)[r]{\strut{}$60$}}%
      \put(1669,484){\makebox(0,0){\strut{}$10$}}%
      \put(2071,484){\makebox(0,0){\strut{}$20$}}%
      \put(2474,484){\makebox(0,0){\strut{}$30$}}%
      \put(2877,484){\makebox(0,0){\strut{}$40$}}%
      \put(3279,484){\makebox(0,0){\strut{}$50$}}%
      \put(3682,484){\makebox(0,0){\strut{}$60$}}%
    }%
    \gplgaddtomacro\gplfronttext{%
      \csname LTb\endcsname
      \put(670,1911){\rotatebox{-270}{\makebox(0,0){\strut{}$j$}}}%
      \put(2494,154){\makebox(0,0){\strut{}$i$}}%
      \put(2494,3449){\makebox(0,0){\strut{}KLOE08 bin contributions to $\chi^2$}}%
      \csname LTb\endcsname
      \put(4014,704){\makebox(0,0)[l]{\strut{}$-5$}}%
      \put(4014,1005){\makebox(0,0)[l]{\strut{}$0$}}%
      \put(4014,1307){\makebox(0,0)[l]{\strut{}$5$}}%
      \put(4014,1609){\makebox(0,0)[l]{\strut{}$10$}}%
      \put(4014,1911){\makebox(0,0)[l]{\strut{}$15$}}%
      \put(4014,2213){\makebox(0,0)[l]{\strut{}$20$}}%
      \put(4014,2515){\makebox(0,0)[l]{\strut{}$25$}}%
      \put(4014,2817){\makebox(0,0)[l]{\strut{}$30$}}%
      \put(4014,3119){\makebox(0,0)[l]{\strut{}$35$}}%
    }%
    \gplbacktext
    \put(0,0){\includegraphics{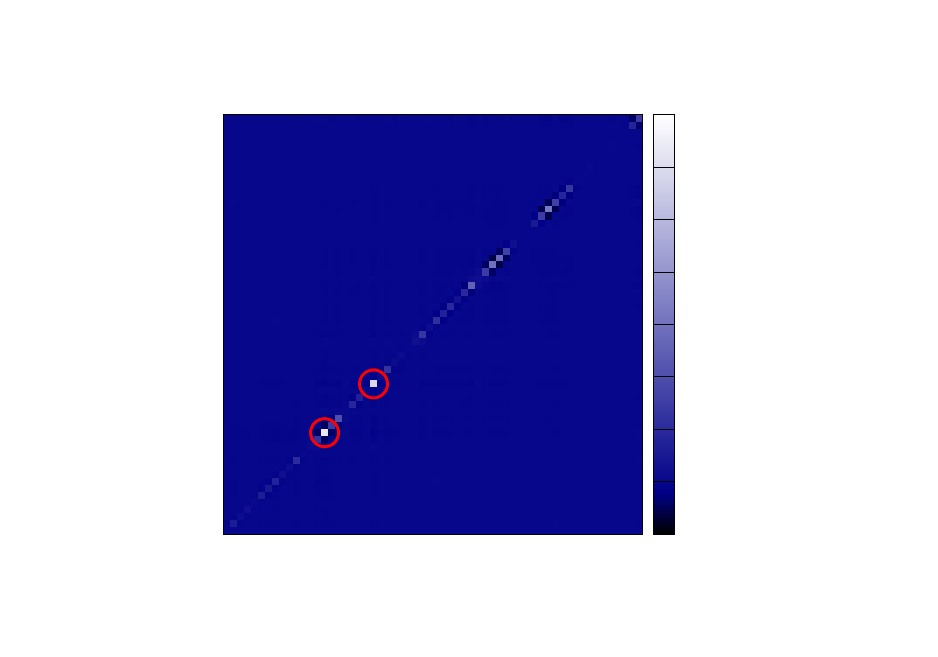}}%
    \gplfronttext
  \end{picture}%
\endgroup

%% file: plots/chi2-BESIII.tex
\begingroup
  \makeatletter
  \providecommand\color[2][]{%
    \GenericError{(gnuplot) \space\space\space\@spaces}{%
      Package color not loaded in conjunction with
      terminal option `colourtext'%
    }{See the gnuplot documentation for explanation.%
    }{Either use 'blacktext' in gnuplot or load the package
      color.sty in LaTeX.}%
    \renewcommand\color[2][]{}%
  }%
  \providecommand\includegraphics[2][]{%
    \GenericError{(gnuplot) \space\space\space\@spaces}{%
      Package graphicx or graphics not loaded%
    }{See the gnuplot documentation for explanation.%
    }{The gnuplot epslatex terminal needs graphicx.sty or graphics.sty.}%
    \renewcommand\includegraphics[2][]{}%
  }%
  \providecommand\rotatebox[2]{#2}%
  \@ifundefined{ifGPcolor}{%
    \newif\ifGPcolor
    \GPcolorfalse
  }{}%
  \@ifundefined{ifGPblacktext}{%
    \newif\ifGPblacktext
    \GPblacktexttrue
  }{}%
  \let\gplgaddtomacro\g@addto@macro
  \gdef\gplbacktext{}%
  \gdef\gplfronttext{}%
  \makeatother
  \ifGPblacktext
    \def\colorrgb#1{}%
    \def\colorgray#1{}%
  \else
    \ifGPcolor
      \def\colorrgb#1{\color[rgb]{#1}}%
      \def\colorgray#1{\color[gray]{#1}}%
      \expandafter\def\csname LTw\endcsname{\color{white}}%
      \expandafter\def\csname LTb\endcsname{\color{black}}%
      \expandafter\def\csname LTa\endcsname{\color{black}}%
      \expandafter\def\csname LT0\endcsname{\color[rgb]{1,0,0}}%
      \expandafter\def\csname LT1\endcsname{\color[rgb]{0,1,0}}%
      \expandafter\def\csname LT2\endcsname{\color[rgb]{0,0,1}}%
      \expandafter\def\csname LT3\endcsname{\color[rgb]{1,0,1}}%
      \expandafter\def\csname LT4\endcsname{\color[rgb]{0,1,1}}%
      \expandafter\def\csname LT5\endcsname{\color[rgb]{1,1,0}}%
      \expandafter\def\csname LT6\endcsname{\color[rgb]{0,0,0}}%
      \expandafter\def\csname LT7\endcsname{\color[rgb]{1,0.3,0}}%
      \expandafter\def\csname LT8\endcsname{\color[rgb]{0.5,0.5,0.5}}%
    \else
      \def\colorrgb#1{\color{black}}%
      \def\colorgray#1{\color[gray]{#1}}%
      \expandafter\def\csname LTw\endcsname{\color{white}}%
      \expandafter\def\csname LTb\endcsname{\color{black}}%
      \expandafter\def\csname LTa\endcsname{\color{black}}%
      \expandafter\def\csname LT0\endcsname{\color{black}}%
      \expandafter\def\csname LT1\endcsname{\color{black}}%
      \expandafter\def\csname LT2\endcsname{\color{black}}%
      \expandafter\def\csname LT3\endcsname{\color{black}}%
      \expandafter\def\csname LT4\endcsname{\color{black}}%
      \expandafter\def\csname LT5\endcsname{\color{black}}%
      \expandafter\def\csname LT6\endcsname{\color{black}}%
      \expandafter\def\csname LT7\endcsname{\color{black}}%
      \expandafter\def\csname LT8\endcsname{\color{black}}%
    \fi
  \fi
    \setlength{\unitlength}{0.0500bp}%
    \ifx\gptboxheight\undefined%
      \newlength{\gptboxheight}%
      \newlength{\gptboxwidth}%
      \newsavebox{\gptboxtext}%
    \fi%
    \setlength{\fboxrule}{0.5pt}%
    \setlength{\fboxsep}{1pt}%
\begin{picture}(5400.00,3780.00)%
    \gplgaddtomacro\gplbacktext{%
      \csname LTb\endcsname
      \put(1154,1086){\makebox(0,0)[r]{\strut{}$10$}}%
      \put(1154,1489){\makebox(0,0)[r]{\strut{}$20$}}%
      \put(1154,1891){\makebox(0,0)[r]{\strut{}$30$}}%
      \put(1154,2294){\makebox(0,0)[r]{\strut{}$40$}}%
      \put(1154,2696){\makebox(0,0)[r]{\strut{}$50$}}%
      \put(1154,3099){\makebox(0,0)[r]{\strut{}$60$}}%
      \put(1669,484){\makebox(0,0){\strut{}$10$}}%
      \put(2071,484){\makebox(0,0){\strut{}$20$}}%
      \put(2474,484){\makebox(0,0){\strut{}$30$}}%
      \put(2877,484){\makebox(0,0){\strut{}$40$}}%
      \put(3279,484){\makebox(0,0){\strut{}$50$}}%
      \put(3682,484){\makebox(0,0){\strut{}$60$}}%
    }%
    \gplgaddtomacro\gplfronttext{%
      \csname LTb\endcsname
      \put(670,1911){\rotatebox{-270}{\makebox(0,0){\strut{}$j$}}}%
      \put(2494,154){\makebox(0,0){\strut{}$i$}}%
      \put(2494,3449){\makebox(0,0){\strut{}BESIII bin contributions to $\chi^2$}}%
      \csname LTb\endcsname
      \put(4014,704){\makebox(0,0)[l]{\strut{}$-2000$}}%
      \put(4014,972){\makebox(0,0)[l]{\strut{}$-1500$}}%
      \put(4014,1240){\makebox(0,0)[l]{\strut{}$-1000$}}%
      \put(4014,1509){\makebox(0,0)[l]{\strut{}$-500$}}%
      \put(4014,1777){\makebox(0,0)[l]{\strut{}$0$}}%
      \put(4014,2045){\makebox(0,0)[l]{\strut{}$500$}}%
      \put(4014,2314){\makebox(0,0)[l]{\strut{}$1000$}}%
      \put(4014,2582){\makebox(0,0)[l]{\strut{}$1500$}}%
      \put(4014,2850){\makebox(0,0)[l]{\strut{}$2000$}}%
      \put(4014,3119){\makebox(0,0)[l]{\strut{}$2500$}}%
    }%
    \gplbacktext
    \put(0,0){\includegraphics{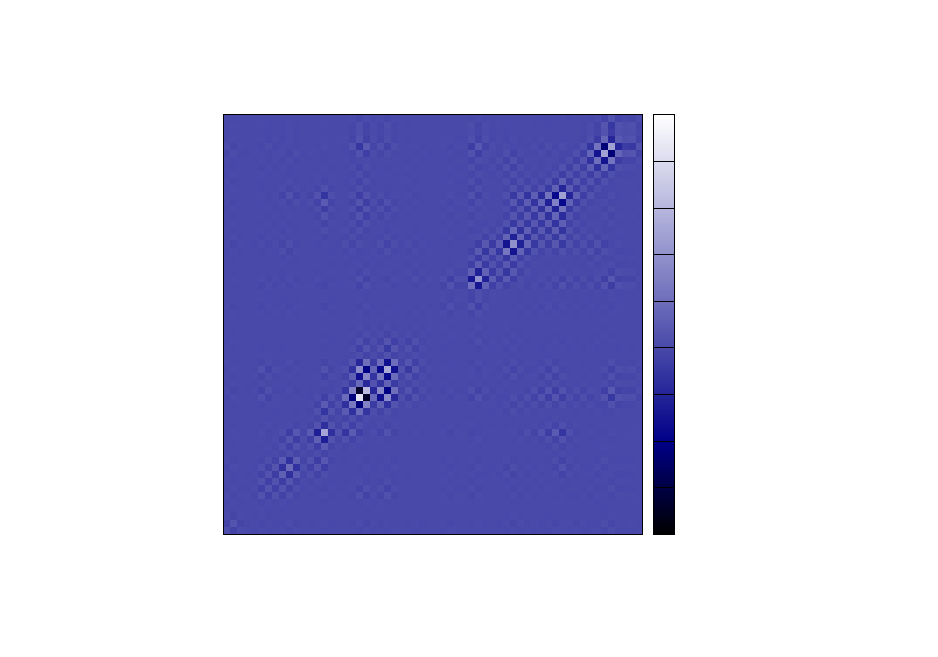}}%
    \gplfronttext
  \end{picture}%
\endgroup

%% file: plots/phasediff.tex
\begingroup
  \makeatletter
  \providecommand\color[2][]{%
    \GenericError{(gnuplot) \space\space\space\@spaces}{%
      Package color not loaded in conjunction with
      terminal option `colourtext'%
    }{See the gnuplot documentation for explanation.%
    }{Either use 'blacktext' in gnuplot or load the package
      color.sty in LaTeX.}%
    \renewcommand\color[2][]{}%
  }%
  \providecommand\includegraphics[2][]{%
    \GenericError{(gnuplot) \space\space\space\@spaces}{%
      Package graphicx or graphics not loaded%
    }{See the gnuplot documentation for explanation.%
    }{The gnuplot epslatex terminal needs graphicx.sty or graphics.sty.}%
    \renewcommand\includegraphics[2][]{}%
  }%
  \providecommand\rotatebox[2]{#2}%
  \@ifundefined{ifGPcolor}{%
    \newif\ifGPcolor
    \GPcolorfalse
  }{}%
  \@ifundefined{ifGPblacktext}{%
    \newif\ifGPblacktext
    \GPblacktexttrue
  }{}%
  \let\gplgaddtomacro\g@addto@macro
  \gdef\gplbacktext{}%
  \gdef\gplfronttext{}%
  \makeatother
  \ifGPblacktext
    \def\colorrgb#1{}%
    \def\colorgray#1{}%
  \else
    \ifGPcolor
      \def\colorrgb#1{\color[rgb]{#1}}%
      \def\colorgray#1{\color[gray]{#1}}%
      \expandafter\def\csname LTw\endcsname{\color{white}}%
      \expandafter\def\csname LTb\endcsname{\color{black}}%
      \expandafter\def\csname LTa\endcsname{\color{black}}%
      \expandafter\def\csname LT0\endcsname{\color[rgb]{1,0,0}}%
      \expandafter\def\csname LT1\endcsname{\color[rgb]{0,1,0}}%
      \expandafter\def\csname LT2\endcsname{\color[rgb]{0,0,1}}%
      \expandafter\def\csname LT3\endcsname{\color[rgb]{1,0,1}}%
      \expandafter\def\csname LT4\endcsname{\color[rgb]{0,1,1}}%
      \expandafter\def\csname LT5\endcsname{\color[rgb]{1,1,0}}%
      \expandafter\def\csname LT6\endcsname{\color[rgb]{0,0,0}}%
      \expandafter\def\csname LT7\endcsname{\color[rgb]{1,0.3,0}}%
      \expandafter\def\csname LT8\endcsname{\color[rgb]{0.5,0.5,0.5}}%
    \else
      \def\colorrgb#1{\color{black}}%
      \def\colorgray#1{\color[gray]{#1}}%
      \expandafter\def\csname LTw\endcsname{\color{white}}%
      \expandafter\def\csname LTb\endcsname{\color{black}}%
      \expandafter\def\csname LTa\endcsname{\color{black}}%
      \expandafter\def\csname LT0\endcsname{\color{black}}%
      \expandafter\def\csname LT1\endcsname{\color{black}}%
      \expandafter\def\csname LT2\endcsname{\color{black}}%
      \expandafter\def\csname LT3\endcsname{\color{black}}%
      \expandafter\def\csname LT4\endcsname{\color{black}}%
      \expandafter\def\csname LT5\endcsname{\color{black}}%
      \expandafter\def\csname LT6\endcsname{\color{black}}%
      \expandafter\def\csname LT7\endcsname{\color{black}}%
      \expandafter\def\csname LT8\endcsname{\color{black}}%
    \fi
  \fi
    \setlength{\unitlength}{0.0500bp}%
    \ifx\gptboxheight\undefined%
      \newlength{\gptboxheight}%
      \newlength{\gptboxwidth}%
      \newsavebox{\gptboxtext}%
    \fi%
    \setlength{\fboxrule}{0.5pt}%
    \setlength{\fboxsep}{1pt}%
\begin{picture}(7200.00,5040.00)%
    \gplgaddtomacro\gplbacktext{%
      \csname LTb\endcsname
      \put(814,704){\makebox(0,0)[r]{\strut{}$-10$}}%
      \put(814,1623){\makebox(0,0)[r]{\strut{}$-5$}}%
      \put(814,2542){\makebox(0,0)[r]{\strut{}$0$}}%
      \put(814,3460){\makebox(0,0)[r]{\strut{}$5$}}%
      \put(814,4379){\makebox(0,0)[r]{\strut{}$10$}}%
      \put(946,484){\makebox(0,0){\strut{}$0.9$}}%
      \put(2117,484){\makebox(0,0){\strut{}$0.95$}}%
      \put(3289,484){\makebox(0,0){\strut{}$1$}}%
      \put(4460,484){\makebox(0,0){\strut{}$1.05$}}%
      \put(5632,484){\makebox(0,0){\strut{}$1.1$}}%
      \put(6803,484){\makebox(0,0){\strut{}$1.15$}}%
    }%
    \gplgaddtomacro\gplfronttext{%
      \csname LTb\endcsname
      \put(198,2541){\rotatebox{-270}{\makebox(0,0){\strut{}$\mathrm{arg}\big(G_\mathrm{in}^N(s)\big)$ [${}^\circ$]}}}%
      \put(3874,154){\makebox(0,0){\strut{}$\sqrt{s}$ [GeV]}}%
      \put(3874,4709){\makebox(0,0){\strut{}Phase difference due to inelasticity, $N-1=4$}}%
      \csname LTb\endcsname
      \put(2134,4206){\makebox(0,0)[r]{\strut{}SND}}%
      \csname LTb\endcsname
      \put(2134,3986){\makebox(0,0)[r]{\strut{}CMD-2}}%
      \csname LTb\endcsname
      \put(2134,3766){\makebox(0,0)[r]{\strut{}BaBar}}%
      \csname LTb\endcsname
      \put(2134,3546){\makebox(0,0)[r]{\strut{}KLOE}}%
      \csname LTb\endcsname
      \put(2134,3326){\makebox(0,0)[r]{\strut{}E\L{} bound}}%
    }%
    \gplbacktext
    \put(0,0){\includegraphics{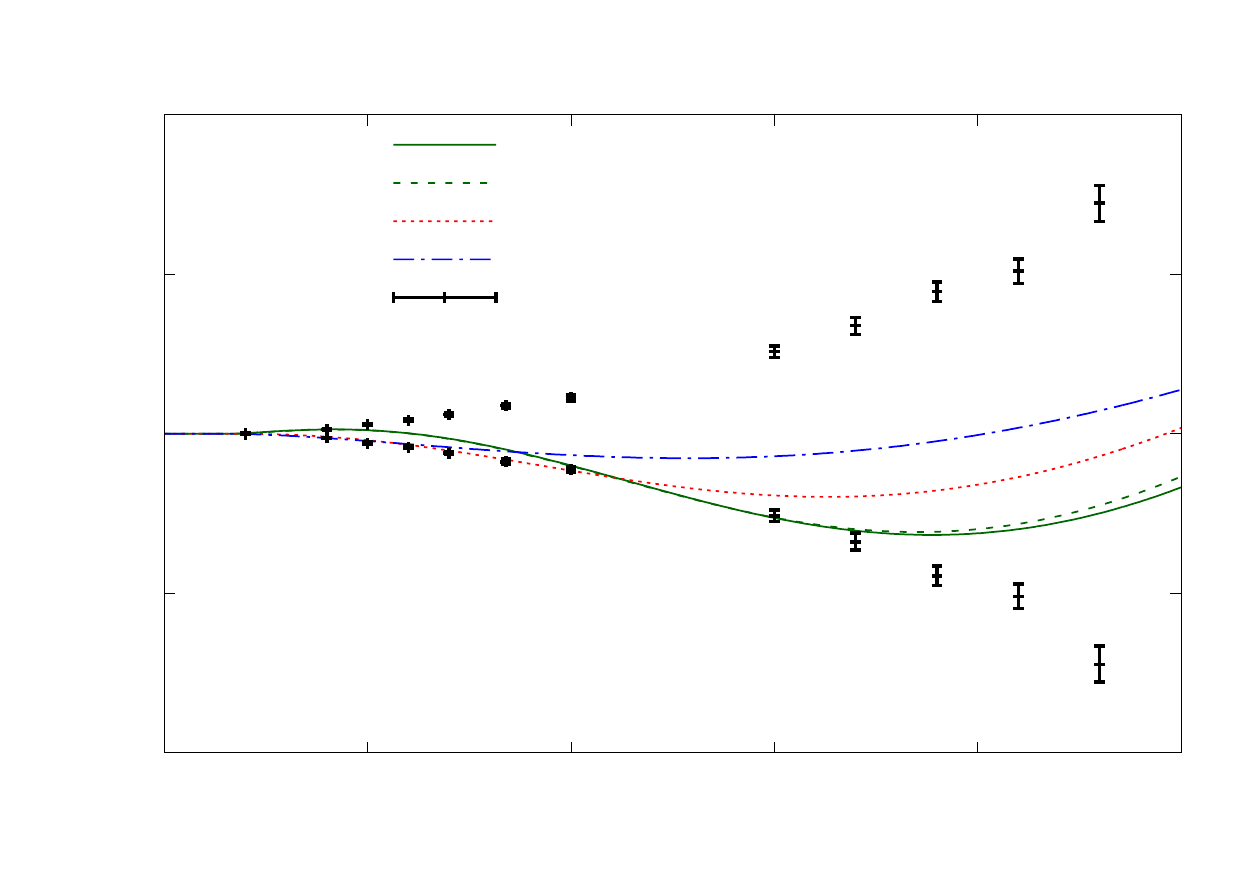}}%
    \gplfronttext
  \end{picture}%
\endgroup

%% file: plots/vff.tex
\begingroup
  \makeatletter
  \providecommand\color[2][]{%
    \GenericError{(gnuplot) \space\space\space\@spaces}{%
      Package color not loaded in conjunction with
      terminal option `colourtext'%
    }{See the gnuplot documentation for explanation.%
    }{Either use 'blacktext' in gnuplot or load the package
      color.sty in LaTeX.}%
    \renewcommand\color[2][]{}%
  }%
  \providecommand\includegraphics[2][]{%
    \GenericError{(gnuplot) \space\space\space\@spaces}{%
      Package graphicx or graphics not loaded%
    }{See the gnuplot documentation for explanation.%
    }{The gnuplot epslatex terminal needs graphicx.sty or graphics.sty.}%
    \renewcommand\includegraphics[2][]{}%
  }%
  \providecommand\rotatebox[2]{#2}%
  \@ifundefined{ifGPcolor}{%
    \newif\ifGPcolor
    \GPcolorfalse
  }{}%
  \@ifundefined{ifGPblacktext}{%
    \newif\ifGPblacktext
    \GPblacktexttrue
  }{}%
  \let\gplgaddtomacro\g@addto@macro
  \gdef\gplbacktext{}%
  \gdef\gplfronttext{}%
  \makeatother
  \ifGPblacktext
    \def\colorrgb#1{}%
    \def\colorgray#1{}%
  \else
    \ifGPcolor
      \def\colorrgb#1{\color[rgb]{#1}}%
      \def\colorgray#1{\color[gray]{#1}}%
      \expandafter\def\csname LTw\endcsname{\color{white}}%
      \expandafter\def\csname LTb\endcsname{\color{black}}%
      \expandafter\def\csname LTa\endcsname{\color{black}}%
      \expandafter\def\csname LT0\endcsname{\color[rgb]{1,0,0}}%
      \expandafter\def\csname LT1\endcsname{\color[rgb]{0,1,0}}%
      \expandafter\def\csname LT2\endcsname{\color[rgb]{0,0,1}}%
      \expandafter\def\csname LT3\endcsname{\color[rgb]{1,0,1}}%
      \expandafter\def\csname LT4\endcsname{\color[rgb]{0,1,1}}%
      \expandafter\def\csname LT5\endcsname{\color[rgb]{1,1,0}}%
      \expandafter\def\csname LT6\endcsname{\color[rgb]{0,0,0}}%
      \expandafter\def\csname LT7\endcsname{\color[rgb]{1,0.3,0}}%
      \expandafter\def\csname LT8\endcsname{\color[rgb]{0.5,0.5,0.5}}%
    \else
      \def\colorrgb#1{\color{black}}%
      \def\colorgray#1{\color[gray]{#1}}%
      \expandafter\def\csname LTw\endcsname{\color{white}}%
      \expandafter\def\csname LTb\endcsname{\color{black}}%
      \expandafter\def\csname LTa\endcsname{\color{black}}%
      \expandafter\def\csname LT0\endcsname{\color{black}}%
      \expandafter\def\csname LT1\endcsname{\color{black}}%
      \expandafter\def\csname LT2\endcsname{\color{black}}%
      \expandafter\def\csname LT3\endcsname{\color{black}}%
      \expandafter\def\csname LT4\endcsname{\color{black}}%
      \expandafter\def\csname LT5\endcsname{\color{black}}%
      \expandafter\def\csname LT6\endcsname{\color{black}}%
      \expandafter\def\csname LT7\endcsname{\color{black}}%
      \expandafter\def\csname LT8\endcsname{\color{black}}%
    \fi
  \fi
    \setlength{\unitlength}{0.0500bp}%
    \ifx\gptboxheight\undefined%
      \newlength{\gptboxheight}%
      \newlength{\gptboxwidth}%
      \newsavebox{\gptboxtext}%
    \fi%
    \setlength{\fboxrule}{0.5pt}%
    \setlength{\fboxsep}{1pt}%
\begin{picture}(7200.00,5040.00)%
    \gplgaddtomacro\gplbacktext{%
      \csname LTb\endcsname
      \put(682,704){\makebox(0,0)[r]{\strut{}$0$}}%
      \put(682,1072){\makebox(0,0)[r]{\strut{}$5$}}%
      \put(682,1439){\makebox(0,0)[r]{\strut{}$10$}}%
      \put(682,1807){\makebox(0,0)[r]{\strut{}$15$}}%
      \put(682,2174){\makebox(0,0)[r]{\strut{}$20$}}%
      \put(682,2542){\makebox(0,0)[r]{\strut{}$25$}}%
      \put(682,2909){\makebox(0,0)[r]{\strut{}$30$}}%
      \put(682,3277){\makebox(0,0)[r]{\strut{}$35$}}%
      \put(682,3644){\makebox(0,0)[r]{\strut{}$40$}}%
      \put(682,4012){\makebox(0,0)[r]{\strut{}$45$}}%
      \put(682,4379){\makebox(0,0)[r]{\strut{}$50$}}%
      \put(1275,484){\makebox(0,0){\strut{}$-0.2$}}%
      \put(2196,484){\makebox(0,0){\strut{}$0$}}%
      \put(3117,484){\makebox(0,0){\strut{}$0.2$}}%
      \put(4039,484){\makebox(0,0){\strut{}$0.4$}}%
      \put(4960,484){\makebox(0,0){\strut{}$0.6$}}%
      \put(5882,484){\makebox(0,0){\strut{}$0.8$}}%
      \put(6803,484){\makebox(0,0){\strut{}$1$}}%
    }%
    \gplgaddtomacro\gplfronttext{%
      \csname LTb\endcsname
      \put(198,2541){\rotatebox{-270}{\makebox(0,0){\strut{}$|F_\pi^V(s)|^2$}}}%
      \put(3808,154){\makebox(0,0){\strut{}$s$ [GeV${}^2$]}}%
      \put(3808,4709){\makebox(0,0){\strut{}Fit result for the VFF $|F_\pi^V(s)|^2$}}%
      \csname LTb\endcsname
      \put(2002,4206){\makebox(0,0)[r]{\strut{}Total error}}%
      \csname LTb\endcsname
      \put(2002,3986){\makebox(0,0)[r]{\strut{}Fit error}}%
      \csname LTb\endcsname
      \put(2002,3766){\makebox(0,0)[r]{\strut{}NA7}}%
      \csname LTb\endcsname
      \put(2002,3546){\makebox(0,0)[r]{\strut{}SND}}%
      \csname LTb\endcsname
      \put(2002,3326){\makebox(0,0)[r]{\strut{}CMD-2}}%
      \csname LTb\endcsname
      \put(2002,3106){\makebox(0,0)[r]{\strut{}BaBar}}%
      \csname LTb\endcsname
      \put(2002,2886){\makebox(0,0)[r]{\strut{}KLOE08}}%
      \csname LTb\endcsname
      \put(2002,2666){\makebox(0,0)[r]{\strut{}KLOE10}}%
      \csname LTb\endcsname
      \put(2002,2446){\makebox(0,0)[r]{\strut{}KLOE12}}%
    }%
    \gplbacktext
    \put(0,0){\includegraphics{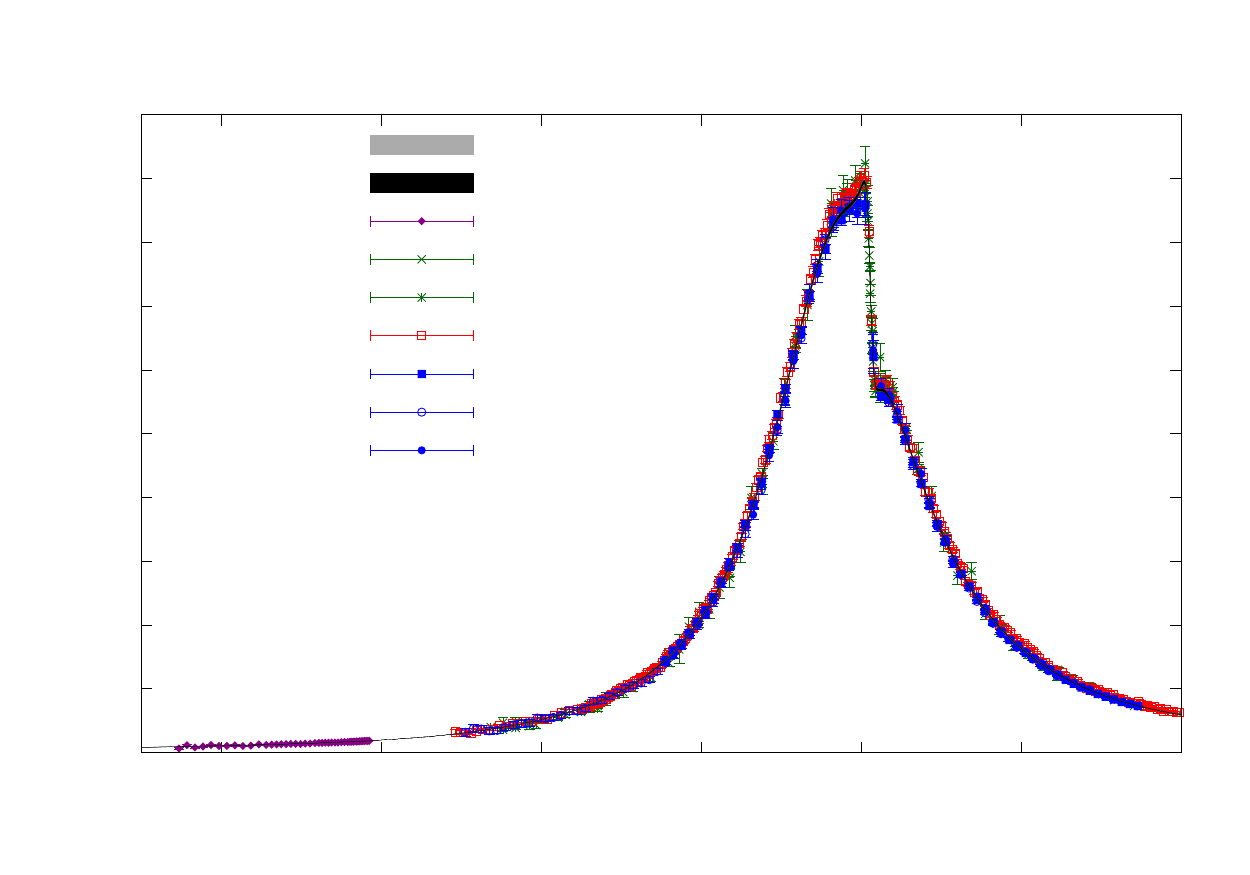}}%
    \gplfronttext
  \end{picture}%
\endgroup

%% file: plots/vff-spacelike.tex
\begingroup
  \makeatletter
  \providecommand\color[2][]{%
    \GenericError{(gnuplot) \space\space\space\@spaces}{%
      Package color not loaded in conjunction with
      terminal option `colourtext'%
    }{See the gnuplot documentation for explanation.%
    }{Either use 'blacktext' in gnuplot or load the package
      color.sty in LaTeX.}%
    \renewcommand\color[2][]{}%
  }%
  \providecommand\includegraphics[2][]{%
    \GenericError{(gnuplot) \space\space\space\@spaces}{%
      Package graphicx or graphics not loaded%
    }{See the gnuplot documentation for explanation.%
    }{The gnuplot epslatex terminal needs graphicx.sty or graphics.sty.}%
    \renewcommand\includegraphics[2][]{}%
  }%
  \providecommand\rotatebox[2]{#2}%
  \@ifundefined{ifGPcolor}{%
    \newif\ifGPcolor
    \GPcolorfalse
  }{}%
  \@ifundefined{ifGPblacktext}{%
    \newif\ifGPblacktext
    \GPblacktexttrue
  }{}%
  \let\gplgaddtomacro\g@addto@macro
  \gdef\gplbacktext{}%
  \gdef\gplfronttext{}%
  \makeatother
  \ifGPblacktext
    \def\colorrgb#1{}%
    \def\colorgray#1{}%
  \else
    \ifGPcolor
      \def\colorrgb#1{\color[rgb]{#1}}%
      \def\colorgray#1{\color[gray]{#1}}%
      \expandafter\def\csname LTw\endcsname{\color{white}}%
      \expandafter\def\csname LTb\endcsname{\color{black}}%
      \expandafter\def\csname LTa\endcsname{\color{black}}%
      \expandafter\def\csname LT0\endcsname{\color[rgb]{1,0,0}}%
      \expandafter\def\csname LT1\endcsname{\color[rgb]{0,1,0}}%
      \expandafter\def\csname LT2\endcsname{\color[rgb]{0,0,1}}%
      \expandafter\def\csname LT3\endcsname{\color[rgb]{1,0,1}}%
      \expandafter\def\csname LT4\endcsname{\color[rgb]{0,1,1}}%
      \expandafter\def\csname LT5\endcsname{\color[rgb]{1,1,0}}%
      \expandafter\def\csname LT6\endcsname{\color[rgb]{0,0,0}}%
      \expandafter\def\csname LT7\endcsname{\color[rgb]{1,0.3,0}}%
      \expandafter\def\csname LT8\endcsname{\color[rgb]{0.5,0.5,0.5}}%
    \else
      \def\colorrgb#1{\color{black}}%
      \def\colorgray#1{\color[gray]{#1}}%
      \expandafter\def\csname LTw\endcsname{\color{white}}%
      \expandafter\def\csname LTb\endcsname{\color{black}}%
      \expandafter\def\csname LTa\endcsname{\color{black}}%
      \expandafter\def\csname LT0\endcsname{\color{black}}%
      \expandafter\def\csname LT1\endcsname{\color{black}}%
      \expandafter\def\csname LT2\endcsname{\color{black}}%
      \expandafter\def\csname LT3\endcsname{\color{black}}%
      \expandafter\def\csname LT4\endcsname{\color{black}}%
      \expandafter\def\csname LT5\endcsname{\color{black}}%
      \expandafter\def\csname LT6\endcsname{\color{black}}%
      \expandafter\def\csname LT7\endcsname{\color{black}}%
      \expandafter\def\csname LT8\endcsname{\color{black}}%
    \fi
  \fi
    \setlength{\unitlength}{0.0500bp}%
    \ifx\gptboxheight\undefined%
      \newlength{\gptboxheight}%
      \newlength{\gptboxwidth}%
      \newsavebox{\gptboxtext}%
    \fi%
    \setlength{\fboxrule}{0.5pt}%
    \setlength{\fboxsep}{1pt}%
\begin{picture}(7200.00,5040.00)%
    \gplgaddtomacro\gplbacktext{%
      \csname LTb\endcsname
      \put(814,704){\makebox(0,0)[r]{\strut{}$0.2$}}%
      \put(814,1163){\makebox(0,0)[r]{\strut{}$0.3$}}%
      \put(814,1623){\makebox(0,0)[r]{\strut{}$0.4$}}%
      \put(814,2082){\makebox(0,0)[r]{\strut{}$0.5$}}%
      \put(814,2542){\makebox(0,0)[r]{\strut{}$0.6$}}%
      \put(814,3001){\makebox(0,0)[r]{\strut{}$0.7$}}%
      \put(814,3460){\makebox(0,0)[r]{\strut{}$0.8$}}%
      \put(814,3920){\makebox(0,0)[r]{\strut{}$0.9$}}%
      \put(814,4379){\makebox(0,0)[r]{\strut{}$1$}}%
      \put(946,484){\makebox(0,0){\strut{}$-0.3$}}%
      \put(1922,484){\makebox(0,0){\strut{}$-0.25$}}%
      \put(2898,484){\makebox(0,0){\strut{}$-0.2$}}%
      \put(3874,484){\makebox(0,0){\strut{}$-0.15$}}%
      \put(4851,484){\makebox(0,0){\strut{}$-0.1$}}%
      \put(5827,484){\makebox(0,0){\strut{}$-0.05$}}%
      \put(6803,484){\makebox(0,0){\strut{}$0$}}%
    }%
    \gplgaddtomacro\gplfronttext{%
      \csname LTb\endcsname
      \put(198,2541){\rotatebox{-270}{\makebox(0,0){\strut{}$|F_\pi^V(s)|^2$}}}%
      \put(3874,154){\makebox(0,0){\strut{}$s$ [GeV${}^2$]}}%
      \put(3874,4709){\makebox(0,0){\strut{}Fit result for the VFF $|F_\pi^V(s)|^2$}}%
      \csname LTb\endcsname
      \put(2134,4206){\makebox(0,0)[r]{\strut{}Total error}}%
      \csname LTb\endcsname
      \put(2134,3986){\makebox(0,0)[r]{\strut{}Fit error}}%
      \csname LTb\endcsname
      \put(2134,3766){\makebox(0,0)[r]{\strut{}NA7}}%
    }%
    \gplbacktext
    \put(0,0){\includegraphics{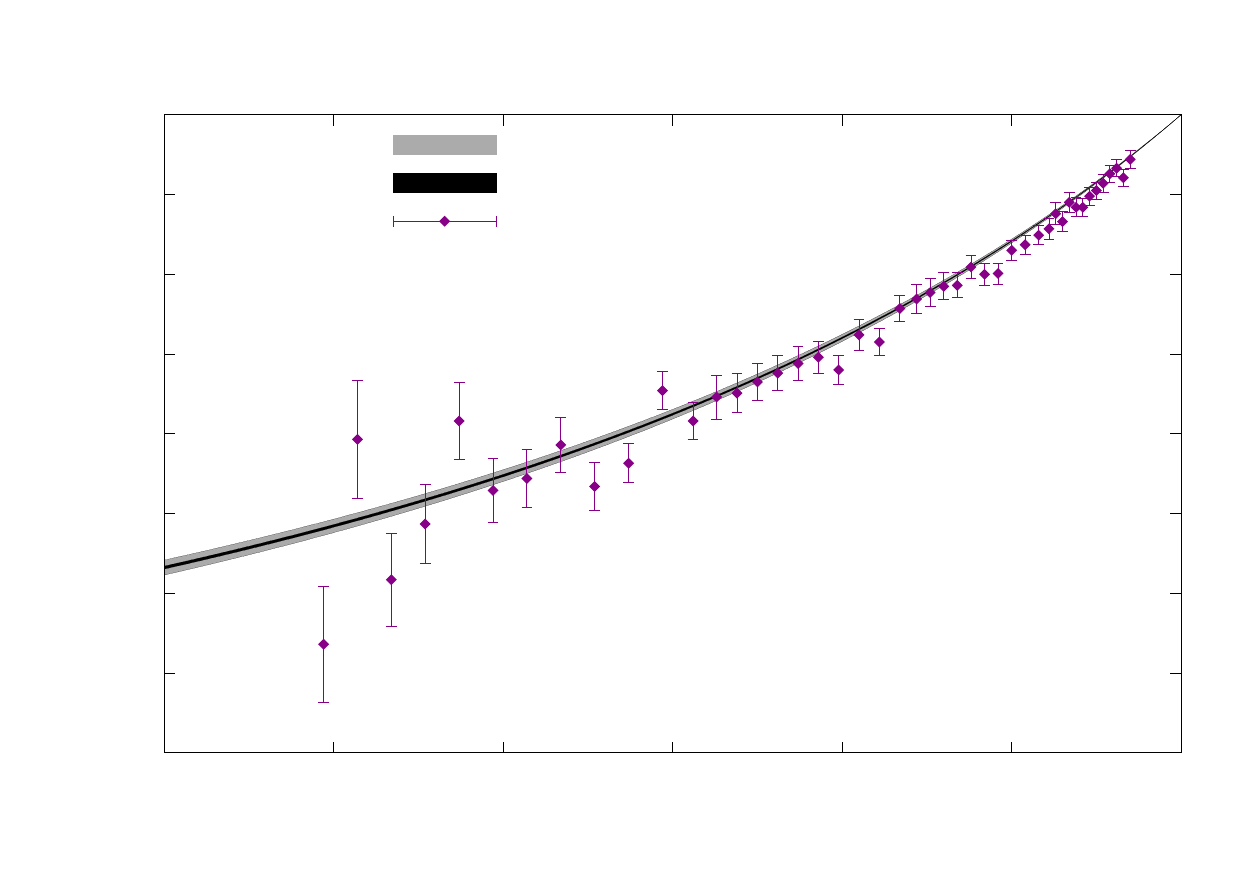}}%
    \gplfronttext
  \end{picture}%
\endgroup

%% file: plots/vff-zoom-Xi.tex
\begingroup
  \makeatletter
  \providecommand\color[2][]{%
    \GenericError{(gnuplot) \space\space\space\@spaces}{%
      Package color not loaded in conjunction with
      terminal option `colourtext'%
    }{See the gnuplot documentation for explanation.%
    }{Either use 'blacktext' in gnuplot or load the package
      color.sty in LaTeX.}%
    \renewcommand\color[2][]{}%
  }%
  \providecommand\includegraphics[2][]{%
    \GenericError{(gnuplot) \space\space\space\@spaces}{%
      Package graphicx or graphics not loaded%
    }{See the gnuplot documentation for explanation.%
    }{The gnuplot epslatex terminal needs graphicx.sty or graphics.sty.}%
    \renewcommand\includegraphics[2][]{}%
  }%
  \providecommand\rotatebox[2]{#2}%
  \@ifundefined{ifGPcolor}{%
    \newif\ifGPcolor
    \GPcolorfalse
  }{}%
  \@ifundefined{ifGPblacktext}{%
    \newif\ifGPblacktext
    \GPblacktexttrue
  }{}%
  \let\gplgaddtomacro\g@addto@macro
  \gdef\gplbacktext{}%
  \gdef\gplfronttext{}%
  \makeatother
  \ifGPblacktext
    \def\colorrgb#1{}%
    \def\colorgray#1{}%
  \else
    \ifGPcolor
      \def\colorrgb#1{\color[rgb]{#1}}%
      \def\colorgray#1{\color[gray]{#1}}%
      \expandafter\def\csname LTw\endcsname{\color{white}}%
      \expandafter\def\csname LTb\endcsname{\color{black}}%
      \expandafter\def\csname LTa\endcsname{\color{black}}%
      \expandafter\def\csname LT0\endcsname{\color[rgb]{1,0,0}}%
      \expandafter\def\csname LT1\endcsname{\color[rgb]{0,1,0}}%
      \expandafter\def\csname LT2\endcsname{\color[rgb]{0,0,1}}%
      \expandafter\def\csname LT3\endcsname{\color[rgb]{1,0,1}}%
      \expandafter\def\csname LT4\endcsname{\color[rgb]{0,1,1}}%
      \expandafter\def\csname LT5\endcsname{\color[rgb]{1,1,0}}%
      \expandafter\def\csname LT6\endcsname{\color[rgb]{0,0,0}}%
      \expandafter\def\csname LT7\endcsname{\color[rgb]{1,0.3,0}}%
      \expandafter\def\csname LT8\endcsname{\color[rgb]{0.5,0.5,0.5}}%
    \else
      \def\colorrgb#1{\color{black}}%
      \def\colorgray#1{\color[gray]{#1}}%
      \expandafter\def\csname LTw\endcsname{\color{white}}%
      \expandafter\def\csname LTb\endcsname{\color{black}}%
      \expandafter\def\csname LTa\endcsname{\color{black}}%
      \expandafter\def\csname LT0\endcsname{\color{black}}%
      \expandafter\def\csname LT1\endcsname{\color{black}}%
      \expandafter\def\csname LT2\endcsname{\color{black}}%
      \expandafter\def\csname LT3\endcsname{\color{black}}%
      \expandafter\def\csname LT4\endcsname{\color{black}}%
      \expandafter\def\csname LT5\endcsname{\color{black}}%
      \expandafter\def\csname LT6\endcsname{\color{black}}%
      \expandafter\def\csname LT7\endcsname{\color{black}}%
      \expandafter\def\csname LT8\endcsname{\color{black}}%
    \fi
  \fi
    \setlength{\unitlength}{0.0500bp}%
    \ifx\gptboxheight\undefined%
      \newlength{\gptboxheight}%
      \newlength{\gptboxwidth}%
      \newsavebox{\gptboxtext}%
    \fi%
    \setlength{\fboxrule}{0.5pt}%
    \setlength{\fboxsep}{1pt}%
\begin{picture}(7200.00,5040.00)%
    \gplgaddtomacro\gplbacktext{%
      \csname LTb\endcsname
      \put(682,704){\makebox(0,0)[r]{\strut{}$20$}}%
      \put(682,1317){\makebox(0,0)[r]{\strut{}$25$}}%
      \put(682,1929){\makebox(0,0)[r]{\strut{}$30$}}%
      \put(682,2542){\makebox(0,0)[r]{\strut{}$35$}}%
      \put(682,3154){\makebox(0,0)[r]{\strut{}$40$}}%
      \put(682,3767){\makebox(0,0)[r]{\strut{}$45$}}%
      \put(682,4379){\makebox(0,0)[r]{\strut{}$50$}}%
      \put(814,484){\makebox(0,0){\strut{}$0.74$}}%
      \put(1563,484){\makebox(0,0){\strut{}$0.75$}}%
      \put(2311,484){\makebox(0,0){\strut{}$0.76$}}%
      \put(3060,484){\makebox(0,0){\strut{}$0.77$}}%
      \put(3809,484){\makebox(0,0){\strut{}$0.78$}}%
      \put(4557,484){\makebox(0,0){\strut{}$0.79$}}%
      \put(5306,484){\makebox(0,0){\strut{}$0.8$}}%
      \put(6054,484){\makebox(0,0){\strut{}$0.81$}}%
      \put(6803,484){\makebox(0,0){\strut{}$0.82$}}%
    }%
    \gplgaddtomacro\gplfronttext{%
      \csname LTb\endcsname
      \put(198,2541){\rotatebox{-270}{\makebox(0,0){\strut{}$|F_\pi^V(s)|^2$}}}%
      \put(3808,154){\makebox(0,0){\strut{}$\sqrt{s}$ [GeV]}}%
      \put(3808,4709){\makebox(0,0){\strut{}VFF fit result and data with energy rescaling}}%
      \csname LTb\endcsname
      \put(5816,4206){\makebox(0,0)[r]{\strut{}Total error}}%
      \csname LTb\endcsname
      \put(5816,3986){\makebox(0,0)[r]{\strut{}Fit error}}%
      \csname LTb\endcsname
      \put(5816,3766){\makebox(0,0)[r]{\strut{}SND}}%
      \csname LTb\endcsname
      \put(5816,3546){\makebox(0,0)[r]{\strut{}CMD-2}}%
      \csname LTb\endcsname
      \put(5816,3326){\makebox(0,0)[r]{\strut{}BaBar}}%
      \csname LTb\endcsname
      \put(5816,3106){\makebox(0,0)[r]{\strut{}KLOE08}}%
      \csname LTb\endcsname
      \put(5816,2886){\makebox(0,0)[r]{\strut{}KLOE10}}%
      \csname LTb\endcsname
      \put(5816,2666){\makebox(0,0)[r]{\strut{}KLOE12}}%
    }%
    \gplbacktext
    \put(0,0){\includegraphics{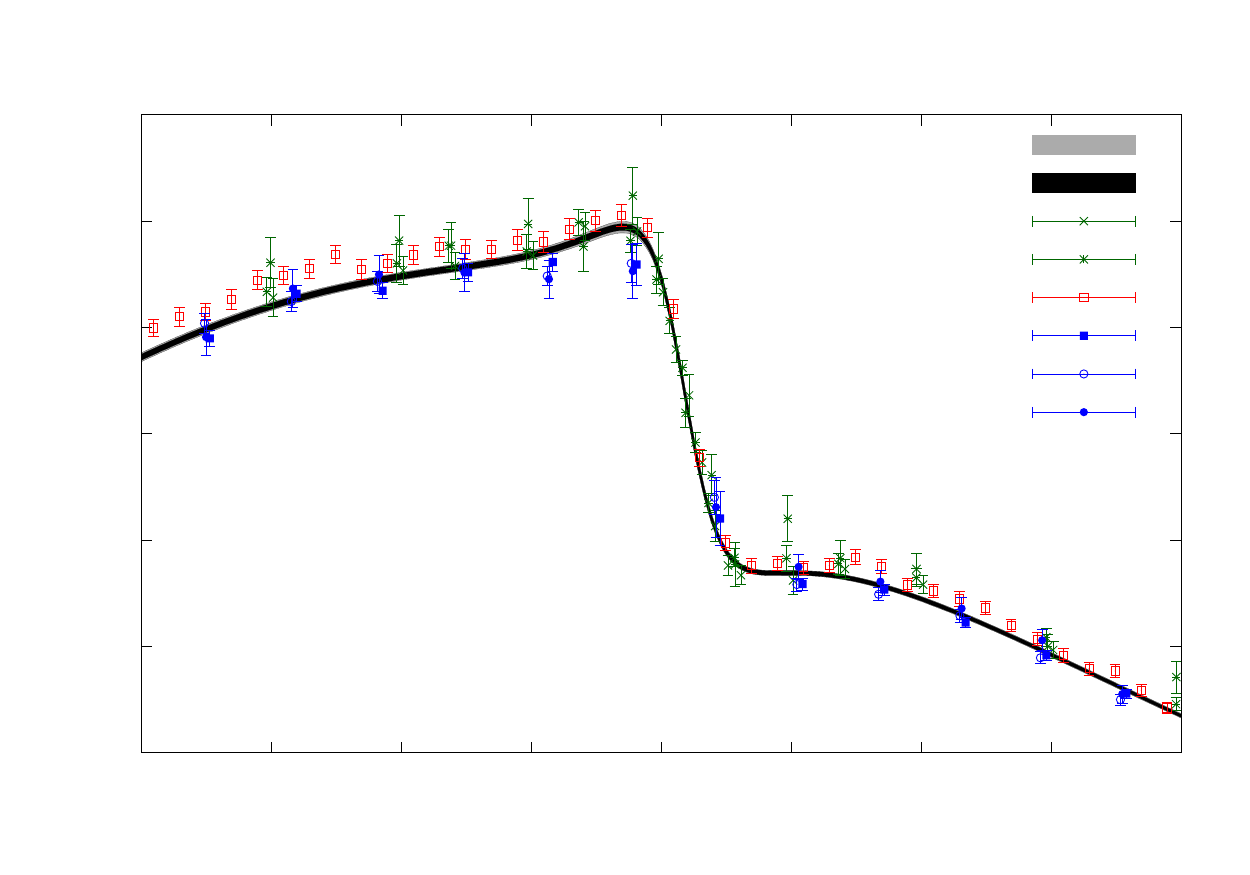}}%
    \gplfronttext
  \end{picture}%
\endgroup

%% file: plots/vff-zoom.tex
\begingroup
  \makeatletter
  \providecommand\color[2][]{%
    \GenericError{(gnuplot) \space\space\space\@spaces}{%
      Package color not loaded in conjunction with
      terminal option `colourtext'%
    }{See the gnuplot documentation for explanation.%
    }{Either use 'blacktext' in gnuplot or load the package
      color.sty in LaTeX.}%
    \renewcommand\color[2][]{}%
  }%
  \providecommand\includegraphics[2][]{%
    \GenericError{(gnuplot) \space\space\space\@spaces}{%
      Package graphicx or graphics not loaded%
    }{See the gnuplot documentation for explanation.%
    }{The gnuplot epslatex terminal needs graphicx.sty or graphics.sty.}%
    \renewcommand\includegraphics[2][]{}%
  }%
  \providecommand\rotatebox[2]{#2}%
  \@ifundefined{ifGPcolor}{%
    \newif\ifGPcolor
    \GPcolorfalse
  }{}%
  \@ifundefined{ifGPblacktext}{%
    \newif\ifGPblacktext
    \GPblacktexttrue
  }{}%
  \let\gplgaddtomacro\g@addto@macro
  \gdef\gplbacktext{}%
  \gdef\gplfronttext{}%
  \makeatother
  \ifGPblacktext
    \def\colorrgb#1{}%
    \def\colorgray#1{}%
  \else
    \ifGPcolor
      \def\colorrgb#1{\color[rgb]{#1}}%
      \def\colorgray#1{\color[gray]{#1}}%
      \expandafter\def\csname LTw\endcsname{\color{white}}%
      \expandafter\def\csname LTb\endcsname{\color{black}}%
      \expandafter\def\csname LTa\endcsname{\color{black}}%
      \expandafter\def\csname LT0\endcsname{\color[rgb]{1,0,0}}%
      \expandafter\def\csname LT1\endcsname{\color[rgb]{0,1,0}}%
      \expandafter\def\csname LT2\endcsname{\color[rgb]{0,0,1}}%
      \expandafter\def\csname LT3\endcsname{\color[rgb]{1,0,1}}%
      \expandafter\def\csname LT4\endcsname{\color[rgb]{0,1,1}}%
      \expandafter\def\csname LT5\endcsname{\color[rgb]{1,1,0}}%
      \expandafter\def\csname LT6\endcsname{\color[rgb]{0,0,0}}%
      \expandafter\def\csname LT7\endcsname{\color[rgb]{1,0.3,0}}%
      \expandafter\def\csname LT8\endcsname{\color[rgb]{0.5,0.5,0.5}}%
    \else
      \def\colorrgb#1{\color{black}}%
      \def\colorgray#1{\color[gray]{#1}}%
      \expandafter\def\csname LTw\endcsname{\color{white}}%
      \expandafter\def\csname LTb\endcsname{\color{black}}%
      \expandafter\def\csname LTa\endcsname{\color{black}}%
      \expandafter\def\csname LT0\endcsname{\color{black}}%
      \expandafter\def\csname LT1\endcsname{\color{black}}%
      \expandafter\def\csname LT2\endcsname{\color{black}}%
      \expandafter\def\csname LT3\endcsname{\color{black}}%
      \expandafter\def\csname LT4\endcsname{\color{black}}%
      \expandafter\def\csname LT5\endcsname{\color{black}}%
      \expandafter\def\csname LT6\endcsname{\color{black}}%
      \expandafter\def\csname LT7\endcsname{\color{black}}%
      \expandafter\def\csname LT8\endcsname{\color{black}}%
    \fi
  \fi
    \setlength{\unitlength}{0.0500bp}%
    \ifx\gptboxheight\undefined%
      \newlength{\gptboxheight}%
      \newlength{\gptboxwidth}%
      \newsavebox{\gptboxtext}%
    \fi%
    \setlength{\fboxrule}{0.5pt}%
    \setlength{\fboxsep}{1pt}%
\begin{picture}(7200.00,5040.00)%
    \gplgaddtomacro\gplbacktext{%
      \csname LTb\endcsname
      \put(682,704){\makebox(0,0)[r]{\strut{}$20$}}%
      \put(682,1317){\makebox(0,0)[r]{\strut{}$25$}}%
      \put(682,1929){\makebox(0,0)[r]{\strut{}$30$}}%
      \put(682,2542){\makebox(0,0)[r]{\strut{}$35$}}%
      \put(682,3154){\makebox(0,0)[r]{\strut{}$40$}}%
      \put(682,3767){\makebox(0,0)[r]{\strut{}$45$}}%
      \put(682,4379){\makebox(0,0)[r]{\strut{}$50$}}%
      \put(814,484){\makebox(0,0){\strut{}$0.74$}}%
      \put(1563,484){\makebox(0,0){\strut{}$0.75$}}%
      \put(2311,484){\makebox(0,0){\strut{}$0.76$}}%
      \put(3060,484){\makebox(0,0){\strut{}$0.77$}}%
      \put(3809,484){\makebox(0,0){\strut{}$0.78$}}%
      \put(4557,484){\makebox(0,0){\strut{}$0.79$}}%
      \put(5306,484){\makebox(0,0){\strut{}$0.8$}}%
      \put(6054,484){\makebox(0,0){\strut{}$0.81$}}%
      \put(6803,484){\makebox(0,0){\strut{}$0.82$}}%
    }%
    \gplgaddtomacro\gplfronttext{%
      \csname LTb\endcsname
      \put(198,2541){\rotatebox{-270}{\makebox(0,0){\strut{}$|F_\pi^V(s)|^2$}}}%
      \put(3808,154){\makebox(0,0){\strut{}$\sqrt{s}$ [GeV]}}%
      \put(3808,4709){\makebox(0,0){\strut{}VFF fit result with $M_\omega^\text{PDG}$ and data without energy rescaling}}%
      \csname LTb\endcsname
      \put(5816,4206){\makebox(0,0)[r]{\strut{}Total error}}%
      \csname LTb\endcsname
      \put(5816,3986){\makebox(0,0)[r]{\strut{}Fit error}}%
      \csname LTb\endcsname
      \put(5816,3766){\makebox(0,0)[r]{\strut{}SND}}%
      \csname LTb\endcsname
      \put(5816,3546){\makebox(0,0)[r]{\strut{}CMD-2}}%
      \csname LTb\endcsname
      \put(5816,3326){\makebox(0,0)[r]{\strut{}BaBar}}%
      \csname LTb\endcsname
      \put(5816,3106){\makebox(0,0)[r]{\strut{}KLOE08}}%
      \csname LTb\endcsname
      \put(5816,2886){\makebox(0,0)[r]{\strut{}KLOE10}}%
      \csname LTb\endcsname
      \put(5816,2666){\makebox(0,0)[r]{\strut{}KLOE12}}%
    }%
    \gplbacktext
    \put(0,0){\includegraphics{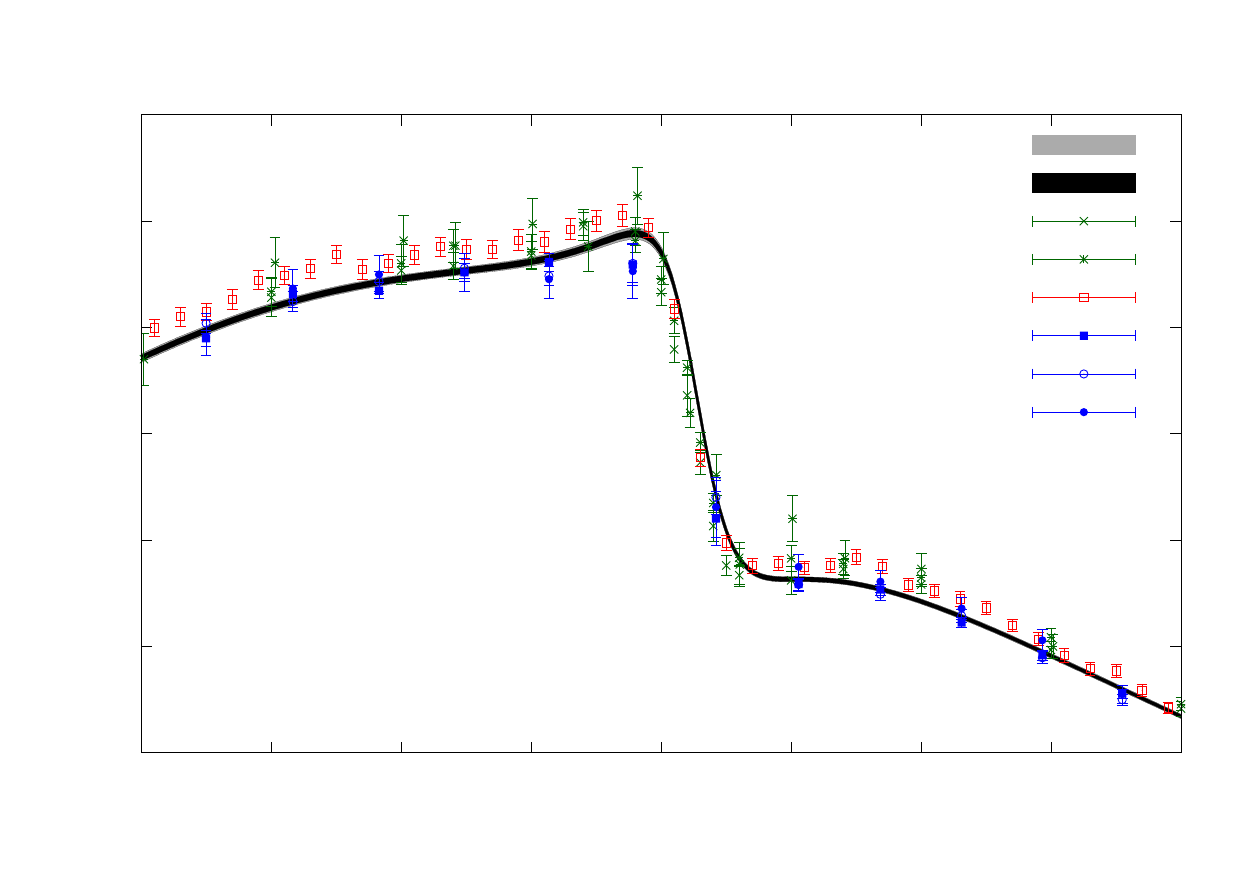}}%
    \gplfronttext
  \end{picture}%
\endgroup

%% file: plots/vff-rel-Xi.tex
\begingroup
  \makeatletter
  \providecommand\color[2][]{%
    \GenericError{(gnuplot) \space\space\space\@spaces}{%
      Package color not loaded in conjunction with
      terminal option `colourtext'%
    }{See the gnuplot documentation for explanation.%
    }{Either use 'blacktext' in gnuplot or load the package
      color.sty in LaTeX.}%
    \renewcommand\color[2][]{}%
  }%
  \providecommand\includegraphics[2][]{%
    \GenericError{(gnuplot) \space\space\space\@spaces}{%
      Package graphicx or graphics not loaded%
    }{See the gnuplot documentation for explanation.%
    }{The gnuplot epslatex terminal needs graphicx.sty or graphics.sty.}%
    \renewcommand\includegraphics[2][]{}%
  }%
  \providecommand\rotatebox[2]{#2}%
  \@ifundefined{ifGPcolor}{%
    \newif\ifGPcolor
    \GPcolorfalse
  }{}%
  \@ifundefined{ifGPblacktext}{%
    \newif\ifGPblacktext
    \GPblacktexttrue
  }{}%
  \let\gplgaddtomacro\g@addto@macro
  \gdef\gplbacktext{}%
  \gdef\gplfronttext{}%
  \makeatother
  \ifGPblacktext
    \def\colorrgb#1{}%
    \def\colorgray#1{}%
  \else
    \ifGPcolor
      \def\colorrgb#1{\color[rgb]{#1}}%
      \def\colorgray#1{\color[gray]{#1}}%
      \expandafter\def\csname LTw\endcsname{\color{white}}%
      \expandafter\def\csname LTb\endcsname{\color{black}}%
      \expandafter\def\csname LTa\endcsname{\color{black}}%
      \expandafter\def\csname LT0\endcsname{\color[rgb]{1,0,0}}%
      \expandafter\def\csname LT1\endcsname{\color[rgb]{0,1,0}}%
      \expandafter\def\csname LT2\endcsname{\color[rgb]{0,0,1}}%
      \expandafter\def\csname LT3\endcsname{\color[rgb]{1,0,1}}%
      \expandafter\def\csname LT4\endcsname{\color[rgb]{0,1,1}}%
      \expandafter\def\csname LT5\endcsname{\color[rgb]{1,1,0}}%
      \expandafter\def\csname LT6\endcsname{\color[rgb]{0,0,0}}%
      \expandafter\def\csname LT7\endcsname{\color[rgb]{1,0.3,0}}%
      \expandafter\def\csname LT8\endcsname{\color[rgb]{0.5,0.5,0.5}}%
    \else
      \def\colorrgb#1{\color{black}}%
      \def\colorgray#1{\color[gray]{#1}}%
      \expandafter\def\csname LTw\endcsname{\color{white}}%
      \expandafter\def\csname LTb\endcsname{\color{black}}%
      \expandafter\def\csname LTa\endcsname{\color{black}}%
      \expandafter\def\csname LT0\endcsname{\color{black}}%
      \expandafter\def\csname LT1\endcsname{\color{black}}%
      \expandafter\def\csname LT2\endcsname{\color{black}}%
      \expandafter\def\csname LT3\endcsname{\color{black}}%
      \expandafter\def\csname LT4\endcsname{\color{black}}%
      \expandafter\def\csname LT5\endcsname{\color{black}}%
      \expandafter\def\csname LT6\endcsname{\color{black}}%
      \expandafter\def\csname LT7\endcsname{\color{black}}%
      \expandafter\def\csname LT8\endcsname{\color{black}}%
    \fi
  \fi
    \setlength{\unitlength}{0.0500bp}%
    \ifx\gptboxheight\undefined%
      \newlength{\gptboxheight}%
      \newlength{\gptboxwidth}%
      \newsavebox{\gptboxtext}%
    \fi%
    \setlength{\fboxrule}{0.5pt}%
    \setlength{\fboxsep}{1pt}%
\begin{picture}(7200.00,5040.00)%
    \gplgaddtomacro\gplbacktext{%
      \csname LTb\endcsname
      \put(1078,704){\makebox(0,0)[r]{\strut{}$-0.1$}}%
      \put(1078,1317){\makebox(0,0)[r]{\strut{}$-0.05$}}%
      \put(1078,1929){\makebox(0,0)[r]{\strut{}$0$}}%
      \put(1078,2542){\makebox(0,0)[r]{\strut{}$0.05$}}%
      \put(1078,3154){\makebox(0,0)[r]{\strut{}$0.1$}}%
      \put(1078,3766){\makebox(0,0)[r]{\strut{}$0.15$}}%
      \put(1078,4379){\makebox(0,0)[r]{\strut{}$0.2$}}%
      \put(1210,484){\makebox(0,0){\strut{}$0.6$}}%
      \put(2142,484){\makebox(0,0){\strut{}$0.65$}}%
      \put(3074,484){\makebox(0,0){\strut{}$0.7$}}%
      \put(4007,484){\makebox(0,0){\strut{}$0.75$}}%
      \put(4939,484){\makebox(0,0){\strut{}$0.8$}}%
      \put(5871,484){\makebox(0,0){\strut{}$0.85$}}%
      \put(6803,484){\makebox(0,0){\strut{}$0.9$}}%
    }%
    \gplgaddtomacro\gplfronttext{%
      \csname LTb\endcsname
      \put(198,2541){\rotatebox{-270}{\makebox(0,0){\strut{}$\frac{|F_\pi^V(s)|^2_\mathrm{data}}{|F_\pi^V(s)|^2_\mathrm{fit}}-1$}}}%
      \put(4006,154){\makebox(0,0){\strut{}$\sqrt{s}$ [GeV]}}%
      \put(4006,4709){\makebox(0,0){\strut{}Relative difference between data sets and fit result}}%
      \csname LTb\endcsname
      \put(2530,4184){\makebox(0,0)[r]{\strut{}Total error}}%
      \csname LTb\endcsname
      \put(2530,3964){\makebox(0,0)[r]{\strut{}Fit error}}%
      \csname LTb\endcsname
      \put(2530,3744){\makebox(0,0)[r]{\strut{}SND}}%
      \csname LTb\endcsname
      \put(2530,3524){\makebox(0,0)[r]{\strut{}CMD-2}}%
      \csname LTb\endcsname
      \put(4573,4184){\makebox(0,0)[r]{\strut{}BaBar}}%
      \csname LTb\endcsname
      \put(4573,3964){\makebox(0,0)[r]{\strut{}KLOE08}}%
      \csname LTb\endcsname
      \put(4573,3744){\makebox(0,0)[r]{\strut{}KLOE10}}%
      \csname LTb\endcsname
      \put(4573,3524){\makebox(0,0)[r]{\strut{}KLOE12}}%
    }%
    \gplbacktext
    \put(0,0){\includegraphics{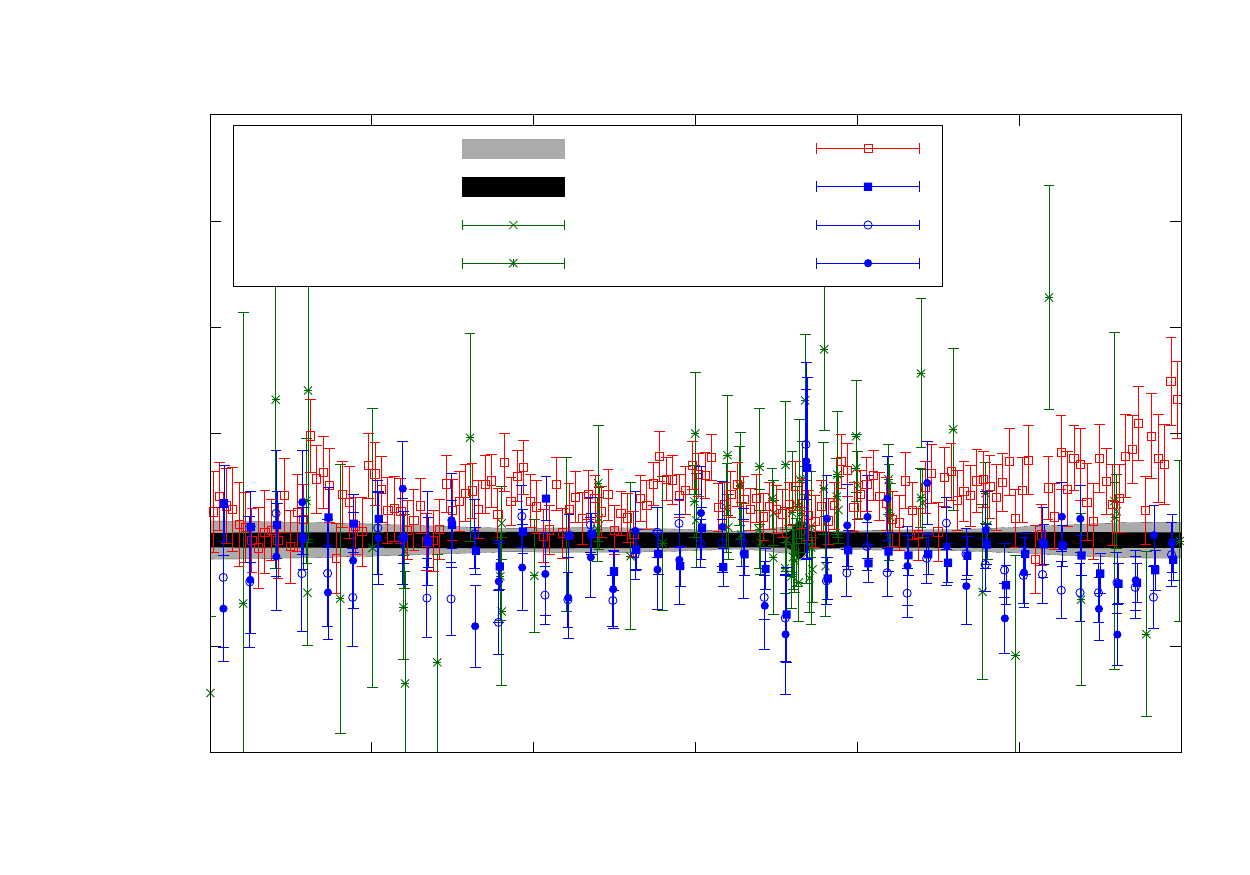}}%
    \gplfronttext
  \end{picture}%
\endgroup

%% file: plots/ellipses.tex
\begingroup
  \makeatletter
  \providecommand\color[2][]{%
    \GenericError{(gnuplot) \space\space\space\@spaces}{%
      Package color not loaded in conjunction with
      terminal option `colourtext'%
    }{See the gnuplot documentation for explanation.%
    }{Either use 'blacktext' in gnuplot or load the package
      color.sty in LaTeX.}%
    \renewcommand\color[2][]{}%
  }%
  \providecommand\includegraphics[2][]{%
    \GenericError{(gnuplot) \space\space\space\@spaces}{%
      Package graphicx or graphics not loaded%
    }{See the gnuplot documentation for explanation.%
    }{The gnuplot epslatex terminal needs graphicx.sty or graphics.sty.}%
    \renewcommand\includegraphics[2][]{}%
  }%
  \providecommand\rotatebox[2]{#2}%
  \@ifundefined{ifGPcolor}{%
    \newif\ifGPcolor
    \GPcolorfalse
  }{}%
  \@ifundefined{ifGPblacktext}{%
    \newif\ifGPblacktext
    \GPblacktexttrue
  }{}%
  \let\gplgaddtomacro\g@addto@macro
  \gdef\gplbacktext{}%
  \gdef\gplfronttext{}%
  \makeatother
  \ifGPblacktext
    \def\colorrgb#1{}%
    \def\colorgray#1{}%
  \else
    \ifGPcolor
      \def\colorrgb#1{\color[rgb]{#1}}%
      \def\colorgray#1{\color[gray]{#1}}%
      \expandafter\def\csname LTw\endcsname{\color{white}}%
      \expandafter\def\csname LTb\endcsname{\color{black}}%
      \expandafter\def\csname LTa\endcsname{\color{black}}%
      \expandafter\def\csname LT0\endcsname{\color[rgb]{1,0,0}}%
      \expandafter\def\csname LT1\endcsname{\color[rgb]{0,1,0}}%
      \expandafter\def\csname LT2\endcsname{\color[rgb]{0,0,1}}%
      \expandafter\def\csname LT3\endcsname{\color[rgb]{1,0,1}}%
      \expandafter\def\csname LT4\endcsname{\color[rgb]{0,1,1}}%
      \expandafter\def\csname LT5\endcsname{\color[rgb]{1,1,0}}%
      \expandafter\def\csname LT6\endcsname{\color[rgb]{0,0,0}}%
      \expandafter\def\csname LT7\endcsname{\color[rgb]{1,0.3,0}}%
      \expandafter\def\csname LT8\endcsname{\color[rgb]{0.5,0.5,0.5}}%
    \else
      \def\colorrgb#1{\color{black}}%
      \def\colorgray#1{\color[gray]{#1}}%
      \expandafter\def\csname LTw\endcsname{\color{white}}%
      \expandafter\def\csname LTb\endcsname{\color{black}}%
      \expandafter\def\csname LTa\endcsname{\color{black}}%
      \expandafter\def\csname LT0\endcsname{\color{black}}%
      \expandafter\def\csname LT1\endcsname{\color{black}}%
      \expandafter\def\csname LT2\endcsname{\color{black}}%
      \expandafter\def\csname LT3\endcsname{\color{black}}%
      \expandafter\def\csname LT4\endcsname{\color{black}}%
      \expandafter\def\csname LT5\endcsname{\color{black}}%
      \expandafter\def\csname LT6\endcsname{\color{black}}%
      \expandafter\def\csname LT7\endcsname{\color{black}}%
      \expandafter\def\csname LT8\endcsname{\color{black}}%
    \fi
  \fi
    \setlength{\unitlength}{0.0500bp}%
    \ifx\gptboxheight\undefined%
      \newlength{\gptboxheight}%
      \newlength{\gptboxwidth}%
      \newsavebox{\gptboxtext}%
    \fi%
    \setlength{\fboxrule}{0.5pt}%
    \setlength{\fboxsep}{1pt}%
\begin{picture}(7200.00,5040.00)%
    \gplgaddtomacro\gplbacktext{%
      \csname LTb\endcsname
      \put(946,704){\makebox(0,0)[r]{\strut{}$1.8$}}%
      \put(946,1368){\makebox(0,0)[r]{\strut{}$1.85$}}%
      \put(946,2031){\makebox(0,0)[r]{\strut{}$1.9$}}%
      \put(946,2695){\makebox(0,0)[r]{\strut{}$1.95$}}%
      \put(946,3359){\makebox(0,0)[r]{\strut{}$2$}}%
      \put(946,4023){\makebox(0,0)[r]{\strut{}$2.05$}}%
      \put(946,4686){\makebox(0,0)[r]{\strut{}$2.1$}}%
      \put(1518,484){\makebox(0,0){\strut{}$781.2$}}%
      \put(2399,484){\makebox(0,0){\strut{}$781.4$}}%
      \put(3280,484){\makebox(0,0){\strut{}$781.6$}}%
      \put(4161,484){\makebox(0,0){\strut{}$781.8$}}%
      \put(5041,484){\makebox(0,0){\strut{}$782$}}%
      \put(5922,484){\makebox(0,0){\strut{}$782.2$}}%
      \put(6803,484){\makebox(0,0){\strut{}$782.4$}}%
    }%
    \gplgaddtomacro\gplfronttext{%
      \csname LTb\endcsname
      \put(198,2761){\rotatebox{-270}{\makebox(0,0){\strut{}$10^3 \times \epsilon_\omega$}}}%
      \put(3940,154){\makebox(0,0){\strut{}$M_\omega$ [MeV]}}%
      \csname LTb\endcsname
      \put(5816,4525){\makebox(0,0)[r]{\strut{}All $e^+e^-$ (KLOE$''$), NA7}}%
      \csname LTb\endcsname
      \put(5816,4305){\makebox(0,0)[r]{\strut{}SND}}%
      \csname LTb\endcsname
      \put(5816,4085){\makebox(0,0)[r]{\strut{}CMD-2}}%
      \csname LTb\endcsname
      \put(5816,3865){\makebox(0,0)[r]{\strut{}BaBar}}%
      \csname LTb\endcsname
      \put(5816,3645){\makebox(0,0)[r]{\strut{}KLOE$''$}}%
    }%
    \gplbacktext
    \put(0,0){\includegraphics{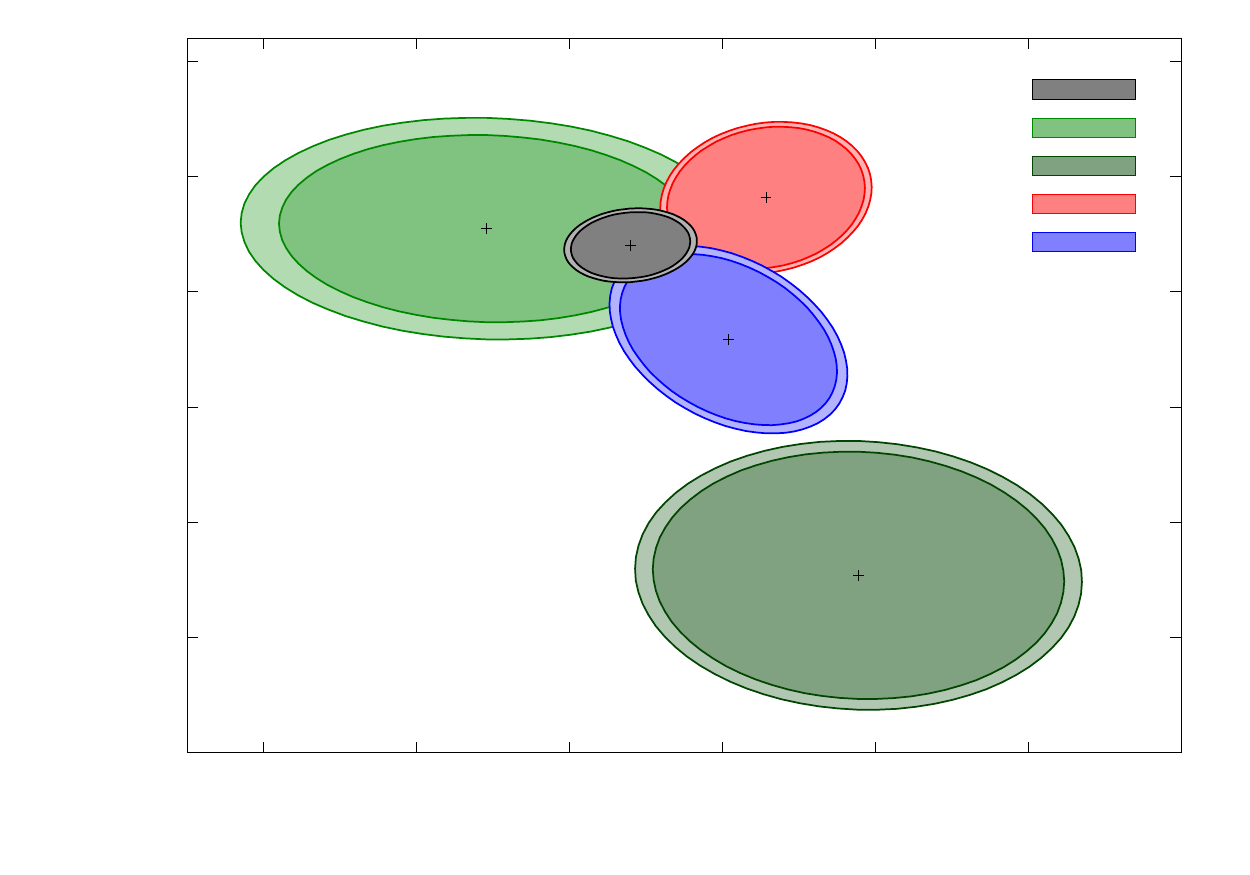}}%
    \gplfronttext
  \end{picture}%
\endgroup

%% file: sections/amu.tex

\section{Consequences for the anomalous magnetic moment of the muon}

\begin{table}[t]
	\centering
	\footnotesize
	\begin{tabularx}{16.8cm}{p{3.25cm} c c c c c}
	\toprule
	$10^{10} \times a_\mu^{\pi\pi}$ \\
	Energy region [GeV]	& $\le0.6$	& $\le0.7$	& $\le0.8$	& $\le0.9$    & $\le1.0$ \\
	\midrule
	SND		&	$110.3(1.2)(1.4)$  	&	$215.8(2.9)(2.5)$	&	$416.3(5.7)(3.6)$	&	$484.0(6.7)(4.0)$	&	$499.7(6.9)(4.1)$	\\
	CMD-2	&	$109.1(1.0)(1.3)$  	&	$212.8(2.1)(2.1)$	&	$413.2(3.4)(2.1)$	&	$481.4(3.9)(2.3)$	&	$496.9(4.0)(2.3)$	\\
	BaBar	&	$110.8(6)(8)$  		&	$216.8(1.4)(1.3)$	&	$418.2(2.8)(1.8)$	&	$486.1(3.2)(2.0)$	&	$501.9(3.3)(2.0)$	\\
	KLOE	&	$110.1(5)(5)$  		&	$214.6(1.1)(1.2)$	&	$411.2(1.9)(1.6)$	&	$477.0(2.2)(1.8)$	&	$492.0(2.2)(1.8)$	\\
	KLOE$''$	&	$110.2(5)(5)$  		&	$214.6(1.0)(1.0)$	&	$410.9(1.8)(1.4)$	&	$476.7(2.0)(1.7)$	&	$491.8(2.1)(1.8)$	\\
	\end{tabularx}
	\begin{tabularx}{16.8cm}{p{3.25cm} c c c c c c c c c}
	\midrule
	Energy region [GeV]	& \phantom{$\le0.6$}	& $[0.6,0.7]$	& $[0.7,0.8]$	& $[0.8,0.9]$    & $[0.9,1.0]$ \\
	\midrule
	SND		&	\phantom{$110.3(1.2)(1.4)$} 	&	$105.5(1.7)(1.1)$	&	$200.4(3.0)(1.3)$	&	$67.7(1.1)(0.5)$	&	$15.7(3)(2)$	\\
	CMD-2	&	\phantom{$109.1(1.0)(1.3)$}  	&	$103.7(1.2)(0.8)$	&	$200.4(1.7)(0.0)$	&	$68.2(7)(2)$		&	$15.6(2)(0)$	\\
	BaBar	&	\phantom{$110.5(6)(8)$}  		&	$106.0(8)(5)$		&	$201.3(1.5)(0.6)$	&	$68.0(6)(3)$		&	$15.8(2)(1)$	\\
	KLOE	&	\phantom{$110.0(5)(6)$}  		&	$104.5(6)(5)$		&	$196.6(1.0)(0.8)$	&	$65.8(3)(3)$		&	$15.1(1)(1)$	\\
	KLOE$''$	&	\phantom{$110.0(5)(5)$}  		&	$104.4(6)(5)$		&	$196.3(9)(8)$	&	$65.8(3)(3)$		&	$15.1(1)(1)$	\\
	\end{tabularx}
	\begin{tabularx}{16.8cm}{p{3.25cm} c c c c c c c c c}
	\midrule
	Energy region [GeV]  & $\le0.63$	& $[0.6,0.9]$ & Ref.~\cite{Anastasi:2017eio} & $\big[\sqrt{0.1},\sqrt{0.95}\,\big]$ & Ref.~\cite{Anastasi:2017eio} \\
	\midrule
	SND		&	$133.2(1.6)(1.7)$	&	$373.6(5.6)(2.6)$  	&	$371.7(5.0)$	&	$495.3(6.9)(4.0)$	&				\\
	CMD-2	&	$131.6(1.2)(1.6)$	&	$372.2(3.1)(1.0)$  	&	$372.4(3.0)$	&	$492.6(3.9)(2.3)$	&				\\
	BaBar	&	$133.8(8)(9)$		&	$375.3(2.7)(1.2)$  	&	$376.7(2.7)$	&	$497.5(3.3)(2.0)$	&				\\
	KLOE	&	$132.8(6)(8)$		&	$366.8(1.8)(1.5)$  	&	$366.9(2.1)$	&	$487.7(2.2)(1.8)$	&	$489.8(5.1)$	\\
	KLOE$''$	&	$132.9(6)(6)$		&	$366.5(1.7)(1.6)$  	&	$366.9(2.1)$	&	$487.5(2.1)(1.7)$	&	$489.8(5.1)$	\\
	\bottomrule
	\end{tabularx}
	\caption{Values for $a_\mu^{\pi\pi}$ from our final fits to single $e^+e^-$ experiments. The first error is the fit uncertainty, inflated by $\sqrt{\chi^2/\mathrm{dof}}$, the second error the combination of all systematic uncertainties. We provide the results for several energy regions separately, to enable a detailed comparison with other (future) evaluations. The energy regions in the third block are provided to facilitate comparison with~\cite{Ananthanarayan:2016mns} and the results of the direct integration~\cite{Anastasi:2017eio}.}
	\label{tab:FinalAmuSingleExperiments}
\end{table}

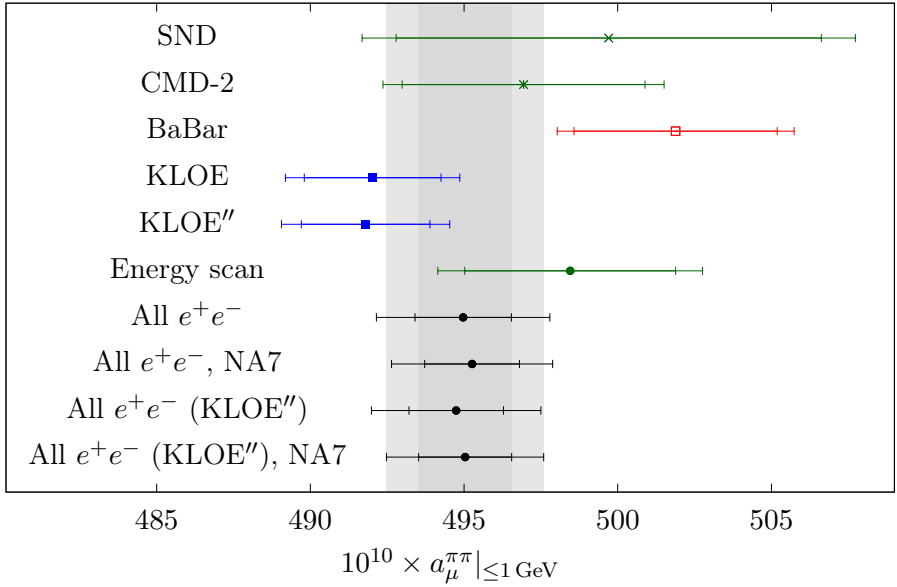
\begin{figure}[t]
	\centering
	\input{plots/amu}
	\caption{Results for $a_\mu^{\pi\pi}$ in the energy range $\le1\GeV$. The smaller error bars are the fit uncertainties, inflated by $\sqrt{\chi^2/\mathrm{dof}}$, the larger error bars are the total uncertainties. The gray bands indicate our final result.}
	\label{img:AmuResults}
\end{figure}

In Table~\ref{tab:FinalAmuSingleExperiments} we collect the results for  $a_\mu^{\pi\pi}$ for single time-like experiments and a variety of different energy regions below $1\GeV$, supplemented by the range $\sqrt{s} \in [0.6, 0.9]\GeV$ as before and the ranges $s\in [0.1, 0.95]\GeV^2$ and $s\in [4M_\pi^2, (0.63\GeV)^2 ]$ that have been considered in previous work. 
The same set of results is shown in Table~\ref{tab:FinalAmuCombinations} for the combination of the energy-scan experiments SND and \mbox{CMD-2}, all time-like data sets, and the full combination including NA7,
see Fig.~\ref{img:AmuResults} for the results for $a_\mu^{\pi\pi}$ below
$1\GeV$. Note that the result for the combined fit does not exactly
coincide with a naive weighted average of the fit results to single
experiments: most importantly, correlations play a role in the same way as discussed in Sect.~\ref{sec:omega}.
Further small deviations are due to the non-linear dependence of the
fit function on the parameters, which leads to distortions of the $\chi^2$. 
We checked that the deviations of the $\chi^2$ from a quadratic
function in the parameters are very small within the standard confidence
regions of the parameter space. Further away from the $\chi^2$
minimum, these deviations become more important and they have an observable
effect in the combination of the BaBar and KLOE data sets, which reflects
the well-known tension between these two experiments, see
Fig.~\ref{img:VFFfitRelativeXi}. 
Taking into account the correlation of the systematic uncertainties, this discrepancy between the BaBar and KLOE results for $a_\mu^{\pi\pi}$ below $1\GeV$ amounts to $2.6\sigma$.

\begin{figure}[t]
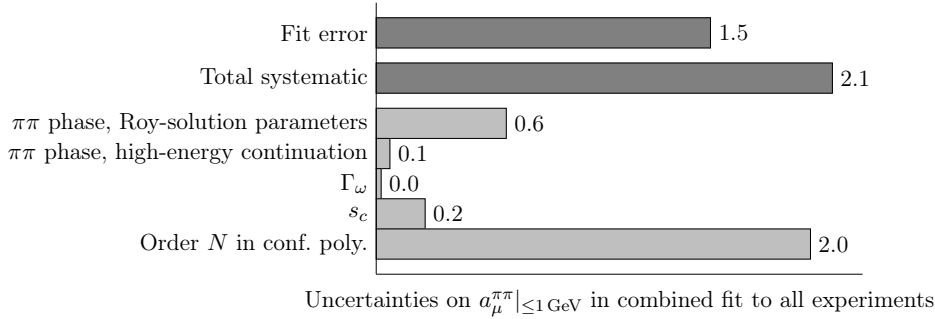

	\centering
	\renewcommand{\bcfontstyle}{\rmfamily}
	\scalebox{0.8}{
	\begin{bchart}[max=2.2,plain]
		\bcbar[label={Fit error},color=gray,value=1.5]{1.5119}
		\smallskip
		\bcbar[label=Total systematic,color=gray,value=2.1]{2.06328}
		\smallskip
		\bcbar[label={$\pi\pi$ phase, Roy-solution parameters},color=lightgray,value=0.6]{0.5883}
		\bcbar[label={$\pi\pi$ phase, high-energy continuation},color=lightgray,value=0.1]{0.0618}
		\bcbar[label={$\Gamma_\omega$},color=lightgray,value=0.0]{0.0222}
		\bcbar[label={$s_c$},color=lightgray,value=0.2]{0.2212}
		\bcbar[label={Order $N$ in conf.\ poly.},color=lightgray,value=2.0]{1.9641}
		\bcxlabel{Uncertainties on $a_\mu^{\pi\pi}|_{\le1\GeV}$ in combined fit to all experiments}
	\end{bchart} }%
	\caption{Contributions to the uncertainty of $a_\mu^{\pi\pi}$ in the energy range $\le1\GeV$ for the combined fit to SND, CMD-2, BaBar, KLOE$''$, and NA7.}
	\label{img:Errors}
\end{figure}

Finally, the relative size of various sources of systematic uncertainties is illustrated in Fig.~\ref{img:Errors}. 
The dominant systematic error is due to the order of the conformal polynomial, followed by the Roy parameters (including $\iota_1$) and $s_c$ from the conformal expansion.
This pattern holds true for most of the fit variants considered.

\begin{table}[t]
	\centering
	\footnotesize
	\begin{tabularx}{16.8cm}{p{3.5cm} c c c c c}
	\toprule
	$10^{10} \times a_\mu^{\pi\pi}$ \\
	Energy region [GeV]	& $\le0.6$	& $\le0.7$	& $\le0.8$	& $\le0.9$    & $\le1.0$ \\
	\midrule
	Energy scan					&	$109.9(0.7)(1.3)$  	&	$214.6(1.6)(2.0)$	&	$414.9(2.9)(2.4)$	&	$482.8(3.3)(2.6)$	&	$498.5(3.4)(2.6)$	\\
	All $e^+e^-$					&	$110.1(3)(9)$  		&	$214.8(0.7)(1.6)$	&	$413.2(1.3)(2.2)$	&	$479.7(1.5)(2.3)$	&	$495.0(1.6)(2.3)$	\\
	All $e^+e^-$, NA7				&	$110.2(3)(9)$  		&	$215.0(0.7)(1.5)$	&	$413.4(1.3)(2.0)$	&	$480.0(1.5)(2.1)$	&	$495.3(1.5)(2.1)$	\\
	All $e^+e^-$ (KLOE$''$)			&	$110.1(3)(9)$  		&	$214.7(0.7)(1.6)$	&	$412.9(1.3)(2.1)$	&	$479.5(1.5)(2.3)$	&	$494.7(1.5)(2.3)$	\\
	All $e^+e^-$ (KLOE$''$), NA7		&	$110.1(3)(9)$  		&	$214.8(0.7)(1.5)$	&	$413.2(1.3)(1.9)$	&	$479.8(1.5)(2.1)$	&	$495.0(1.5)(2.1)$	\\
	\end{tabularx}
	\begin{tabularx}{16.8cm}{p{3.5cm} c c c c c c c c c}
	\midrule
	Energy region [GeV]	& $\phantom{\le0.6}$	& $[0.6,0.7]$	& $[0.7,0.8]$	& $[0.8,0.9]$    & $[0.9,1.0]$ \\
	\midrule
	Energy scan					&	\phantom{$109.9(0.7)(1.3)$}  	&	$104.8(9)(8)$		&	$200.3(1.5)(0.3)$	&	$67.9(6)(2)$	&	$15.7(2)(0)$	\\
	All $e^+e^-$					&	\phantom{$110.1(3)(9)$}  		&	$104.7(4)(7)$		&	$198.4(7)(5)$		&	$66.5(2)(2)$	&	$15.3(1)(0)$	\\
	All $e^+e^-$, NA7				&	\phantom{$110.2(3)(9)$}  		&	$104.8(4)(6)$		&	$198.5(7)(6)$		&	$66.6(2)(3)$	&	$15.3(1)(0)$	\\
	All $e^+e^-$ (KLOE$''$)			&	\phantom{$110.1(3)(9)$}  		&	$104.6(4)(7)$		&	$198.2(7)(5)$		&	$66.6(2)(2)$	&	$15.3(1)(0)$	\\
	All $e^+e^-$ (KLOE$''$), NA7		&	\phantom{$110.1(3)(9)$}  		&	$104.7(4)(6)$		&	$198.3(7)(6)$		&	$66.6(2)(3)$	&	$15.3(1)(0)$	\\
	\end{tabularx}
	\begin{tabularx}{16.8cm}{p{3.5cm} c c c c c c c c c}
	\midrule
	Energy region [GeV]  & $\le0.63$ & Ref.~\cite{Ananthanarayan:2016mns}	& $[0.6,0.9]$ & Ref.~\cite{Keshavarzi:2018mgv} & $\big[\sqrt{0.1},\sqrt{0.95}\,\big]$ \\
	\midrule
	Energy scan					&	$132.6(0.9)(1.5)$ 	&				&	$372.9(2.8)(1.4)$  	&	$370.8(2.6)$	&	$494.1(3.4)(2.6)$					\\
	All $e^+e^-$					&	$132.8(0.4)(1.1)$	&	$133.0(8)$	&	$369.6(1.3)(1.4)$  	&	$369.4(1.3)$	&	$490.6(1.6)(2.3)$					\\
	All $e^+e^-$, NA7				&	$132.9(0.4)(1.1)$	&				&	$369.8(1.3)(1.3)$  	&	$ $			&	$490.9(1.5)(2.1)$					\\
	All $e^+e^-$ (KLOE$''$)			&	$132.8(0.4)(1.1)$	&	$133.0(8)$	&	$369.4(1.3)(1.4)$  	&	$369.4(1.3)$	&	$490.4(1.5)(2.3)$					\\
	All $e^+e^-$ (KLOE$''$), NA7		&	$132.8(0.4)(1.0)$	&				&	$369.6(1.2)(1.2)$  	&	$ $			&	$490.7(1.5)(2.1)$					\\
	\bottomrule
	\end{tabularx}
	\caption{Values for $a_\mu^{\pi\pi}$ from our final fits to combinations of data sets. The first error is the fit uncertainty, inflated by $\sqrt{\chi^2/\mathrm{dof}}$, the second error is the combination of all systematic uncertainties.}
	\label{tab:FinalAmuCombinations}
\end{table}

Where published results are available, we have included the comparison in
the tables, e.g.\ from direct
integration~\cite{Anastasi:2017eio,Keshavarzi:2018mgv} and the dispersive
analysis~\cite{Ananthanarayan:2016mns}. We find that in those cases where
reference values exist, our results appear well compatible, within
uncertainties of a similar size.
An exception is the comparison to the direct integration of the data
between $\sqrt{0.1}$ and $\sqrt{0.95}$ GeV performed by
KLOE~\cite{Anastasi:2017eio} where our method shows a significant reduction
of the uncertainties: this is mainly due to the region below $0.6$ GeV
where KLOE data show a loss of precision. With our approach once precise
data are available in the most sensitive region around the $\rho$ peak they
strongly constrain the curve in the whole low-energy region and the
extrapolation down to the two-pion threshold does not lead to any loss in
precision: this is a clear advantage of our method with respect to the
application of the trapezoidal rule. On the other hand,
Tables~\ref{tab:FinalAmuSingleExperiments} and \ref{tab:FinalAmuCombinations} show
that in the regions where there are high-quality data, these are so precise
and densely spaced that our method does not lead to an increase of
precision, but mainly serves as a check of the consistency of the data with
the principles of analyticity and unitarity. 
Reversing the argument, the fact that our uncertainties are of similar size
 as those of other analyses shows that the systematic
uncertainties in the dispersive representation, which we have
investigated in detail, are well under control in the whole region below $1\GeV$.
We stress that our uncertainty estimates, illustrated and summarized in
Fig.~\ref{img:Errors}, rely on minimal assumptions, the dispersive
parametrization as a consequence of QCD and the covariances matrices
provided by experiment, where the latter then effectively determine the
relative weight of each data set in the combined fit. In particular, a
local inflation of the uncertainties would be difficult to justify 
in this formalism, which emphasizes the importance of the finding that each
data set allows for a statistically acceptable fit once potential
uncertainties in the energy calibration  
are taken into account (and the two outliers in KLOE08 removed).\footnote{As demonstrated by Table~\ref{tab:FinalAmuCombinations}, the central values in the combined fit to all experiments barely change when the two KLOE08 outliers are retained: the value~\eqref{a_mu_final} for $a_\mu^{\pi\pi}$ below $1\GeV$ increases by $0.2\times10^{-10}$, but the total $\chi^2$ is worse by about 30 units and leads to a slightly larger scale factor~\eqref{eq:Chi2ScaleFactor} of $1.13$ instead of $1.11$.} 
We look forward to more detailed comparisons 
to direct integration~\cite{Davier:2017zfy,Keshavarzi:2018mgv}, which
should lead to a better understanding of the uncertainties in the critical
$\pi\pi$ channel and thereby to a consolidation of the overall uncertainty
estimate for HVP.

Our most comprehensive result gives the full contribution below $1\GeV$ in a combination of all available time- and space-like constraints
\beq
\label{a_mu_final}
a_\mu^{\pi\pi}|_{\leq 1\GeV}=495.0(1.5)(2.1)\times 10^{-10}=495.0(2.6)\times 10^{-10},
\eeq
where the inclusion of the space-like data does allow for a modest reduction of the uncertainty from $2.8$ to $2.6$ units.
As noted before~\cite{Ananthanarayan:2016mns}, the main advantage over direct integration occurs in energy regions where data are still scarce, most notably the low-energy region
\beq
a_\mu^{\pi\pi}|_{\leq 0.63\GeV}=132.8(0.4)(1.0)\times 10^{-10}=132.8(1.1)\times 10^{-10}.
\eeq
Our result agrees with the combination of $e^+e^-$ data sets from~\cite{Ananthanarayan:2016mns}, $a_\mu^{\pi\pi}|_{\leq 0.63\GeV}=133.0(8)\times 10^{-10}$, which provides
another important cross check given several conceptual differences compared to our study.\footnote{Note that the final number quoted in~\cite{Ananthanarayan:2016mns}, $a_\mu^{\pi\pi}|_{\leq 0.63\GeV}=133.3(7)\times 10^{-10}$, 
also includes information from $\tau$ data, but given the difficulties in controlling the required isospin-breaking corrections we only consider $e^+e^-$ data here.} 
The main difference to our approach concerns the fact that the $\pi\pi$ phase shift is not fit to the data, but taken as an input. The dispersive formalism is then set up in such a way that 
the phase shift in the elastic region alone, in combination with data for the modulus of the VFF in the energy region $\sqrt{s}\in[0.65,0.71]\GeV$, constrains the VFF in the low-energy region $\sqrt{s}\leq 0.63\GeV$.
On the one hand, in this way the systematic uncertainties related to the high-energy continuation of the phase shift and the inelastic corrections no longer need to be considered, 
but on the other hand the method is then necessarily restricted to rather low energies. In contrast, our representation remains applicable as long as inelastic corrections can still be controlled, 
within the formalism that we have employed here at least up to $1\GeV$. Moreover, our approach avoids a circularity problem that arises because the $\pi\pi$ phase shifts used as input 
have been extracted from previous form factor fits themselves, even though the numerical impact of this effect might be negligible in the end. 
The HVP result for the low-energy region agrees in both implementations of dispersive constraints on the pion VFF.


%% file: plots/amu.tex
\begingroup
  \makeatletter
  \providecommand\color[2][]{%
    \GenericError{(gnuplot) \space\space\space\@spaces}{%
      Package color not loaded in conjunction with
      terminal option `colourtext'%
    }{See the gnuplot documentation for explanation.%
    }{Either use 'blacktext' in gnuplot or load the package
      color.sty in LaTeX.}%
    \renewcommand\color[2][]{}%
  }%
  \providecommand\includegraphics[2][]{%
    \GenericError{(gnuplot) \space\space\space\@spaces}{%
      Package graphicx or graphics not loaded%
    }{See the gnuplot documentation for explanation.%
    }{The gnuplot epslatex terminal needs graphicx.sty or graphics.sty.}%
    \renewcommand\includegraphics[2][]{}%
  }%
  \providecommand\rotatebox[2]{#2}%
  \@ifundefined{ifGPcolor}{%
    \newif\ifGPcolor
    \GPcolorfalse
  }{}%
  \@ifundefined{ifGPblacktext}{%
    \newif\ifGPblacktext
    \GPblacktexttrue
  }{}%
  \let\gplgaddtomacro\g@addto@macro
  \gdef\gplbacktext{}%
  \gdef\gplfronttext{}%
  \makeatother
  \ifGPblacktext
    \def\colorrgb#1{}%
    \def\colorgray#1{}%
  \else
    \ifGPcolor
      \def\colorrgb#1{\color[rgb]{#1}}%
      \def\colorgray#1{\color[gray]{#1}}%
      \expandafter\def\csname LTw\endcsname{\color{white}}%
      \expandafter\def\csname LTb\endcsname{\color{black}}%
      \expandafter\def\csname LTa\endcsname{\color{black}}%
      \expandafter\def\csname LT0\endcsname{\color[rgb]{1,0,0}}%
      \expandafter\def\csname LT1\endcsname{\color[rgb]{0,1,0}}%
      \expandafter\def\csname LT2\endcsname{\color[rgb]{0,0,1}}%
      \expandafter\def\csname LT3\endcsname{\color[rgb]{1,0,1}}%
      \expandafter\def\csname LT4\endcsname{\color[rgb]{0,1,1}}%
      \expandafter\def\csname LT5\endcsname{\color[rgb]{1,1,0}}%
      \expandafter\def\csname LT6\endcsname{\color[rgb]{0,0,0}}%
      \expandafter\def\csname LT7\endcsname{\color[rgb]{1,0.3,0}}%
      \expandafter\def\csname LT8\endcsname{\color[rgb]{0.5,0.5,0.5}}%
    \else
      \def\colorrgb#1{\color{black}}%
      \def\colorgray#1{\color[gray]{#1}}%
      \expandafter\def\csname LTw\endcsname{\color{white}}%
      \expandafter\def\csname LTb\endcsname{\color{black}}%
      \expandafter\def\csname LTa\endcsname{\color{black}}%
      \expandafter\def\csname LT0\endcsname{\color{black}}%
      \expandafter\def\csname LT1\endcsname{\color{black}}%
      \expandafter\def\csname LT2\endcsname{\color{black}}%
      \expandafter\def\csname LT3\endcsname{\color{black}}%
      \expandafter\def\csname LT4\endcsname{\color{black}}%
      \expandafter\def\csname LT5\endcsname{\color{black}}%
      \expandafter\def\csname LT6\endcsname{\color{black}}%
      \expandafter\def\csname LT7\endcsname{\color{black}}%
      \expandafter\def\csname LT8\endcsname{\color{black}}%
    \fi
  \fi
    \setlength{\unitlength}{0.0500bp}%
    \ifx\gptboxheight\undefined%
      \newlength{\gptboxheight}%
      \newlength{\gptboxwidth}%
      \newsavebox{\gptboxtext}%
    \fi%
    \setlength{\fboxrule}{0.5pt}%
    \setlength{\fboxsep}{1pt}%
\begin{picture}(7200.00,5040.00)%
    \gplgaddtomacro\gplbacktext{%
      \csname LTb\endcsname
      \put(1300,484){\makebox(0,0){\strut{}$485$}}%
      \put(2446,484){\makebox(0,0){\strut{}$490$}}%
      \put(3593,484){\makebox(0,0){\strut{}$495$}}%
      \put(4739,484){\makebox(0,0){\strut{}$500$}}%
      \put(5886,484){\makebox(0,0){\strut{}$505$}}%
    }%
    \gplgaddtomacro\gplfronttext{%
      \csname LTb\endcsname
      \put(198,2541){\rotatebox{-270}{\makebox(0,0){\strut{}}}}%
      \put(3489,154){\makebox(0,0){\strut{}$10^{10} \times a_\mu^{\pi\pi}|_{\le 1\GeV}$}}%
      \put(3489,4709){\makebox(0,0){\strut{}Result for $a_\mu^{\pi\pi}|_{\le 1\GeV}$ from the VFF fits to single experiments and combinations}}%
      \csname LTb\endcsname
      \put(1529,967){\makebox(0,0){\strut{}All $e^+e^-$ (KLOE$''$), NA7}}%
      \put(1529,1317){\makebox(0,0){\strut{}All $e^+e^-$ (KLOE$''$)}}%
      \put(1529,1667){\makebox(0,0){\strut{}All $e^+e^-$, NA7}}%
      \put(1529,2017){\makebox(0,0){\strut{}All $e^+e^-$}}%
      \put(1529,2367){\makebox(0,0){\strut{}Energy scan}}%
      \put(1529,2717){\makebox(0,0){\strut{}KLOE$''$}}%
      \put(1529,3067){\makebox(0,0){\strut{}KLOE}}%
      \put(1529,3417){\makebox(0,0){\strut{}BaBar}}%
      \put(1529,3767){\makebox(0,0){\strut{}CMD-2}}%
      \put(1529,4117){\makebox(0,0){\strut{}SND}}%
    }%
    \gplbacktext
    \put(0,0){\includegraphics{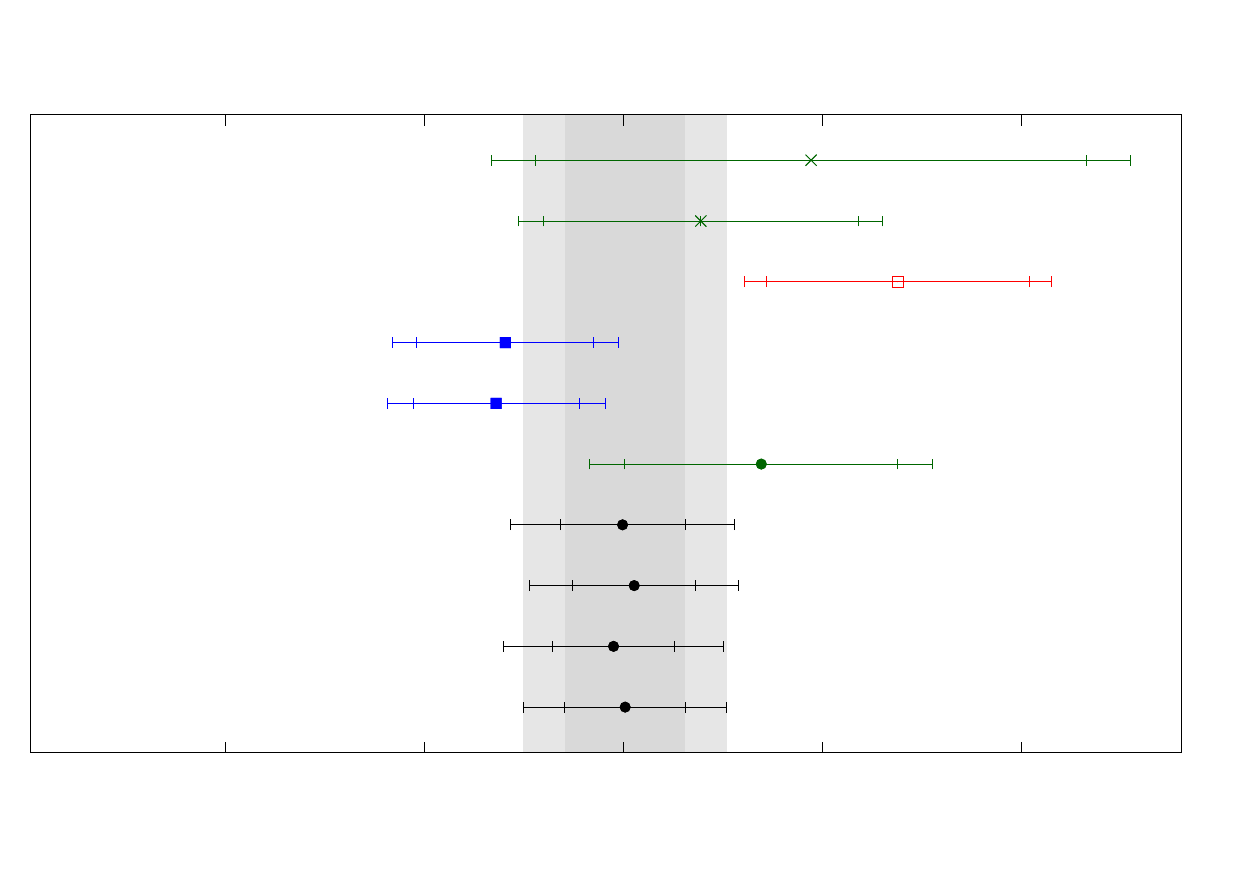}}%
    \gplfronttext
  \end{picture}%
\endgroup

%% file: sections/PhaseShifts.tex

\section{\boldmath Improved determination of the $\pi\pi$ $P$-wave phase shift $\delta_1^1$}

\begin{figure}[t]
	\centering
	\input{plots/d11fit}
	\caption{Fit result for the elastic $P$-wave $\pi\pi$ scattering phase shift $\delta_1^1$. The gray band shows the systematic uncertainty due to the parameters in the dispersive form factor representation, while the black error band representing the fit uncertainty is hardly visible. Note that, in contrast to Fig.~\ref{img:Delta11}, the band includes the systematic uncertainties related to
	the asymptotic continuation of the phase shift.}
	\label{img:Delta11Fit}
\end{figure}
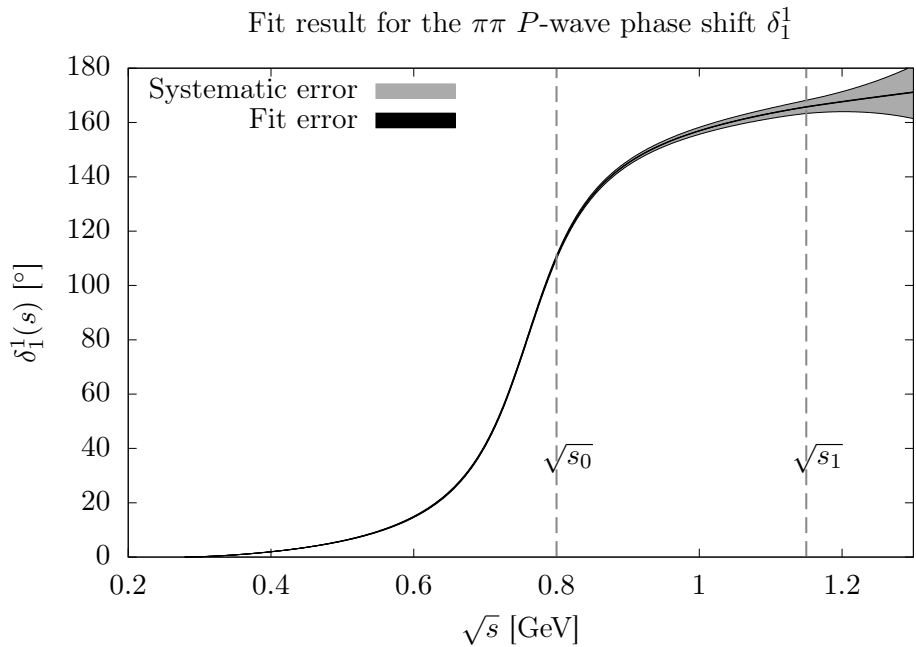

The final results for the $P$-wave phase shifts at $0.8$ and $1.15\GeV$ have already been given in Table~\ref{tab:FinalFitsCombinations}
\beq
\label{pipi_phase_final}
\delta^1_1(s_0)=110.4(1)(7)^\circ=110.4(7)^\circ,\qquad 
\delta^1_1(s_1)=165.7(0.1)(2.4)^\circ=165.7(2.4)^\circ.
\eeq
The correlations corresponding to the fit uncertainties and systematic errors are given by
\beq
	\mathrm{Corr}_\mathrm{fit}\big(\delta_1^1(s_0) , \delta_1^1(s_1) \big) = 0.66 , \qquad
	\mathrm{Corr}_\mathrm{syst}\big(\delta_1^1(s_0) , \delta_1^1(s_1) \big) = 0.83 .
\eeq
Both phase shift values are fully compatible with the ranges from~\eqref{eq:Delta11ParametersRoy}, with statistical uncertainties well below these errors.
In all cases, we observe that the fit results are extremely stable among different data sets, in such a way that by far the dominant uncertainty
now arises from systematic effects.

To arrive at~\eqref{pipi_phase_final}, we considered separately each of the 25 additional parameters in the Roy solution, see Sect.~\ref{sec:PhaseShiftInput},
and added all uncertainties in quadrature only at the very end of the calculation. However, very similar results emerge 
if instead one defines a smooth band around the central Roy solution by adding in quadrature all uncertainties other than those from $\delta^1_1(s_0)$ and $\delta^1_1(s_1)$. 
The propagation of the individual parameter uncertainties also allows one to identify the source of the relatively large systematic effects in $\delta^1_1(s_1)$, which are dominated by 
the asymptotics of the imaginary part of the partial wave, $\Im t^1_1$, as well as a low-energy parameter from the isospin-$0$ $S$-wave. 
This interrelation shows that for a global analysis of low-energy $\pi\pi$ phase shifts the 
role of these systematic effects, in particular the interplay with the Roy parameters corresponding to other isospin channels, needs to be carefully investigated. 
This will be addressed in future work.

In this regard, it might appear curious that the final error quoted for $\delta^1_1(s_1)$ is actually slightly larger 
than in~\eqref{eq:Delta11ParametersRoy}. However, one should keep in mind that in the solution of the Roy equations~\cite{Caprini:2011ky}, all the parameters are to be varied independently within their uncertainty ranges. With our fit of the VFF to data, the phase values~\eqref{pipi_phase_final} are no longer independent parameters but correlated with the remaining Roy parameters $p_i$. Linearizing the fit result around their central values $p_i^c$, we write
\begin{align}
	\delta_1^1(s_i) = \tilde \delta_1^1(s_i) + \sum_{k=1}^{25} a_{ik} ( p_k - p_k^c ) , \qquad i=0,1
\end{align}
in order to make the systematic dependence on the additional Roy solution parameters explicit. The values of $\tilde\delta_1^1(s_i)$ now only contain the systematic effects that are independent of the 25 additional parameters of the Roy solution:
\beq
	\label{eq:pipi_phase_independent}
	\tilde\delta^1_1(s_0)=110.4(1)(3)^\circ=110.4(3)^\circ,\qquad 
	\tilde\delta^1_1(s_1)=165.7(1)(5)^\circ=165.7(5)^\circ,
\eeq
and only these much smaller errors constitute the irreducible systematic effects derived from the VFF, while the rest, at least in principle, can
be improved with additional input for the 25 remaining Roy parameters. 
In particular, this separation clearly shows the improvement in the determination of the phase shift compared to the independent parameter ranges~\eqref{eq:Delta11ParametersRoy}.

As illustrated by Fig.~\ref{img:Errors}, these issues are immaterial for the HVP contribution, so that for the present application
we do not attempt to reduce the systematic errors further. The present status of the $P$-wave phase shift, corresponding to~\eqref{pipi_phase_final},
is illustrated in Fig.~\ref{img:Delta11Fit}. As expected, the band characterizing the systematic uncertainties widens rapidly above $1.15\GeV$, while
throughout the contribution of the statistical error is completely negligible.

%% file: plots/d11fit.tex
\begingroup
  \makeatletter
  \providecommand\color[2][]{%
    \GenericError{(gnuplot) \space\space\space\@spaces}{%
      Package color not loaded in conjunction with
      terminal option `colourtext'%
    }{See the gnuplot documentation for explanation.%
    }{Either use 'blacktext' in gnuplot or load the package
      color.sty in LaTeX.}%
    \renewcommand\color[2][]{}%
  }%
  \providecommand\includegraphics[2][]{%
    \GenericError{(gnuplot) \space\space\space\@spaces}{%
      Package graphicx or graphics not loaded%
    }{See the gnuplot documentation for explanation.%
    }{The gnuplot epslatex terminal needs graphicx.sty or graphics.sty.}%
    \renewcommand\includegraphics[2][]{}%
  }%
  \providecommand\rotatebox[2]{#2}%
  \@ifundefined{ifGPcolor}{%
    \newif\ifGPcolor
    \GPcolorfalse
  }{}%
  \@ifundefined{ifGPblacktext}{%
    \newif\ifGPblacktext
    \GPblacktexttrue
  }{}%
  \let\gplgaddtomacro\g@addto@macro
  \gdef\gplbacktext{}%
  \gdef\gplfronttext{}%
  \makeatother
  \ifGPblacktext
    \def\colorrgb#1{}%
    \def\colorgray#1{}%
  \else
    \ifGPcolor
      \def\colorrgb#1{\color[rgb]{#1}}%
      \def\colorgray#1{\color[gray]{#1}}%
      \expandafter\def\csname LTw\endcsname{\color{white}}%
      \expandafter\def\csname LTb\endcsname{\color{black}}%
      \expandafter\def\csname LTa\endcsname{\color{black}}%
      \expandafter\def\csname LT0\endcsname{\color[rgb]{1,0,0}}%
      \expandafter\def\csname LT1\endcsname{\color[rgb]{0,1,0}}%
      \expandafter\def\csname LT2\endcsname{\color[rgb]{0,0,1}}%
      \expandafter\def\csname LT3\endcsname{\color[rgb]{1,0,1}}%
      \expandafter\def\csname LT4\endcsname{\color[rgb]{0,1,1}}%
      \expandafter\def\csname LT5\endcsname{\color[rgb]{1,1,0}}%
      \expandafter\def\csname LT6\endcsname{\color[rgb]{0,0,0}}%
      \expandafter\def\csname LT7\endcsname{\color[rgb]{1,0.3,0}}%
      \expandafter\def\csname LT8\endcsname{\color[rgb]{0.5,0.5,0.5}}%
    \else
      \def\colorrgb#1{\color{black}}%
      \def\colorgray#1{\color[gray]{#1}}%
      \expandafter\def\csname LTw\endcsname{\color{white}}%
      \expandafter\def\csname LTb\endcsname{\color{black}}%
      \expandafter\def\csname LTa\endcsname{\color{black}}%
      \expandafter\def\csname LT0\endcsname{\color{black}}%
      \expandafter\def\csname LT1\endcsname{\color{black}}%
      \expandafter\def\csname LT2\endcsname{\color{black}}%
      \expandafter\def\csname LT3\endcsname{\color{black}}%
      \expandafter\def\csname LT4\endcsname{\color{black}}%
      \expandafter\def\csname LT5\endcsname{\color{black}}%
      \expandafter\def\csname LT6\endcsname{\color{black}}%
      \expandafter\def\csname LT7\endcsname{\color{black}}%
      \expandafter\def\csname LT8\endcsname{\color{black}}%
    \fi
  \fi
    \setlength{\unitlength}{0.0500bp}%
    \ifx\gptboxheight\undefined%
      \newlength{\gptboxheight}%
      \newlength{\gptboxwidth}%
      \newsavebox{\gptboxtext}%
    \fi%
    \setlength{\fboxrule}{0.5pt}%
    \setlength{\fboxsep}{1pt}%
\begin{picture}(7200.00,5040.00)%
    \gplgaddtomacro\gplbacktext{%
      \csname LTb\endcsname
      \put(814,704){\makebox(0,0)[r]{\strut{}$0$}}%
      \put(814,1112){\makebox(0,0)[r]{\strut{}$20$}}%
      \put(814,1521){\makebox(0,0)[r]{\strut{}$40$}}%
      \put(814,1929){\makebox(0,0)[r]{\strut{}$60$}}%
      \put(814,2337){\makebox(0,0)[r]{\strut{}$80$}}%
      \put(814,2746){\makebox(0,0)[r]{\strut{}$100$}}%
      \put(814,3154){\makebox(0,0)[r]{\strut{}$120$}}%
      \put(814,3562){\makebox(0,0)[r]{\strut{}$140$}}%
      \put(814,3971){\makebox(0,0)[r]{\strut{}$160$}}%
      \put(814,4379){\makebox(0,0)[r]{\strut{}$180$}}%
      \put(946,484){\makebox(0,0){\strut{}$0.2$}}%
      \put(2011,484){\makebox(0,0){\strut{}$0.4$}}%
      \put(3076,484){\makebox(0,0){\strut{}$0.6$}}%
      \put(4141,484){\makebox(0,0){\strut{}$0.8$}}%
      \put(5206,484){\makebox(0,0){\strut{}$1$}}%
      \put(6271,484){\makebox(0,0){\strut{}$1.2$}}%
      \put(4034,1459){\makebox(0,0)[l]{\strut{}$\sqrt{s_0}$}}%
      \put(5898,1459){\makebox(0,0)[l]{\strut{}$\sqrt{s_1}$}}%
    }%
    \gplgaddtomacro\gplfronttext{%
      \csname LTb\endcsname
      \put(198,2541){\rotatebox{-270}{\makebox(0,0){\strut{}$\delta_1^1(s)$ [${}^\circ$]}}}%
      \put(3874,154){\makebox(0,0){\strut{}$\sqrt{s}$ [GeV]}}%
      \put(3874,4709){\makebox(0,0){\strut{}Fit result for the $\pi\pi$ $P$-wave phase shift $\delta_1^1$}}%
      \csname LTb\endcsname
      \put(2662,4206){\makebox(0,0)[r]{\strut{}Systematic error}}%
      \csname LTb\endcsname
      \put(2662,3986){\makebox(0,0)[r]{\strut{}Fit error}}%
    }%
    \gplbacktext
    \put(0,0){\includegraphics{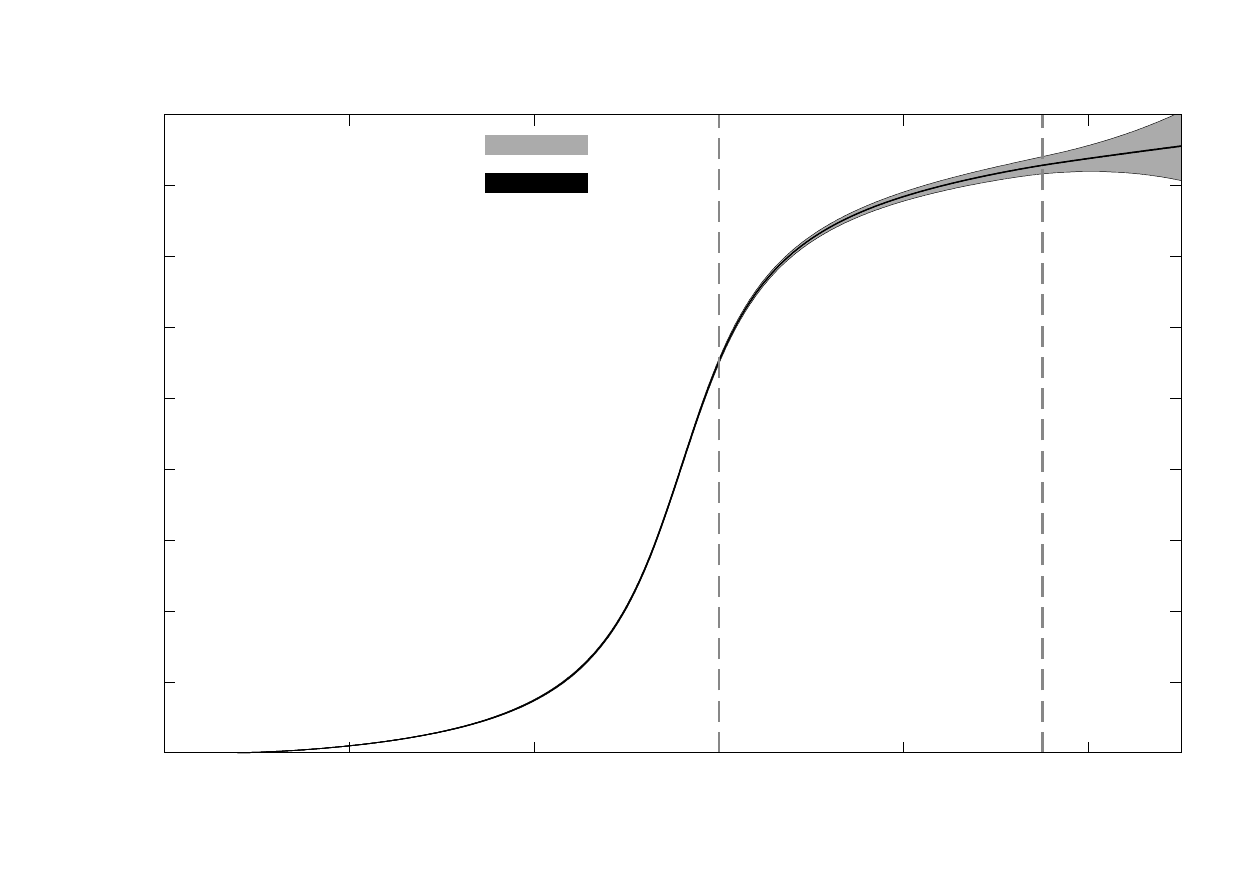}}%
    \gplfronttext
  \end{picture}%
\endgroup

%% file: sections/ChargeRadius.tex

\section{Charge radius of the pion}

\begin{table}[t]
	\centering
	\begin{tabular}{l c c c c c l}
	\toprule
	$\langle r^2_\pi\rangle\,[\fm^2]$	\\
	$N-1$			&	$1$	&	$2$	&	$3$	&	$4$	&	$5$	&	\;\;\;Central	\\
	\midrule
	SND 					&	$0.431(1)$		&	$0.434(4)$	&	$0.435(5)$	&	$0.426(6)$	&	$0.424(7)$	&	$0.431(1)(9)$	\\
	CMD-2					&	$0.429(1)$		&	$0.430(3)$	&	$0.430(4)$	&	$0.421(5)$	&	$0.421(6)$	&	$0.429(2)(9)$	\\
	BaBar					&	$0.432(1)$		&	$0.434(2)$	&	$0.433(2)$	&	$0.429(4)$	&	$0.427(5)$	&	$0.432(1)(7)$	\\
	KLOE					&	$0.428(1)$		&	$0.431(1)$	&	$0.430(1)$	&	$0.427(3)$	&	$0.428(4)$	&	$0.428(1)(4)$	\\
	KLOE$''$					&	$0.428(1)$		&	$0.431(1)$	&	$0.430(1)$	&	$0.427(3)$	&	$0.429(4)$	&	$0.428(1)(4)$	\\
	\midrule
	Energy scan				&	$0.431(1)$		&	$0.432(2)$	&	$0.433(3)$	&	$0.424(4)$	&	$0.423(4)$	&	$0.431(1)(9)$	\\
	All $e^+e^-$ (KLOE)			&	$0.429(1)$		&	$0.432(1)$	&	$0.431(1)$	&	$0.426(2)$	&	$0.425(3)$	&	$0.429(1)(5)$	\\
	All $e^+e^-$ (KLOE), NA7		&	$0.429(1)$		&	$0.432(1)$	&	$0.431(1)$	&	$0.426(1)$	&	$0.426(3)$	&	$0.429(1)(4)$	\\
	All $e^+e^-$ (KLOE$''$)		&	$0.429(1)$		&	$0.432(1)$	&	$0.431(1)$	&	$0.426(1)$	&	$0.425(3)$	&	$0.429(1)(5)$	\\
	All $e^+e^-$ (KLOE$''$), NA7	&	$0.429(1)$		&	$0.432(1)$	&	$0.431(1)$	&	$0.426(1)$	&	$0.427(3)$	&	$0.429(1)(4)$	\\
	\bottomrule
	\end{tabular}
	\caption{Charge radius corresponding to the fits to single time-like experiments and to combinations of data sets. The errors in the first five columns are the fit uncertainties, inflated by $\sqrt{\chi^2/\mathrm{dof}}$. The results in the last column correspond to $N-1=1$. The first error is the inflated fit uncertainty, the second error is the total uncertainty (which includes the variation $N-1=1\ldots5$).}
	\label{tab:Radius}
\end{table}

The charge radius of the pion, $\langle r_\pi^2\rangle$, is defined by the derivative of the VFF at $s=0$
\beq
	\label{eq:rpiSumRule}
	\langle r_\pi^2\rangle=6\frac{d F_\pi^V(s)}{d s}\bigg|_{s=0}=\frac{6}{\pi}\int_{4\mpi^2}^\infty d s\frac{\Im F_\pi^V(s)}{s^2},
\eeq
where the derivative is again evaluated via a dispersion relation. With the VFF determined from the fit to $e^+e^-\to\pi^+\pi^-$ data, this integral produces the results
collected in Table~\ref{tab:Radius}.
The uncertainties are dominated by the variation of the order of the conformal polynomial $N$. In particular, in contrast to the HVP contribution, the sum rule~\eqref{eq:rpiSumRule}
is directly sensitive to the phase of the conformal polynomial, which is only constrained by the \EL{} bound up to $1.15\GeV$. The oscillations of this phase for large values of $N$
impede a convergence of the extracted value for $\langle r_\pi^2\rangle$ in $N$, to the extent that the most stable results are obtained for small values of $N$ and, as seen from 
Table~\ref{tab:Radius}, $N-1=4$ and $5$ already begin to go astray. We still keep the 
full systematic variations for $N-1=1\ldots 5$, otherwise one would have to investigate in more detail the potential role of inelastic effects above the energy range constrained by the \EL{} bound.
As central values we quote the results for $N-1=1$, 
both motivated by the fact that the extrapolation uncertainties of the conformal polynomial beyond $1.15\GeV$ are expected to be smallest for the lowest order
and since these values happen to lie around the middle of the range given in Table~\ref{tab:Radius}.\footnote{For a central value defined by $N-1=4$, the final result would change to $\langle r_\pi^2 \rangle = 0.426(1)(6)\fm^2$, where the systematic error points entirely in the upward direction.} 
Our final result, including both time- and space-like data sets, reads
\beq
	\label{rpi_final}
	\langle r_\pi^2 \rangle = 0.429(1)(4)\fm^2=0.429(4)\fm^2.
\eeq
Within uncertainties, this value is consistent with the previous dispersive extraction $\langle r_\pi^2 \rangle=0.432(4)\fm^2$ from~\cite{Ananthanarayan:2017efc}, but the tension with the PDG average
$\langle r_\pi^2 \rangle=0.452(11)\fm^2$~\cite{PDG2018} is further exacerbated. However, as noted before~\cite{Hanhart:2016pcd},\footnote{We observe that the results in Table~\ref{tab:Radius} do not change within the fit uncertainty of $0.001\fm^2$ if VP is absorbed into the definition of the VFF, whereas significant effects do occur in the evaluation of $a_\mu^{\pi\pi}$.} this average does not contain any modern $e^+e^-\to\pi^+\pi^-$ data sets
and, if potentially model-dependent extractions from $e N\to e \pi N$~\cite{Bebek:1977pe,Liesenfeld:1999mv} were excluded, would be dominated by NA7 
$\langle r_\pi^2 \rangle=0.439(8)\fm^2$~\cite{Amendolia:1986wj}, in better agreement with~\eqref{rpi_final}. Indeed, if the NA7 data were in conflict with our dispersive determination, a simultaneous fit of time- and space-like data would not be possible. Our calculation therefore provides further evidence that the PDG average for $\langle r_\pi^2 \rangle$ needs to be revised.

%% file: sections/Conclusion.tex

\section{Summary and outlook}

Analyticity and unitarity imply strong constraints both on $\pi\pi$
scattering and the pion VFF. In this paper, we analyzed these constraints
comprehensively as regards consequences for the HVP contribution to the
anomalous magnetic moment of the muon, including both a consistent
implementation of the experimental uncertainties as well as the systematic
uncertainties associated with the dispersive representation.  The central
outcome of this study~\eqref{a_mu_final}
\[
a_\mu^{\pi\pi}|_{\leq 1\GeV}=495.0(1.5)(2.1)\times 10^{-10}=495.0(2.6)\times 10^{-10},
\]
shows that the main complications in such a representation,
arising from inelastic corrections and high-energy contributions, can be
controlled at a level that renders the dispersive approach a valuable
complementary perspective to the direct integration of the experimental
data. In particular, it provides the best controlled extrapolation down
to the two-pion threshold where data are less precise or just absent.

With the present analysis we have therefore laid the ground work to
consolidate the uncertainty estimate for the $\pi\pi$ channel in HVP. In
contrast to the direct integration, we cannot allow for the local inflation
of uncertainties since the dispersive fit function defines a global
constraint. For that reason it is critical that once possible uncertainties
in the energy calibration are taken into account all present data sets 
can be described in a statistically acceptable way, providing a strong
check on their internal consistency.  The combination of data sets then
follows in a straightforward way from the propagation of the uncertainties
incorporated in the covariance matrices provided by experiment, up to a
small inflation of the final uncertainties by
$\sqrt{\chi^2/\mathrm{dof}}\sim 1.1$ in the standard manner, a global scale
factor that is much smaller than the local scale factors up to 3 that are
required otherwise.  In this way, we have obtained a combination of the
available $e^+e^-$ data sets with minimal assumptions, relying only on the
global fit function that follows from QCD and the stated experimental
uncertainties. We expect that a future more detailed comparison with direct
integration should lead to a better understanding of the uncertainties in
the $\pi\pi$ channel and eventually make the overall error estimate more
robust.  As an added benefit, a dispersive approach has to be able to
accommodate space-like data sets at the same time, which not only provides
a further consistency check both on the data and the formalism, but in this
case actually leads to a modest reduction in uncertainty.

At this point, the experimental data on $e^+e^-\to\pi^+\pi^-$ are so
precise that the systematics of the dispersive representation begin to
dominate, an observation that becomes most apparent for the values of the
$\pi\pi$ phase shift extracted from the fit~\eqref{pipi_phase_final}
and~\eqref{eq:pipi_phase_independent}. For HVP the mismatch between
statistical and systematic errors is still relatively small, but for future
data sets improved variants of the dispersive representation could be
investigated. For instance, Fig.~\ref{img:Errors} shows that by far the
dominant effect arises from the order $N$ of the conformal polynomial that
describes the inelastic corrections, which we estimated very conservatively
by the maximum deviation found among all statistically meaningful fits.
Here, more precise data, in combination with the \EL{} bound, might allow
one to actually identify an optimal value or range of $N$ or even attempt
an explicit description within the dispersive approach of inelastic effects
in terms of physical processes and thereby significantly reduce the
associated uncertainty. Another issue that warrants further investigation concerns
the mass of the $\omega$, for which it would be important to clarify the origin of the
current mismatch between extractions from the $2\pi$ and $3\pi$ channels.

Beyond the HVP contribution, our results for the VFF are important for an
improved understanding of low-energy $\pi\pi$ scattering.  While most
recent work has focused on improving the isospin-$0$ $S$-wave, the input
used for the isospin-$1$ $P$-wave in solving the full system intertwined by
crossing symmetry actually goes back to by now outdated analyses of the
pion VFF.  Based upon the present work it will become possible to perform a
global analysis of low-energy $\pi\pi$ phase shifts including the stringent
constraints on the $P$-wave from the modern high-statistics
$e^+e^-\to\pi^+\pi^-$ experiments.